\documentclass[aps,prl,showpacs,footinbib,superscriptaddress,twocolumn,longbibliography,10pt]{revtex4-2}

\usepackage{pifont}
\usepackage{graphicx}
\usepackage{bm}
\usepackage{xcolor}
\usepackage[colorlinks=true,citecolor=blue]{hyperref}
\usepackage{natbib}
\usepackage{hyperref}
\usepackage{verbatim}
\usepackage{amsmath}
\usepackage{amssymb}
\usepackage{enumitem}

\usepackage[caption=false]{subfig}

\newcommand{\Rmnum}[1]{\expandafter\@slowromancap\romannumeral #1@}
\makeatother

\begin{document}

\title{Generalized Brillouin Zone Fragmentation}

\author{Haiyu Meng}
\affiliation{School of Physics and Optoelectronics, Xiangtan University, Xiangtan 411100, China}
\affiliation{Department of Physics, National University of Singapore, Singapore 117542}
\affiliation{Science, Mathematics and Technology (SMT), Singapore University of Technology and Design, Singapore 487372}

\author{Yee Sin Ang}
\email{yeesin\_ang@sutd.edu.sg}
\affiliation{Science, Mathematics and Technology (SMT), Singapore University of Technology and Design, Singapore 487372}

\author{Ching Hua Lee}
\email{phylch@nus.edu.sg}
\affiliation{Department of Physics, National University of Singapore, Singapore 117542}

\begin{abstract}
The Generalized Brillouin Zone (GBZ) encodes how lattice momentum is complex-deformed due to non-Hermitian skin accumulation, and has proved essential in restoring bulk-boundary correspondences.
However, we find that generically, the GBZ is neither unique nor well-defined if more than one skin localization direction or strength exists, even in systems with no asymmetric hoppings.
Instead, open boundary condition (OBC) eigenstates become complicated superpositions of multiple competing skin modes from ``fragments" of all possible GBZs solutions. 
We develop a formalism that computes the fragmented GBZ in a scalable manner, with fragmentation extent quantified through our newly-defined composition IPR and spectral relative entropy.
GBZ fragmentation is revealed to fundamentally challenge the notion of discontinuous phase transitions, since topological winding contributions from different GBZ fragments can ``melt away" at different rates. 
Phenomenologically, GBZ fragmentation also leads to edge localization in \emph{all} observables in energetically weighted ensembles such as thermal ensembles. This contrasts with conventional GBZs where the skin localization completely cancels in biorthogonal expectations.
Occurring universally in multi-mode non-Hermitian media, as we concretely demonstrate with photonic crystal simulations, GBZ fragmentation points towards a new paradigm that is essential for understanding the band structure and the topological and dynamical properties of diverse generic non-Hermitian systems. 
\end{abstract}

\date{\today}

\maketitle

\noindent\textcolor{blue}{\noindent{\it  Introduction.--} }
The non-Hermitian skin effect (NHSE) is a prominent demonstration of the emergence of long-ranged state amplification from the symbiosis of non-Hermiticity and non-reciprocity~\cite{yao2018edge,kunst2018biorthogonal,lee2019anatomy,xiong2018does,song2019non,longhi2019probing,xiao2020non,helbig2020generalized,okuma2020topological,li2020critical,arouca2020unconventional,lee2021many,shen2022non,li2021impurity,li2022non,zou2021observation,shimomura2024general,zhang2021observation,qin2022non,Jiang2023,yang2025beyond, yoshida2024non,li2024observation,wang2024non,bergholtz2021exceptional,lin2023topological,qin2024kinked,qin2023universal,shen2023observation,koh2025interacting,shen2024enhanced,zhao2025two,liu2024non,Yang2024eud,li2024observation,gliozzi2024many,liu2024localization,yoshida2024non,liu2024non,Gliozzi2024a,Li2022c,Sun2024a}. This non-locality is most evidently embodied in the modified bulk-boundary correspondence for lattices, particularly topological systems whose topological invariants have to be redefined on the generalized Brillouin zone (GBZ)~\cite{lee2019anatomy,yokomizo2019non,Jiang2023,zhang2021tidal,yang2019auxiliary,okuma2020topological,song2019non,zhang2023bulk,yao2018edge,yao2018non,yokomizo2019non,lee2019anatomy,zhang2020correspondence,xiong2018does,longhi2019probing,kawabata2020non,lee2020ultrafast,song2019realspace,wang2024amoeba,liu2024non}. Defined as the complex-deformed momentum path in which exponentially localized skin states can be treated on equal footing with conventional Bloch eigenstates, the GBZ paradigm has seen great success in the physical understanding of non-Hermitian band structures, including those encompassing impurities~\cite{li2021impurity,fang2023point,zhang2024pt,li2024universal,fu2023hybrid,yang2022designing}, disorders~\cite{liu2023modified,wang2021detecting,li2024topological,huang2025universal,mao2021boundary} or smooth spatial inhomogeneities~\cite{li2025phase,zheng2024experimental,sun2025lyapunov,lu2021magnetic,lee2021many,jiang2024dual,rafi2022unconventional,longhi2022self,jia2025unconventional}.



Despite its widespread adoption, the existing GBZ formulation is found to be highly inaccurate in increasingly many settings~\cite{li2020critical,yokomizo2021scaling, liu2020helical, guo2021exact,rafi2022critical,qin2023universal,liu2024non,song2022surprise,Jiang2023,zhang2024edge,zhang2025algebraic,li2025algebraic}. This is because it presupposes a well-defined exponential state accumulation, and will not strictly hold when distinct NHSE pumping channels compete. Such limitations of conventional GBZ theory is not just confined to coupled critical NHSE systems~\cite{li2020critical,yokomizo2021scaling, liu2020helical, guo2021exact,rafi2022critical,qin2023universal,liu2024non} -- NHSE competition becomes inevitable with more sophisticated unit cells or network connectivity. 
Such competition is indeed particularly rampant for higher-dimensional lattices with notoriously non-exponential state accumulation behavior\cite{song2022surprise,Jiang2023,zhang2024edge,zhang2025algebraic,li2025algebraic}, which can be viewed as extensive arrays of coupled chains. 


In this work, we develop a new ``GBZ fragmentation" framework, which permits the rigorous characterization of OBC bands \emph{without} assuming the conventional GBZ constraint $|z_\mu|=|z_\nu|$ on the solutions of $\text{Det}[H(z)-E\,\mathbb{I}]=0$, where $H(z)$, $z=e^{ik}$ is the momentum-space Hamiltonian. With surprising ubiquity, well-defined complex GBZ loops hence disintegrate into superpositions of ``fragmented" pieces in moderately complicated multi-mode media, as demonstrated in our photonic crystal simulations. New phenomenology from GBZ fragmentation include the edge localization of \emph{all} observable expectations in energetically weighted ensembles, as well as the fundamental erosion of sharp phase transitions.


\begin{figure}
    \includegraphics[width=\linewidth]{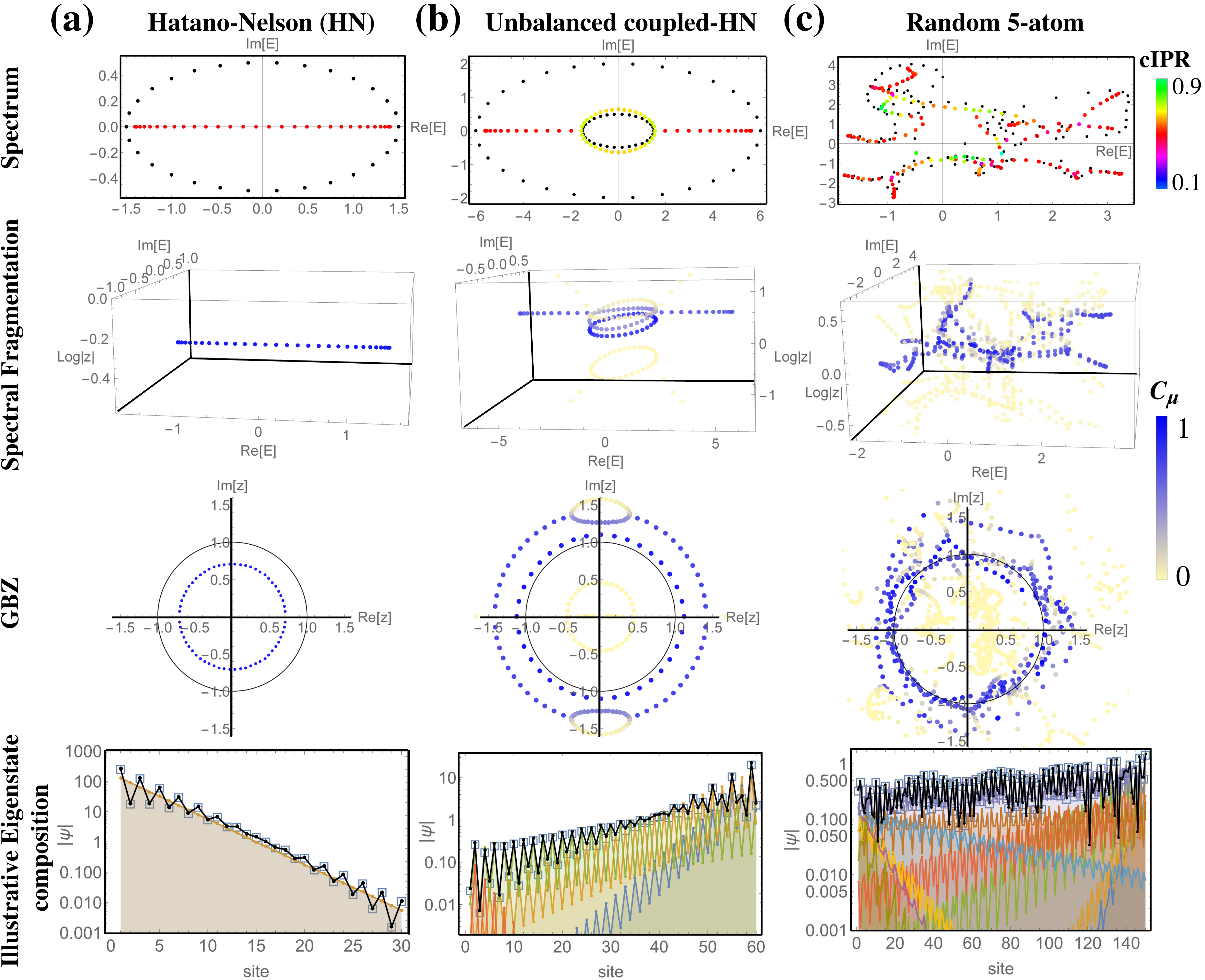}
    \caption{
    Synopsis of GBZ fragmentation in archetypal models, all at $N=30$. Top Row: OBC spectra colored by cIPR [Eq.~\ref{cIPRmain}] (departure from red indicates GBZ fragmentation), compared against PBC spectra (black). 2nd Row: Fragmentation of spectra into different GBZ contributions with distinct $\log|z_\mu|$, colored by composition weight $|c_\mu|^2$ [Eq.~\ref{psix}] evaluated at  $x_0=N/2$. 3rd Row: Corresponding GBZs, similarly colored. Bottom Row: Illustrative highest Im$(E)$ eigenstates $\psi_\text{OBC}(x)$ (black) and and their constituent $\phi_\mu(x)$ (colored). Models are (a) $H_\text{HN}(z)=z+h/z$ with $h=2$; 
   $H_\text{coupled-HN}$ [Eq.~\ref{cNHSEg}] with $g=2$, $\Delta=0.1$, $h=2$ respectively; (c) $5\times 5$ random hopping model $H_\text{rand}$ with coefficients in~\cite{suppmat}. 
    GBZ fragmentation is evident in (b) from the distinct broken GBZ rings (blue)
    , and becomes inevitable with sufficiently complicated random hoppings (d).
          }
    \label{Fig1}
\end{figure}

\noindent\textcolor{blue}{{\it Formalism for GBZ fragmentation.--} }
In conventional treatments of the NHSE, the effects of skin mode accumulation are encapsulated by complex-deforming the lattice momentum, yielding the so-called GBZ. Key to this construction is the assumption that each OBC eigenstate is characterized by a unique real-space decay length. 
Yet, there is no fundamental reason why multiple different skin decay lengths cannot simultaneously exist, in a phenomenon which we call "GBZ fragmentation". 

To understand the inevitability of GBZ fragmentation in generic lattices, 
let us recap how OBC eigenstates can be represented as superpositions of bulk eigensolutions. Consider an arbitrary lattice Hamiltonian
\begin{equation}
H(z)= \sum_{j=-q}^{p} H_j z^j,
\label{Hzpqmain}
\end{equation}
where $z=e^{ik}$, $k$ the lattice momentum.  For $b$ atoms per unit cell, $H_j$ is a $b\times b$ matrix representing the hopping amplitudes between atoms $j$ unit cells apart, with maximal left and right hopping ranges $q$ and $p$. In a finite $N$-unit cell setting, $H(z)$ is represented in real space 
by a $bN\times bN$ block Toeplitz matrix with $j$-th diagonals given by $H_j$; for periodic boundary conditions (PBCs), $j$ is taken modulo $N$. 

A key feature of OBC eigenstates $\psi_\text{OBC}(x)$ is that they are superpositions of 
bulk (PBC) eigensolutions $\phi_\mu(x) = z_\mu^x \varphi_\mu$, with inverse decay rates given by $-\log|z_\mu|$. Specifically, for a given $E_\text{OBC}$ eigenvalue,  
\begin{equation}
\psi_\text{OBC}(x)=\sum_\mu c_\mu \phi_\mu(x-x_0)=\sum_\mu c_\mu z_\mu^{x-x_0} \varphi_\mu,
\label{psix}
\end{equation}
where $z=z_\mu$ is a root of the characteristic bulk dispersion polynomial
\begin{equation}
\text{Det}[H(z)-E_\text{OBC}\,\mathbb{I}]=0,
\label{DetHzmain}
\end{equation}
and $\varphi_\mu$ is its corresponding $b$-component intra unit-cell eigenstate satisfying $H(z_\mu)\varphi_\mu=E\varphi_\mu$. The offset $x_0$ controls the position at which the relative bulk weights $c_\mu$ are evaluated: Setting $x_0=0$ or $N$ reveals the bulk eigensolution composition at the edges, while setting $x_0=N/2$ yields a symmetrically-defined bulk composition. 

Importantly, since each $\phi_\mu(x)$ is already a bulk eigensolution, the OBC eigenvalue problem is reduced to that of determining coefficients $c_\mu$ that satisfy the constraints from boundary hopping truncations, a problem that does scale with $N$. 
As detailed in Sect. I of~\cite{suppmat}, these constraints can be succinctly written as $\sum_\mu M_{\nu\mu}c_\mu=0$, where
 \begin{equation}
M_{\nu\mu}=\left\{\begin{matrix}
\sum\limits_{j=\nu}^{q} z_\mu^{\nu-j-1-x_0}H_{-j} \varphi_\mu\, \quad 1\leq \nu \leq q,
\\
\sum\limits_{j=p+q-\nu+1}^{p}z_\mu^{N-x_0-p-q+\nu+j-1} H_{j} \varphi_\mu, \, q+1\leq \nu \leq q+p. 
\end{matrix}\right.
\label{M}
\end{equation}
In particular, $E_\text{OBC}$ belongs to the OBC spectrum iff $\text{det} M=0$. As such, the coefficient vector $\bold c =\{c_\mu\}$ can be obtained by solving for the kernel of $M$, and are \emph{not} fundamentally restricted to having only two non-vanishing entries, as in conventional GBZs. Conventionally, only the pair $c_\mu, c_{\mu'}$ satisfying $|z_\mu|=|z_{\mu'}|$ is allowed to exist, since having two eigensolutions with equal decay rates allow OBCs to be satisfied at both boundaries using the same superposition weights $c_\mu, c_{\mu'}$, even in the $N\rightarrow \infty$ limit. In such cases, $M$ is invariably dominated by two particular columns $\mu,\mu'$.


However, beyond the simplest models, the kernel of $M$ can be much more intricate, even depending strongly on $N$~\cite{suppmat}. Physically, that corresponds to competing skin accumulations within the lattice that can non-trivially distort the non-Bloch picture. 
When the non-vanishing $c_\mu$ coefficients correspond to more than one inverse skin decay rate $-\log|z_\mu|$, we say that the GBZ has \emph{fragmented} -- this is vividly illustrated in Fig.~\ref{Fig1}, where the spectra (Top Row) of illustrative models break up into multiple constituent $|z|$ ``layers" (2nd Row). To quantify the extent of GBZ fragmentation at each $E_\text{OBC}$, we introduce the composition-IPR
\begin{equation}
\text{cIPR}=\sum_\mu |c_\mu|^4/\left(\sum_\mu |c_\mu|^2\right)^2,
\label{cIPRmain}
\end{equation}
where $1/\text{cIPR}$ measures the effective number of fragmented GBZ pieces. For conventional GBZs, the cIPR typically lies between $0.5$ and $1$ (it is $0.5$ when the two nonzero coefficients $c_\mu$ are equally large). 

\noindent\textcolor{blue}{{\it Occurrence of fragmented GBZs.--} }
As a warm-up, we first examine the simplest Hatano-Nelson (HN) model $H_\text{HN}(z)=z+h/z$, which does \emph{not} exhibit GBZ fragmentation. Its left/right open 
 boundary constraints can be succinctly encoded as $M\bold c=\bold 0$, with $M$ being the $2\times 2$ matrix $\left(\begin{matrix}
h/z_1^{x_0+1} & h/z_2^{x_0+1} \\
z_1^{N-x_0} & z_2^{N-x_0}\end{matrix}\right)$ in the bulk solution basis [Eq.~\ref{psix}]~\cite{suppmat}. 
Simultaneously satisfying both boundary conditions 
requires the basis coefficients $\bold c=(c_1,c_2)^T$ to satisfy $c_1/c_2= -\left(z_2/z_1\right)^{N-x_0}$. But Det$M=0$ necessitates $z_1^{N+1}=z_2^{N+1}$, which fixes $|c_1/c_2|=1$ for all $N$. Furthermore, as $E_\text{OBC}=z+h/z$ is symmetric under $z\leftrightarrow h/z$, the GBZ is determined to be $z_1=z_2^*=\sqrt{h}e^{ik}$, a classic example of a non-fragmented GBZ~\footnote{This also holds for the non-Hermitian SSH model, which has the same form of characteristic dispersion polynomial.}. 

As shown in Fig.~\ref{Fig1}(a), both the numerical spectrum and GBZ of $H_\text{HN}$ indeed agrees with $|z|=|z_1|=|z_2|=\sqrt{h}$. This is corroborated by the OBC eigenstate profile (black) in the bottom plot, whose bulk eigensolutions (orange) decay exactly at that unique rate $-\log\sqrt h$. Generalizing $H_\text{HN}$ to more bands and further hoppings may distort the GBZ and shift the effective lattice momentum points [See Sect. I.B. of ~\cite{suppmat}], but the GBZ loop structure remains fundamentally preserved.

However, this simple unique GBZ no longer holds once another NHSE accumulation length or direction is introduced. 
Consider two coupled oppositely-directed HN chains weighted by $g,g^{-1}$, as presented in Figs.~\ref{Fig1}(b) with $g=2$: 
\begin{equation}
H_\text{coupled-HN}(z) = \left(\begin{matrix}
g\left(z+h/z\right) & \Delta \\
\Delta & \left(1/z+hz\right)/g
\end{matrix}\right).
\label{cNHSEg}
\end{equation}
With Det$M$ being a degree-4 Laurent polynomial, there are four $z_\mu$ solutions. But they contribute very unequally to $\psi_\text{OBC}$, as shown in the GBZ spectral plots (middle two rows, colored yellow/blue according to low/high composition $c_\mu$). Importantly, these $z_\mu$ are \emph{not} doubly degenerate (overlapping) anymore, such that each $E_\text{OBC}$ band corresponds to at least \emph{two} dominant GBZ loops. 

The inevitability of this fragmentation can be understood from the structure of the $M$ matrix. Focusing on $g=1$ for simplicity, note that the two sublattices $\uparrow,\downarrow$ of $H_\text{coupled-HN}(z)$ are related by inversion $z\rightarrow 1/z$, such that only two independent GBZs remain: $z_\pm$ and $1/z_\pm$. In their $c_\pm$ composition basis, 
\begin{equation}
M=\left(\begin{matrix}
 z_+^{-(x_0+1)}\varphi_+^\uparrow + z_+^{x_0+1}\varphi_+^\downarrow &  z_-^{-(x_0+1)}\varphi_-^\uparrow + z_-^{x_0+1}\varphi_-^\downarrow \\
 z_+^{-(x_0+1)}\varphi_+^\downarrow + z_+^{x_0+1}\varphi_+^\uparrow &  z_-^{-(x_0+1)}\varphi_-^\downarrow + z_-^{x_0+1}\varphi_-^\uparrow 
 \end{matrix}\right)
 \end{equation}
with Det$M=0$ giving the constraint [Sect. I C in~\cite{suppmat}]
\begin{equation}
\tanh^{-1}r_--\tanh^{-1}r_+=\tanh^{-1}\left(z^{N+1}_+\right)-\tanh^{-1}\left(z^{N+1}_-\right),
\label{tanh}
\end{equation}
where $
r_\pm=\frac{\varphi^\uparrow_\pm}{\varphi^\downarrow_\pm}
=\left(E_\text{OBC}-z_\pm-\frac{h}{z_\pm}\right)/\Delta$. 
Importantly, for nonzero $\Delta$ couplings, Det$M=0$ yields $|z_+|\neq |z_-|$, as shown rigorously in the Supplement~\cite{suppmat} by solving Eqs.~\ref{DetHzmain} and~\ref{tanh}. One implication is that $E_\text{OBC}$ (Top Row in Fig.~\ref{Fig1}(b)) can now be complex. This is unlike in isolated HN-chains, where $E_\text{OBC}$ is within the $(-2\sqrt{h},2\sqrt{h})$ real segment, such that both $z_+,z_-$ satisfy $E_\text{OBC}-z_\pm-\frac{h}{z_\pm}=0$, leading to $r_\pm=0$.


Despite their unequal decay profiles $z_\pm^x$, $z_\pm^{-x}$, these four GBZ solutions can still superpose to satisfy OBCs [Eq.~\ref{psix}], as shown in the bottom panel of Fig.~\ref{Fig1}(b). At the left boundary, the leading contribution to $\psi_\text{OBC}$ (square markers) is the $z_-^x$ solution (green). In ordinary GBZs, another bulk solution with the same decay rate must participate to make $\psi_\text{OBC}$ vanish at the boundary. But here, this is achieved through a nontrivial superposition of $z_+^x$ (red) and $1/z_-^x$ (orange), which conspires to make $\psi_\text{OBC}$ vanish. This still holds even if inversion symmetry is broken by $g\neq 1$, as is presented in Fig.~\ref{Fig1}(b). Here the GBZ additionally bifurcates into smaller loops at high $|\text{Im}E|$, with continuous varying compositions $c_\mu$ (yellow to purple). All 4 bulk eigensolutions now participate nontrivially to form $\psi_\text{OBC}$ (bottom), resulting in reduced cIPR (yellow-green spectral loop) i.e. greater GBZ fragmentation. 

Beyond these specific models, GBZ fragmentation generically occurs whenever the non-Hermitian hoppings give rise to more than one NHSE decay lengths. Take for instance $b$-component models with random hoppings between the atoms of adjacent unit cells: $H_\text{rand}(z)=A_pz+A_m/z+A_0$, where $A_p,A_m$ and $A_0$ are $b\times b$ random matrices with complex elements uniformly distributed within the $[-1-i,1+i]$ rectangle. As shown in Fig.~\ref{Fig1}(c) for an illustrative $b=5$-band model (detailed in~\cite{suppmat}), some $E_\text{OBC}$ invariably become green or magenta (top), indicative of significant "smearing" of its GBZ contributions $\varphi_\mu$ across different $|z_\mu|$ (blue, Middle Rows). Evidently, no clear GBZ loop is discernable, and the GBZ has indeed "fragmented" into pieces. Correspondingly, the OBC eigenstates $\psi_\text{OBC}(x)$ (bottom) also become complicated superpositions of many non-Bloch solutions $\varphi_\mu(x)$, with the conventional GBZ picture eroded.

\begin{figure}
    \includegraphics[width=\linewidth]{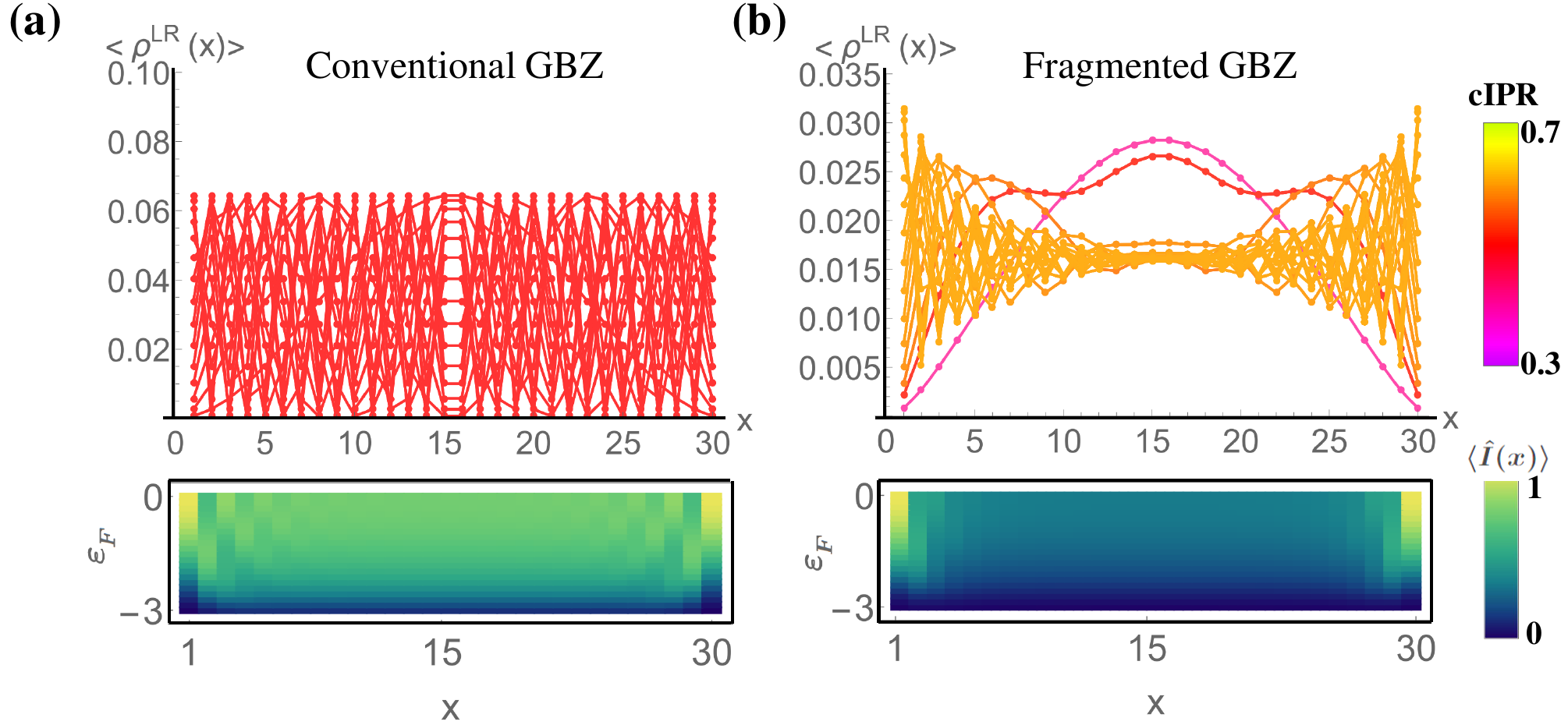}
\caption{Edge skin current unique to GBZ fragmentation: Illustrative models are (a) $H_\text{HN}(z)=z+2/z$ with unfragmented GBZ, and (b) $H_\text{coupled-HN}(z)$ [Eq.~\ref{cNHSEg}] with $h=2,\Delta=0.1,g=1$, that exhibits GBZ fragmentation. 
Top Row: The biorthogonal spatial density $\rho^{LR}(x)$ of all eigenstates, colored by cIPR [Eqs.~\ref{M},~\ref{cIPRmain}] at boundary reference position $x_0=0$. Bottom: Physical tunneling current $\langle \hat I(x)\rangle$ [Eq.~\ref{hatO}] at inverse temperature $\beta=5$, for various Fermi energies $\varepsilon_F$. 
In (a), the NHSE completely cancels in a conventional GBZ, yielding uniform eigenstate density profiles (red) and relatively constant tunneling current. In (b), GBZ fragmentation causes most eigenstates (orange) to acquire large edge-localized densities, such that the physical current  $\langle \hat I(x)\rangle$ in a thermal ensemble also become significantly enhanced at the edges (yellow). 
    }
    \label{fig:biortho}
\end{figure}

\noindent{\it Physical implications of GBZ fragmentation.--}
The fragmentation of a unique non-Bloch GBZ is not just of mathematical significance. It also profoundly affects all biorthogonal expectation values, which apply to all measured quantities that are energetically weighted, i.e. in thermal ensembles. At temperature $\beta^{-1}$, the expectation of a generic observable $\hat{O}(x)$ under OBCs is~\cite{suppmat}
\begin{eqnarray}
&&\langle\hat O(x)\rangle = \frac1{Z}\sum_j e^{-\beta E_j}\langle\psi^L_{j,\text{OBC}}|\hat O|\psi^R_{j,\text{OBC}}\rangle\notag\\
&=& \frac1{Z}\sum_j\sum_{\mu\mu'}  e^{-\beta E_j}c_{\mu,j} c^*_{\mu',j}\left(\frac{z_{\mu,j}}{z_{\mu',j}}\right)^{x}\langle \varphi_{\mu',j}|\hat O|\varphi_{\mu,j}\rangle,
\label{hatO}
\end{eqnarray}
since the left and right eigenvectors $|\psi^{L/R}_{j,\text{OBC}}(x)\rangle = \sum_\mu c_{\mu,j} z_{\mu,j}^{\mp x}|\varphi_{\mu,j}\rangle$ are needed to span the resolution of the identity. Here $Z=\sum_j e^{-\beta E_j}$ where $j$ labels the OBC eigenenergies~\footnote{The combination of the composition coefficients $c_j$ and non-Bloch factors $z_{\mu,j}$ is independent of the reference point $x_0$.}. 

Notably, from the $\left(\frac{z_{\mu,j}}{z_{\mu',j}}\right)^{x}$ factor in Eq.~\ref{hatO}, GBZ fragmentation will give rise to edge accumulation because some $(\mu,j)$, $(\mu',j)$ pairs do not obey $|z_{\mu,j}|=|z_{\mu',j}|$. This contrasts with usual case with conventional GBZs (i.e. HN and nH-SSH models)~\cite{kunst2018biorthogonal,zhang2024biorthogonal}, in which the NHSE cancels off in $\langle\hat O(x)\rangle$. 
Shown in Fig.~\ref{fig:biortho} are the profiles for the local eigenstate densities $\rho^\text{LR}_j(x) = \langle\psi^L_{j,\text{OBC}}|\psi^R_{j,\text{OBC}}\rangle=
\sum_{\mu\mu'}c_{\mu,j} c^*_{\mu',j}\left(\frac{z_{\mu,j}}{z_{\mu',j}}\right)^{x}\langle \varphi_{\mu',j}|\varphi_{\mu,j}\rangle$ and their finite-temperature tunneling current expectation $\langle\hat I(x)\rangle = i\langle b^\dagger_{x+1}b_x-b^\dagger_x b_{x+1}\rangle$, for (a) $H_\text{HN}$ without GBZ fragmentation, and (b) $H_\text{coupled-HN}$ with GBZ fragmentation. Saliently, $H_\text{HN}$ exhibits uniform $\rho^\text{LR}(x)$ density despite the NHSE, and its tunneling current $\langle\hat I(x)\rangle$ is only weakly non-uniform due to edge effects. But for $H_\text{coupled-HN}$, most eigenstates (orange) exhibit GBZ fragmentation and possess exponentially growing edge $\rho^\text{LR}(x)$; correspondingly, its tunneling current is much stronger $\langle\hat I(x)\rangle$ at the edges (yellow).

\noindent\textcolor{blue}{{\it Continuous topo. transitions from GBZ fragmentation.--} }Interestingly, GBZ fragmentation also allows a topological transition to occur in a fundamentally continuous manner. Instead of a distinct topological winding jump, a band can distribute across multiple GBZ contributions, and have its winding simply ``melt" away. Consider weakly coupling two topological chains with opposite hopping asymmetries $\gamma$:
$H^\text{ext-SSH}_{\gamma}(z) = $$\left[(1+t')\cos k +t\left(\gamma+\gamma^{-1}\right)/2\right]\sigma_x+ \left[(1-t')\sin k +it\left(\gamma-\gamma^{-1}\right)/2\right]\sigma_y$ with opposite hopping asymmetries $\gamma$:
\begin{equation}
H_\text{coupled-topo}=H^\text{ext-SSH}_{\gamma}\oplus H^\text{ext-SSH}_{\frac1\gamma}+\Delta\,  \mathbb{I}\otimes \sigma_x.
\label{Cext}
\end{equation}
With the non-perturbative effects of the coupling $\Delta$ captured by the GBZ, the topology of $H_\text{coupled-topo}$ is characterized by the winding of $U(z)= \sqrt{\frac{\frac{t}{\gamma} +\frac1{z}+t'z}{t\gamma + z +\frac{t'}{z}}}$ of $H^\text{ext-SSH}_{\gamma}$~\cite{yao2018edge}, evaluated on the GBZ. To accommodate for GBZ fragmentation, this winding is generalized to 
\begin{equation}
W=\frac1{2\pi i} \sum_\mu\oint_{z=z_\mu(\theta)}^{'}|c_\mu(\theta)|^2d[\log U(z_\mu(\theta))],
\label{Wmain}
\end{equation}
where $z_\mu(\theta)$ are the GBZ trajectories generated by periodically threading a flux $\theta$  [Sect. III of~\cite{suppmat}], weighted by GBZ composition $|c_\mu(\theta)|^2$. The $'$ in $\oint'$ directs the contours such that they reduce to the usual topological winding in the unfragmented limit.

\begin{figure}
    \subfloat[$\Delta=10^{-9}$, $W=0.998$]{\includegraphics[width=.33\linewidth]{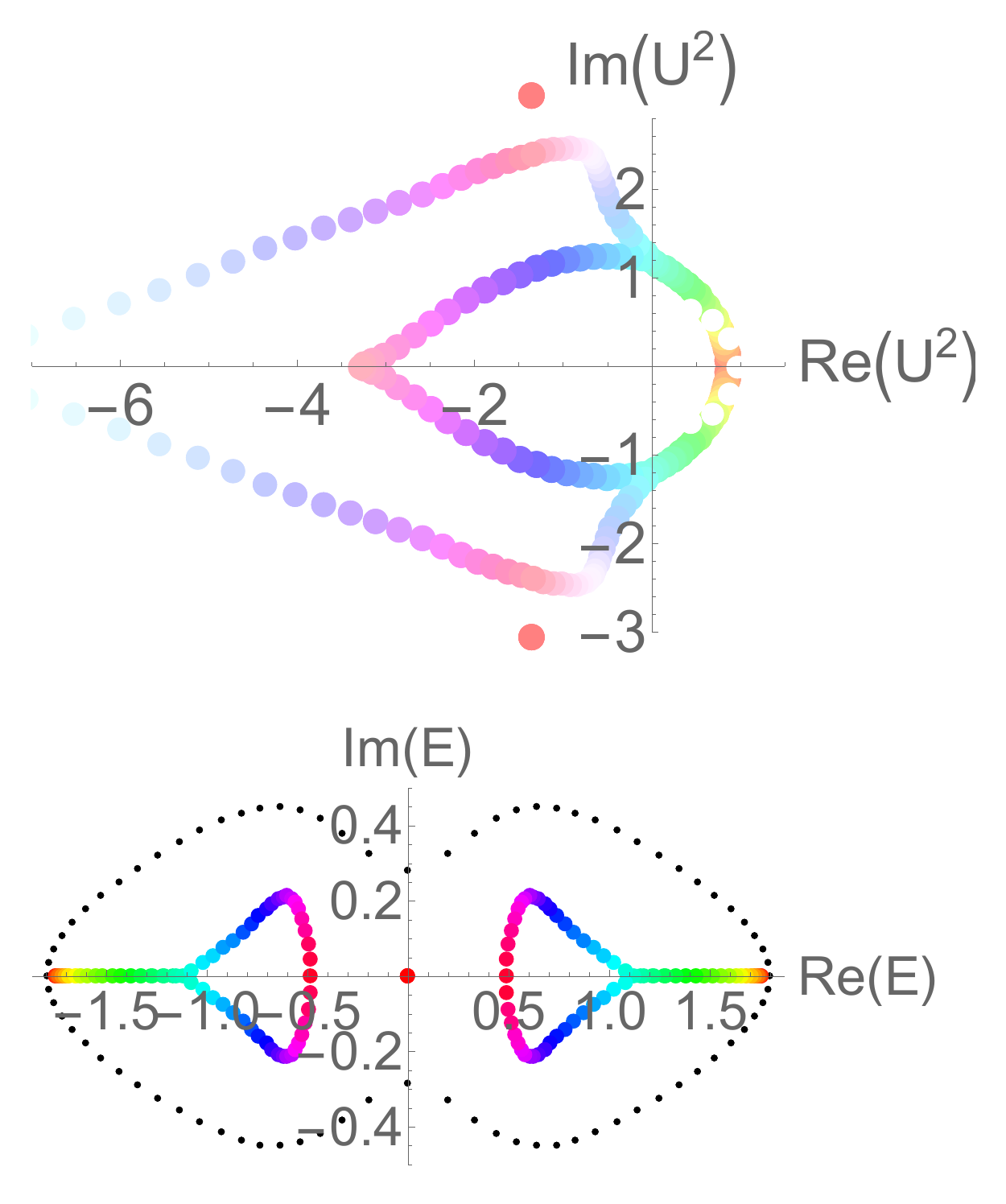}}
    \subfloat[$\Delta=10^{-4}$, $W=0.77$]{\includegraphics[width=.33\linewidth]{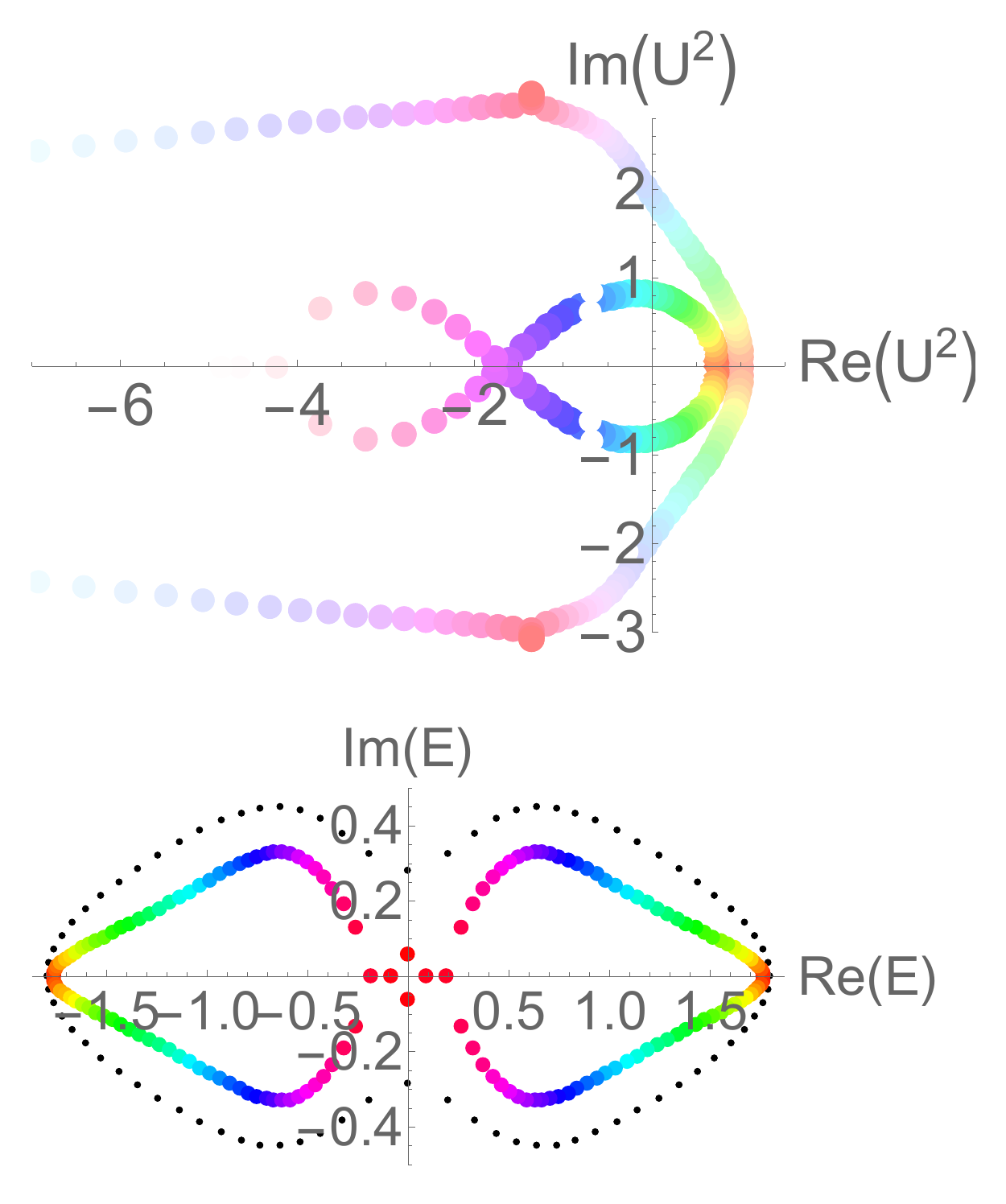}}
    \subfloat[$\Delta=5\times 10^{-3}$, $W=0.18$]{\includegraphics[width=.33\linewidth]{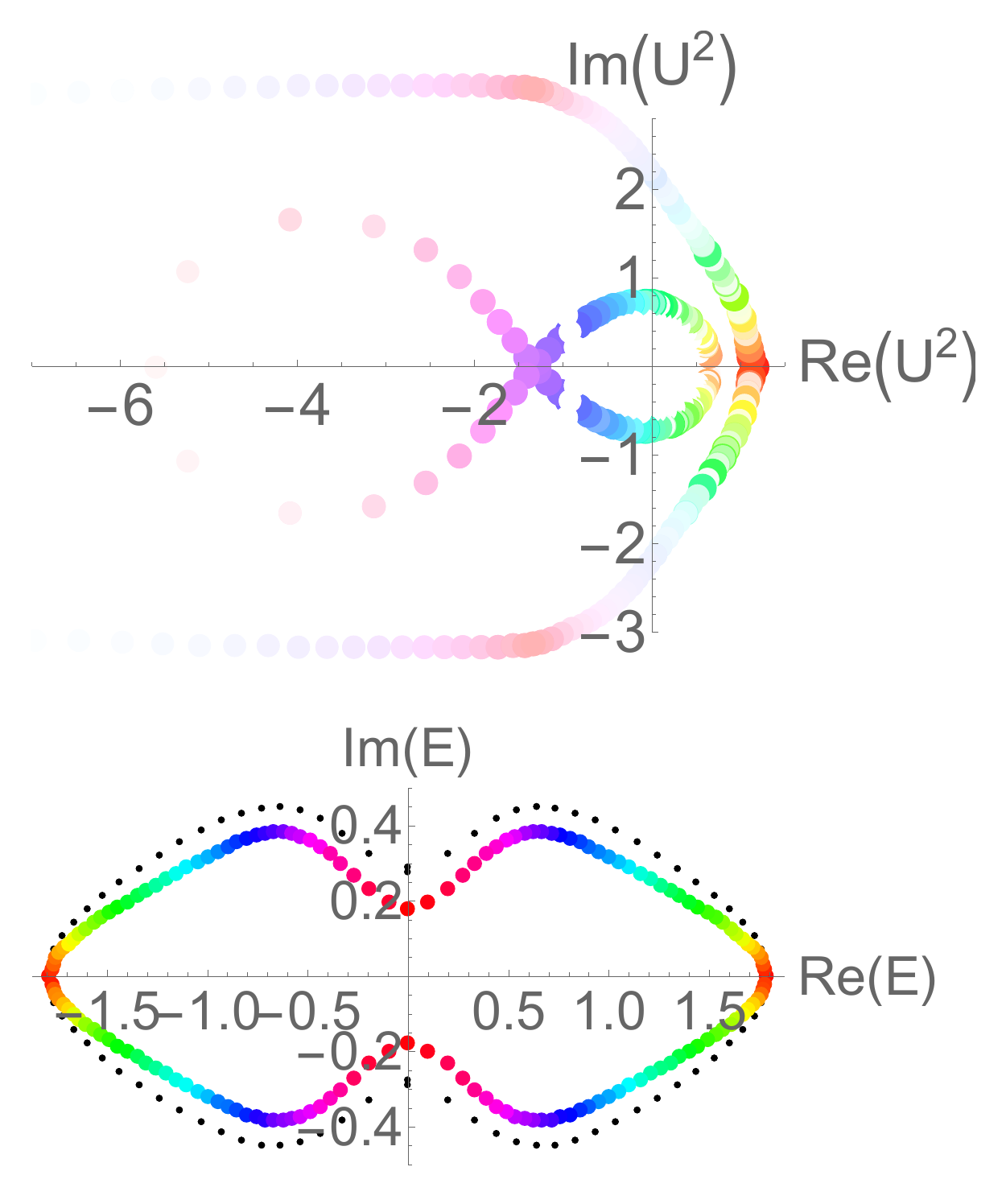}}
    \caption{Intrinsically continuous topological transition of $H_\text{coupled-topo}$ [Eq.~\ref{Cext}] from GBZ fragmentation. The $U^2(z_\mu)$ trajectories (Top Row) are colored by their Re$(E)$ (Bottom Row), with intensity proportional to their GBZ weight $|c_\mu|^2$. Increasing $\Delta$ causes the $U^2(z_\mu)$ windings to slowly fade away, while the spectrum gradually transitions from a real topological gap to local imaginary gap (red cross near zero mode, Center). Parameters are $\gamma=2$,$t=0.6,t'=0.1$. }
    \label{fig:winding}
\end{figure}

Shown in Fig.~\ref{fig:winding} are the squared $U(z_\mu(\theta))$ trajectories (Top) and corresponding OBC spectra (Bottom) of $H_\text{coupled-topo}$ across a topological transition, color coded by Re$(E)$ and weighted by $|c_\mu|^2$. Increasing $\Delta$ causes the gap to close and then reopen, with the topological zero modes destroyed. But notably, $W$ does not decrease abruptly to $0$: Instead, it drops gradually as the the outer GBZ trajectories in $U(z_\mu(\theta))$ fade away, such that no winding eventually remains. Indeed, the numerical spectral gap does not close at any well-defined $\Delta$, with slowly "switching" bands. The intrinsic non-quantization of $W$ stems from the continuous fragmentation weights $|c_\mu|^2$, and is generically unavoidable unless no GBZ fragmentation occurs.

\begin{figure}
    \centering
    \includegraphics[width=1\linewidth]{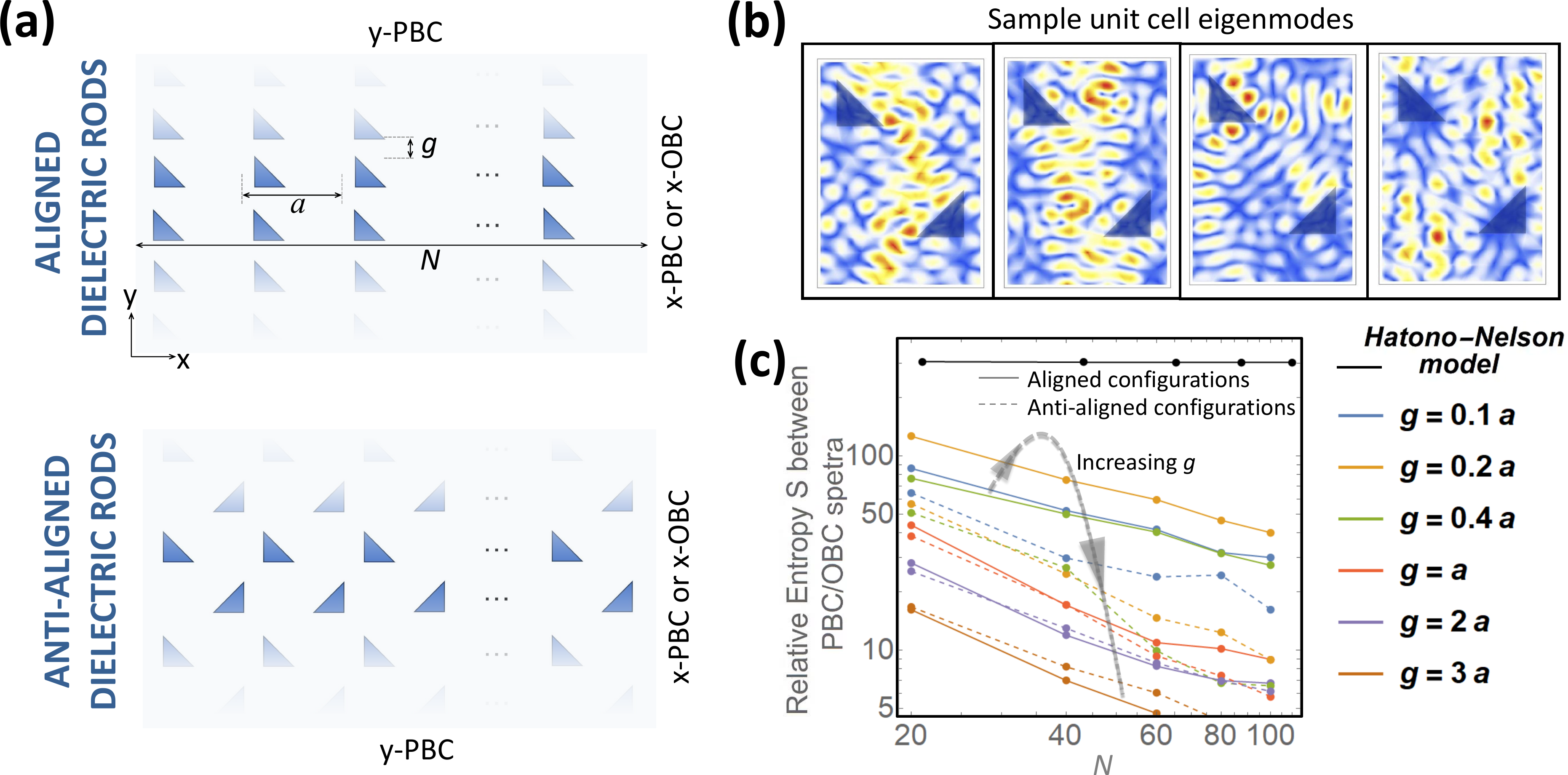}
    \caption{(a) GBZ fragmentation occurs in simple metamaterial structures with multiple eigenmodes per unit cell, as in our illustrative photonic crystals (see~\cite{suppmat} for details). (b) Highly complicated diffuse unit cell eigenmodes lead to multi-component heavily-coupled effective lattices reminiscent of $H_\text{rand}$. (c) The relative entropy $S$, which measures the contrast between PBC/OBC spectra, is much lower for anti-aligned cases (dashed) at small $g$, and also universally decreases with system size $N$. It is far lower than that of $H_\text{HN}$ (black), which experiences clean GBZ deformations. }
    \label{fig:PC}
 \end{figure}

\noindent\textcolor{blue}{{\it Physical realization of fragmented GBZs.--} }
As previously shown, GBZ fragmentation do not just occur in specific coupled models, but arise generically in non-Hermitian media with sufficiently complicated unit cell structure. Lossy photonic crystals (PhCs) [Fig.~\ref{fig:PC}(a)] constitute ideal physical platforms for that, with each unit cell naturally imbued with multiple cavity modes with complicated mutual couplings [Fig.~\ref{fig:PC}(b)]. Breaking the reciprocity of a 2D PhC by specializing to a fixed transverse momentum sector $k_y$, the resultant effective 1D model exhibits NHSE pumping with direction controlled by the mirror asymmetry of the unit cell structure~\cite{hofmann2020reciprocal,zhong2021nontrivial,xu2024skin}. 

We consider PhCs under aligned/anti-aligned configurations [Fig.~\ref{fig:PC}(a)], which reduce to 1D two-chain models with parallel and competing NHSE, coupled more strongly for smaller transverse spacing $g$. As shown in Sect. V of~\cite{suppmat}, their eigenfrequency spectra exhibits stark resemblance to that of $H_\text{rand}$ [Fig.~\ref{Fig1}(d)], with complicated random-looking structures and characteristic dependence on $N$. To quantify the NHSE extent in such intricate GBZ-fragmented spectra, we convert the PBC/OBC spectra to distributions $p_\text{PBC/OBC}(\omega)$, and compute their symmetrized relative entropy~\cite{cover1991entropy} $S=\int \left(p_\text{PBC}(\omega)-p_\text{OBC}(\omega)\right)\log\frac{p_\text{PBC}(\omega)}{p_\text{OBC}(\omega)}d\omega$ [Sect. IV of~\cite{suppmat}]. 
Indeed, from Fig.~\ref{fig:PC}(c), increasing $N$ universally leads to lower $S$ (suppressed NHSE), as expected from stronger NHSE competition under both configurations. At smaller $g$ where inter-chain coupling dominates, anti-aligned cases (dashed) also exhibits much lower $S$, consistent with GBZ fragmented systems~\cite{suppmat}.


\noindent\textcolor{blue}{{\it Discussion.--} }
GBZ fragmentation occurs ubiquitously once we move beyond simple exact models, into realistic multi-mode or polycrystalline non-Hermitian media. Often hidden behind basis transforms~\footnote{As a minimal example, gain/loss terms in $i\sigma_z$ become asymmetric off-diagonal couplings in $i\sigma_y$ upon a Pauli matrix rotation~\cite{xue2022non,zhang2024observation}.}, its requisite competitive mechanism does not require explicitly asymmetric hoppings, and would particularly emerge from extended i.e. Moiré~\cite{shao2024non,esparza2025exceptional} or higher dimensional unit cells~\cite{song2022surprise,Jiang2023,zhang2024edge,zhang2025algebraic,li2025algebraic,zhu2023higher}. 

Physically, GBZ fragmentation manifests as the edge localization of any observable i.e. current, density or magnetization in a thermal ensemble. This contrasts starkly with that in conventional GBZs, where the NHSE profile completely cancels in biorthogonal expectations. Fundamentally, the distributed occupancy of GBZ fragments also erodes the notion of an unique band structure, challenging the existence of discontinuous phase transitions. Open directions include the engineering of parent models with desired GBZ fragments, as well as the prospect of polymorphic ``higher-rank" GBZs when dim(ker($M$))$>1$.




\noindent{{\it Acknowledgements.--} }
HYM acknowledges support by the Natural Science Foundation of Hunan Province (2023JJ40612). CHL acknowledges support from the Ministry of Education, Singapore (MOE Tier-II award number: MOE-T2EP50222-0003) and (Tier-I WBS number: A-8002656-00-00). YSA acknowledges support from the Ministry of Education, Singapore, under the award number MOE-T2EP50224-0021.

\bibliography{ref}

\clearpage

\onecolumngrid
\subsection*{\normalsize Supplementary Materials for ``Generalized Brillouin Zone Fragmentation''}

\setcounter{equation}{0}
\setcounter{figure}{0}
\setcounter{table}{0}
\setcounter{section}{0}
\setcounter{page}{1}
\renewcommand{\theequation}{S\arabic{equation}}
\renewcommand{\thefigure}{S\arabic{figure}}
\renewcommand{\thesection}{S\arabic{section}}
\renewcommand{\thepage}{S\arabic{page}}

This supplement contains the following material in the following sections:
\begin{enumerate}[label=\Roman*.]
\item Elaboration of the $M$ matrix formalism for computing GBZ fragmentation coefficients and the composition IPR; Detailed examples for commonly used non-Hermitian models, as well as paradigmatic models that exhibit GBZ fragmentation; Discussion on the impossibility of extended continua of higher-fold GBZ degeneracies. 
\item Numerical results for the variation of fragmented GBZs with system size, and the inevitability of GBZ fragmentation with sufficiently many random hoppings; Detailed analysis of how the bulk solutions can superposition to satisfy the OBCs.
\item Physical consequences GBZ fragmentation: Its generic impact on biorthogonal quantities, particularly the current; Intrinsically murky topological transitions due to GBZ fragmentation.
\item Converting complex spectra into 2D distributions and characterizing the extent of their difference through their relative entropy; calibration and benchmarking with PBC vs OBC spectra.
\item Photonic crystal with fragmented GBZ: Motivation and construction; trends with respect to system size and feature row separation.
\end{enumerate}

\section{I. Formalism for GBZ fragmentation}

\subsection{A. Reformulation of open boundary conditions in terms of the boundary constraint matrix $M$}

In this section, we detail how an arbitrary large system $H$ under open boundary conditions (OBCs) can be exactly reduced to a much smaller boundary constraint matrix $M$ in the basis of bulk eigenstates. Consider an arbitrary Hamiltonian given in terms of the lattice momentum $k$ by
\begin{equation}
H(z)= \sum_{j=-q}^{p} H_j z^j,
\label{Hzpq}
\end{equation}
where $z=e^{ik}$ and $q,p$ are its hopping ranges in the left and right directions. For $b$ atoms per unit cell (i.e. bands), $H_j$ is a $b\times b$ matrix whose elements represent hoppings between their respective atoms $j$ unit cells apart. 

Conventionally, to solve for the OBC spectrum and eigenstates in a finite OBC setting with $N$ unit cells, one would need to diagonalize the bounded $bN\times bN$ block Toeplitz matrix $H_\text{OBC}$ whose $(x,x')$-th block is given by $H_{x-x'}$. This, however, is computationally expensive since the computation time typically scales polynomially~\cite{lanczos1958iterative} with $bN$, which can be substantial if $N$ is large. Furthermore, in the presence of the non-Hermitian skin effect (NHSE), floating point errors are also amplified exponentially with $N$. 

Below, leveraging on the block Toeplitz structure of this eigenvalue problem, we show how it exactly reduces to finding the kernel of a $B\times B$ \emph{boundary constraint matrix} $M$, which importantly does \emph{not} scale with system size. Typically, when the hoppings are not too sparse, we have $B=b(p+q)$ constraints from the OBCs, which truncate off some hoppings from the first $q$ ($p$) unit cells from the left (right) boundary (and each unit cell contains $b$ atoms). 

This drastic reduction hinges on the fact that, in the bulk eigenstate basis, only $B$ boundary constraints need to be explicitly satisfied, even as the system size $N$ becomes arbitrarily large. Given any reference energy $E$, one can identify bulk eigenstates $\phi_\mu(x)$ that generically assume the form
\begin{equation}
\phi_\mu(x) = z_\mu^x \varphi_\mu,
\end{equation}
where $z=z_\mu$ is a root of the characteristic dispersion polynomial
\begin{equation}
\text{Det}[H(z)-E\,\mathbb{I}]=0,
\label{DetHz}
\end{equation}
and $\varphi_\mu$ is its corresponding $b$-component intra unit-cell eigenstate satisfying $H(z_\mu)\varphi_\mu=E\varphi_\mu$. The degree of $z_\mu$ in Eq.~\ref{DetHz} formally defines $B\leq b(p+q)$, which saturates its upper bound unless certain inter-sublattice hoppings are absent. 
In Hermitian systems, $|z_\mu|=1$ when $E$ lies on the physical energy bands. In non-Hermitian OBC systems, however, we often have  $|z_\mu|=e^{-\text{Im}[k]}\neq 1$ due to the NHSE. The complex nature of $k$ defines the so-called generalized Brillouin zone (GBZ), through which the effects of the NHSE are encapsulated in complex momentum deformation.

By construction, each $\phi_\mu(x)$ already satisfies the OBC eigenequation in the bulk. However, they are not eigensolutions near the OBC boundaries due to translation symmetry breaking. To obtain the OBC eigenstates $\psi_\text{OBC}(x)$, the idea is to build appropriate superpositions of the bulk eigenstates viz.
\begin{equation}
\psi_\text{OBC}(x)=\sum_\mu c_\mu \phi_\mu(x)=\sum_\mu c_\mu z_\mu^x \varphi_\mu.
\label{psixsupp}
\end{equation}
The expansion coefficients $c_\mu$ can be found by demanding that $\psi_\text{OBC}(x)$ satisfies OBCs at both the left and right boundaries. For the closest unit cell to the left boundary, all left hoppings in $H_\text{OBC}$ are truncated, leaving only the onsite and right hoppings to act $\varphi_\mu$ i.e. $\sum_{j=0}^{p} H_j z_\mu^j\varphi_\mu= E\varphi_\mu -\sum_{j=1}^{q}z_\mu^{-j} H_{-j} \varphi_\mu$. Analogously, for the $n$-th unit cell closest to the left boundary, the remaining untruncated hoppings act as $z_\mu^{n-1}\left(E\varphi_\mu -\sum_{j=n}^{q} z_\mu^{-j}H_{-j} \varphi_\mu\right)$. Likewise, for the $n$-th unit cell closest to the right boundary, the untruncated hoppings act as $z_\mu^{N-n}\left(E\varphi_\mu -\sum_{j=n}^{p}z_\mu^{j} H_{j} \varphi_\mu\right)$. 
With some care, and shifting the labeling of the unit cells by $x_0$ such that the coefficients $c_\mu$ reflect the relative weights of the bulk eigensolutions at position $x_0$, we hence arrive at the following boundary constraint formulation of OBCs: For any OBC eigenenergy $E=E_\text{OBC}$, its corresponding eigenequation $H_\text{OBC}\psi_\text{OBC}=E_\text{OBC}\psi_\text{OBC}$ can be recast into the constraint $M\bold c = \bold 0$, where $\bold c = (c_1,c_2,...,c_{B})^T$ and, for $B=b(p+q)$, 
\begin{equation}
M_{\nu\mu}=\left\{\begin{matrix}
\sum\limits_{j=\nu}^{q} z_\mu^{\nu-j-1-x_0}H_{-j} \varphi_\mu\, \qquad\qquad 1\leq \nu \leq q,
\\
\sum\limits_{j=p+q-\nu+1}^{p}z_\mu^{N-x_0-p-q+\nu+j-1} H_{j} \varphi_\mu, \qquad\qquad q+1\leq \nu \leq q+p. 
\end{matrix}\right.
\label{Mmunu}
\end{equation}
In the above, $\mu\in \{1,2,...,b(p+q)\}$ and $\nu\in \{1,2,...,p+q\}$ such that $M_{\mu\nu}$ above is a $b$-component vector and that $M$ is a $b(p+q)\times b(p+q)$ matrix with $B=b(p+q)$. When $B<b(p+q)$, one or more of the above $p+q$ constraints do not exist, and the number of roots $z_\mu$ will be accordingly reduced too.

In this work, we would either set $x_0=0$, such that $c_\mu$ represent the relative contributions of the bulk eigensolutions $\phi_\mu$ at the left boundary, or $x_0=N/2$, such that these relative contributions are taken at the Center of the system, subject to manifestly symmetrical constraints from the left and right boundaries.

\begin{figure*}
    \subfloat[$N=6$]{\includegraphics[width=.31\linewidth]{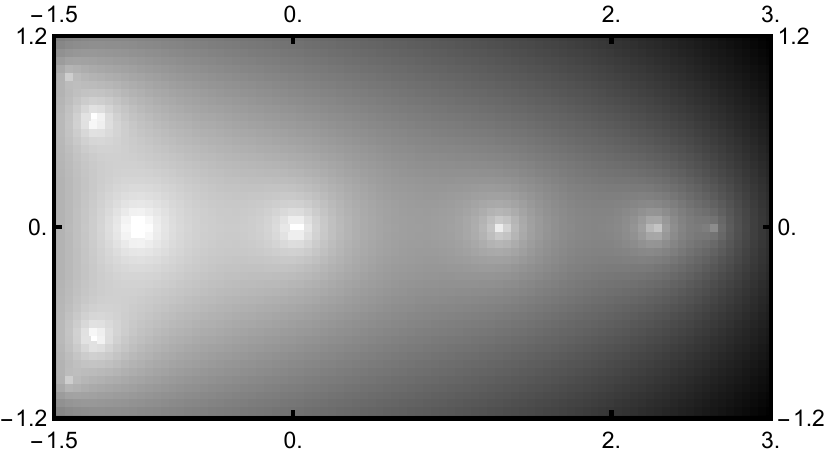}}
    \subfloat[$N=30$]{\includegraphics[width=.31\linewidth]{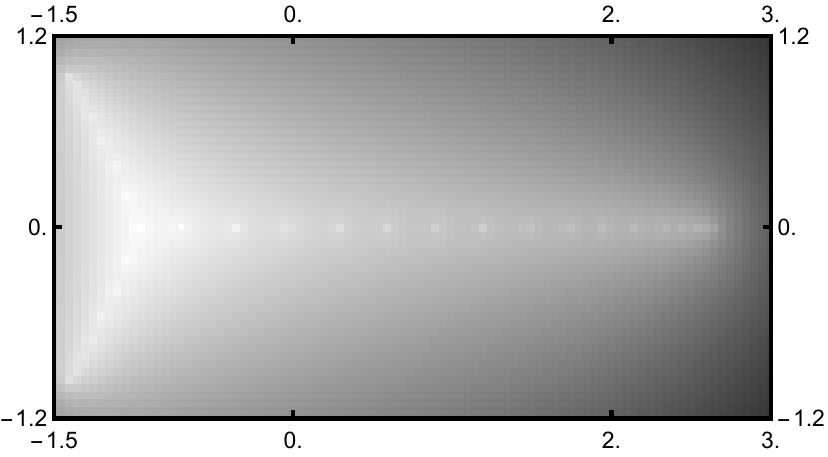}}
    \subfloat[$N=100$]{\includegraphics[width=.31\linewidth]{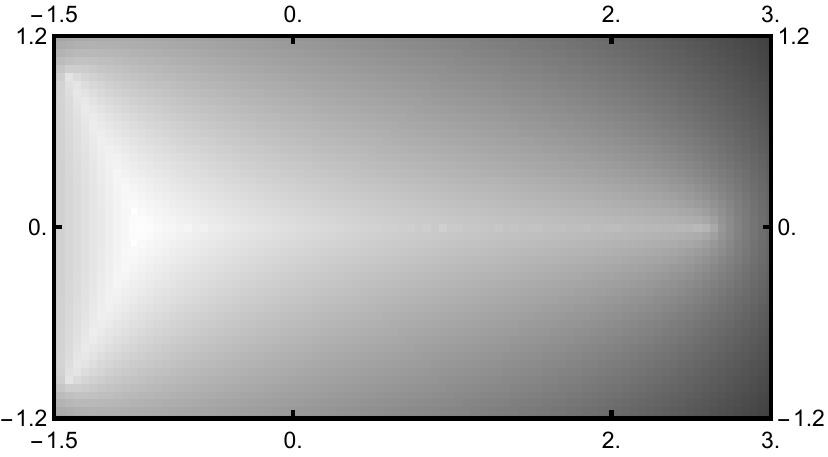}}
    \includegraphics[width=.05\linewidth]{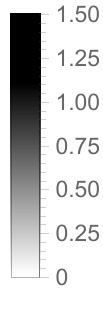}
    \caption{Density plot of $\frac1{N}\log|\text{Det}M|$ for the illustrative model $H(z)=z^2+z+\frac1{z}$, computed from Eq.~\ref{Mmunu} by sweeping across all complex $E$ in the region. The rescaling factor of $\frac1{N}$ exponentially normalizes Det$M$, whose dominant non-vanishing components scale like $z_\mu^N$. In all cases, the $N$ brightest spots with exponentially low Det$M$ correspond to the OBC eigenvalues.  }
    \label{fig:DetM}
\end{figure*}

To summarize, when $E$ is an OBC eigenenergy possessing an eigenstate $\psi_\text{OBC}(x)=\sum_\mu c_\mu z_\mu^x \varphi_\mu$, \textbf{the coefficients $c_\mu$ lies in the kernel of the $M$, which necessitate that $\text{det} M=0$.} Hence, to check whether a given energy $E$ belongs to the OBC spectrum of an arbitrary large $N$ unit-cell system, one just needs to obtain the (non-Bloch) bulk eigenstates $\varphi_\mu$ and their $z_\mu$ [Eq.~\ref{DetHz}], and use them to compute the $B\times B$ matrix $M$ [Eq.~\ref{Mmunu}]. The point $E$ belongs to the OBC spectrum if $\text{Det} M$ vanishes, as illustrated in Fig.~\ref{fig:DetM} (in practice, it just needs to be a sharp local minimum, since departures from $\text{Det} M=0$ (including floating point errors) generally scale with the power of $N$).

\subsection{B. Composition-inverse-participation-ratio (cIPR) and GBZ fragmentation}

Conventionally, a NHSE OBC eigenstate $\psi_\text{OBC}(x)=\sum_\mu c_\mu z_\mu^x \varphi_\mu$ is said to live in a generalized Brillouin zone (GBZ) if it is dominated by only two coefficients $c_\mu,c_{\mu'}$ such that $|z_\mu|=|z_{\mu'}|$. This mirrors the case of Hermitian OBC eigenstates, which are always composed of Bloch bulk eigenstates with  $|z_\mu|=|z_{\mu'}|=1$. The value of $\text{Im}[k]=-\log|z_\mu|=-\log|z_{\mu'}|$ uniquely defines the GBZ, which can be thought of as a complex deformation of the usual BZ. Having a pair of $\phi_\mu(x),\phi_{\mu'}(x)$ solutions that scale equally fast also allows for OBC solutions that remain stable as $N\rightarrow \infty$ \cite{lee2019anatomy,yao2018edge}.

While many well-studied NHSE models possess well-defined GBZs, there is no fundamental reason why the OBC eigenstates must be dominated by only two bulk eigensolutions. Even though the existence of two equally large $|z_\mu|=|z_{\mu'}|$ guarantees a clear (and uneventful) approach towards the thermodynamic limit, there is no reason why three or more $z_\mu$ solutions cannot dominate, particularly at experimentally accessible numbers of unit cells of $N\lesssim \mathcal{O}(10^2)$. When more than one value of $|z_\mu|$ participates with non-negligible weight $c_\mu$ in a certain OBC eigenstate, we say that GBZ fragmentation has occured.



The extent of GBZ fragmentation can be quantified through the so-called composition-inverse-participation-ratio (cIPR), which is defined for each $E_\text{OBC}$ by
\begin{equation}
\text{cIPR}= \frac{\sum_\mu |c_\mu|^4}{\left(\sum_\mu |c_\mu|^2\right)^2},
\label{cIPR}
\end{equation}
where $c_\mu$ is the weightage of the bulk $\phi_\mu$ eigensolution in the corresponding $\psi_\text{OBC}$, and can be obtained by solving $M\bold c=\bold 0$ with $M$ given by Eq.~\ref{Mmunu}. When only two $z_\mu$ solutions have non-negligible weights $|c_\mu|=\cos\theta$ and $|c_\nu|=\sin\theta$, we have $\text{cIPR}=\cos^4\theta+\sin^4\theta = (3 + \cos 4\theta)/4$, which lies between $0.5$ and $1$. In (hypothetical) other extreme where each coefficient is equally weighted, we have $|c_\mu|=1/\sqrt{B}$ and $\text{cIPR}=B^{-1} \cong 1/b(p+q)$. In the various examples shown in this work, we typically have cIPR between 0.2 and 0.8.

\subsection{C. Examples on using the $M$ matrix formalism}

\begin{itemize}
\item \textbf{HN model:} As the most basic example, consider the Hatano-Nelson model
\begin{equation}
H_\text{HN}(z) = z + \frac{h}{z}
\label{HN}
\end{equation}
with trivial unit cell ($b=1$), degrees $p=q=1$ and real scalar coefficients $H_1=1$, $H_{-1}=h$. Through the well-known transformation $z\rightarrow \sqrt{h}z$ that renders the effective hoppings in Eq.~\ref{HN} symmetric, we have $z_1=z_2^*=\sqrt{h}e^{ik}$, where $k=\cos^{-1}\frac{E_\text{OBC}}{2\sqrt{h}}$. This gives the classic GBZ $|z_1|=|z_2|=\sqrt{h}$. From Eq.~\ref{Mmunu} (with $x_0$ set to $N/2$), we have $M_{11}=z_1^{-1-N/2}h=h^{-N/4}e^{-ikN/2}\sqrt{h}e^{-ik}$, $M_{12}=z_2^{-1-N/2}h=h^{-N/4}e^{ikN/2}\sqrt{h}e^{ik}$, $M_{21}=z_1^{N/2}\cdot 1=h^{N/4}e^{ikN/2}$ and $M_{22}=z_2^{N/2}\cdot 1=h^{N/4}e^{-ikN/2}$. Putting them together, we have
\begin{equation}
M=\left(\begin{matrix}
h/z_1^{N/2+1} & h/z_2^{N/2+1} \\
z_1^{N/2} & z_2^{N/2}
\end{matrix}\right)=\left(\begin{matrix}
h^{-N/4}\sqrt{h}e^{-ik(1+N/2)} &h^{-N/4} \sqrt{h}e^{ik(1+N/2)} \\
h^{N/4}e^{ikN/2} & h^{N/4}e^{-ikN/2}
\end{matrix}\right).
\end{equation}
Mandating that $\text{Det}M=-2i\sqrt{h}\sin[k(N+1)]=0$ yields $k=\frac{n\pi}{N+1}$, $n\in \mathbb{Z}$, which we recognize as simply the condition for wavefunctions to vanish at both ends. With this, it follows from $M\bold c = \bold 0$ that $c_1=\frac1{\sqrt{2}}e^{-ikN/2}$ and $c_2 = -\frac1{\sqrt{2}}e^{-ikN/2}$, which gives $\text{cIPR}=\frac1{4}+\frac1{4}=\frac1{2}$ [Eq.~\ref{cIPRmain}], perfectly defining a conventional GBZ.

With these coefficients, we can also write down the OBC eigenstates as $\psi_\text{OBC}(x)=\frac1{\sqrt{2}}\left(e^{-ikN/2}z_1^x-e^{ikN/2}z_1^{-x}\right)=\sqrt{2}ih^{x/2}\sin[k(x-N/2)]$, which exponentially grows as $\sim h^{x/2}$.

Note that, had we not known of the similarity transformation $z\rightarrow \sqrt{h}z$, the condition that det$M=0$ would still have yielded $z_1^{N+1}=z_2^{N+1}$, i.e. $\frac{z_1}{z_2}=e^{2\pi n i /(N+1)}$. Noting from $H_\text{HN}(z) = z + \frac{h}{z}$ that the two solutions should satisfy $z_1=\frac{h}{z_2}$, we arrive at the same conclusion that $z_1=z_2^*=\sqrt{h}e^{ik}$ where $k=\frac{n\pi}{N+1}$. 

\begin{figure*}
     \includegraphics[width=1\linewidth]{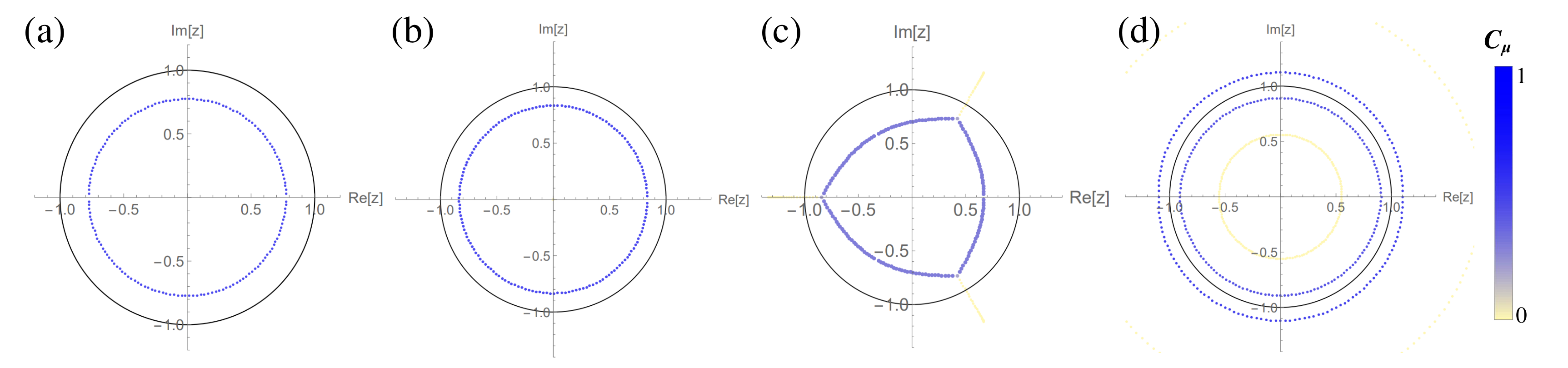}
    \caption{GBZs of the four illustrative models described in this section, colored according to the composition IPR (cIPR) [Eq.~\ref{cIPRmain}] computed at the system midpoint $x_0=N/2$, all for $N=80$ unit cells. Only the last model exhibits significant GBZ fragmentation.     
    (a) Hatano-Nelson model $H_\text{HN}(z)=z+\frac{h}{z}$ [Eq.~\ref{HN}] with $h=0.6$. (b) Non-Hermitian SSH model [Eq.~\ref{SSH}] with $\gamma=1.2$. (c), $H_\text{NNN}(z)=z^2+b/z$ [Eq.~\ref{hNNN}] with $b=0.6$. Note the presence of very weak GBZ fragments (faint yellow) radiating from the cusps.    
    (d) Coupled Hatano-Nelson chains [Eq.~\ref{cNHSEsupp}] with $h=2$ and weak coupling $\Delta=0.01$. Notwithstanding the two faint innermost and outermost rings, the remaining GBZ solutions fragmentate into two non-degenerate loops (blue), each  with significant weight ($c_\mu\approx \frac1{\sqrt{2}}$), different from cases (a-c) where the blue loops are all doubly degenerate. }
    \label{fig:GBZs}
\end{figure*}

\item \textbf{Non-herm SSH model:} Next, consider another prototypical NHSE model, the non-Hermitian SSH model given by
\begin{equation}
H_\text{SSH}(z) = \left(\begin{matrix}
0 & \frac1{z}+t\gamma \\
z+t\frac1{\gamma} & 0
\end{matrix}\right),
\label{SSH}
\end{equation}
with 2-atom unit cells ($b=2$), degrees $p=q=1$ and real matrix coefficients $H_1=\left(\begin{matrix}
0 & 0\\
1 & 0
\end{matrix}\right)$, $H_{-1}=\left(\begin{matrix}
0 & 1\\
0 & 0
\end{matrix}\right)$ and $H_{0}=\left(\begin{matrix}
0 & t\gamma\\
t\frac1{\gamma} & 0
\end{matrix}\right)$. Algebraically, it is of the same class as the Hatano-Nelson model~\cite{tai2023zoology}, with a characteristic polynomial 
\begin{equation}
\text{Det}(H_\text{SSH}(z)-E_\text{OBC})=E_{OBC}^2-(1+t^2)-t\gamma z -\frac{t}{\gamma z} =0
\end{equation}
of similar form that leads to the $B=2$ roots $z_{1,2}=\frac1{\gamma}e^{\pm ik}$, where $E_\text{OBC}^2=1+t^2+2t\cos k$. 

It is instructive to demonstrate how $M$ can be computed, given that $B=2< 2(p+q)=4$.
Henceforth, we fix $t=1$ for notational simplicity. 
To use Eq.~\ref{Mmunu}, we first compute $\varphi_\mu$, which are the eigenvectors of $H(z_\mu)=\left(\begin{matrix}
0 & \gamma(1+e^{\mp i k}) \\
\frac1{\gamma}(1+e^{\pm ik}) & 0
\end{matrix}\right)$. Choosing the positive branch of $E_\text{OBC}$ without loss of generality, we obtain $\varphi_{1,2} = \left(\begin{matrix}\gamma e^{\mp i k/2}\\1 \end{matrix}\right)$. 

The first line of Eq.~\ref{Mmunu} (with $x_0$ set to $N/2$) gives $M_{1\mu}=z^{-1-N/2}_\mu H_{-1}\varphi_\mu\propto z^{-1-N/2}_\mu\left(\begin{matrix}
0 & 1\\
0 & 0
\end{matrix}\right)\left(\begin{matrix}\sqrt{\gamma} \\ \sqrt{z_\mu} \end{matrix}\right)=z_\mu^{-(1+N)/2}\left(\begin{matrix}1 \\ 0 \end{matrix}\right)$. And the second line gives 
$M_{2\mu}=z^{N/2}_\mu H_{1}\varphi_\mu\propto z^{N/2}_\mu\left(\begin{matrix}
0 & 0\\
1 & 0
\end{matrix}\right)\left(\begin{matrix}\sqrt{\gamma} \\ \sqrt{z_\mu} \end{matrix}\right)=\sqrt{\gamma}z_\mu^{N/2}\left(\begin{matrix}0 \\ 1 \end{matrix}\right)$. Saliently, even though each of these $M_{\mu\nu}$ are 2-component vectors, only one component is nonzero. Hence each $M_{\mu\nu}$ is effectively a scalar, and the $M$ matrix is effectively $2\times 2$, consistent with the existence of only $B=2$ bulk eigensolutions $\phi_\mu$. We thus have  
\begin{equation}
M=\left(\begin{matrix}
z_1^{-(N+1)/2} & z_2^{-(N+1)/2} \\
\gamma z_1^{N/2} & \gamma z_2^{N/2}
\end{matrix}\right)=\left(\begin{matrix}
\gamma^{(N+1)/2}e^{-ik(N+1)/2} &\gamma^{(N+1)/2}e^{ik(N+1)/2}  \\
\gamma^{1-N/2}e^{ikN/2} & \gamma^{1-N/2}e^{-ikN/2}
\end{matrix}\right).
\end{equation}
Analogous to the previous Hatano-Nelson model, $\text{Det}M=0$ gives $z_1^{N+1/2}=z_2^{N+1/2}$, i.e. $k=\frac{\pi n}{N+1/2}$, $n\in \mathbb{Z}$. Similarly, $|c_1|=|c_2|=\frac1{\sqrt{2}}$ such  that $\text{cIPR}=\frac1{4}+\frac1{4}=\frac1{2}$ [Eq.~\ref{cIPRmain}], perfectly defining a conventional GBZ.

\item \textbf{NNN model:} For pedagogical purposes, let's examine a model that is more complicated than the non-Hermitian SSH and Hatano-Nelson models, but which still does not exhibit GBZ fragmentation:
\begin{equation}
H_\text{NNN}(z) =z^2+\frac{b}{z}. 
\label{hNNN}
\end{equation}
Here it has two dissimilar hoppings, one across to the next nearest neighbor (NNN), and the other with amplitude $b$ to the opposite nearest neighbor. We shall see that $b$ plays a minor role compared to the fact that the two hoppings are across different distances, but GBZ fragmentation still generically cannot exist.

This model has degrees $p=2$, $q=1$, and matrix coefficients $H_2=1$, $H_{-1}=b$, and $H_1=H_0=0$. For a given $E_\text{OBC}$, the eigenequation $H_\text{NNN}(z)=E_\text{OBC}$ gives 3 roots $z=z_1,z_2,z_3$. Its $M$ matrix is given by [Eq.~\ref{Mmunu}]:
\begin{equation}
M =\left(\begin{matrix}
b/z_1^{1+x_0} & b/z_2^{1+x_0} & b/z_3^{1+x_0} \\
z_1^{N-x_0} & z_2^{N-x_0} & z_3^{N-x_0} \\
z_1^{N+1-x_0} & z_2^{N+1-x_0} & z_3^{N+1-x_0} \\
\end{matrix}\right)\rightarrow 
\left(\begin{matrix}
1/z_1 & 1/z_2 & 1/z_3 \\
z_1^{N} & z_2^{N} & z_3^{N} \\
z_1^{N+1} & z_2^{N+1} & z_3^{N+1} \\
\end{matrix}\right)
\label{MNNN}
\end{equation}
where on the right, we have eliminated constant multipliers that play no part in the subsequent derivations. Even without any further calculation, it should be intuitively clear that the GBZ must be non-fragmented, being dominated by the two roots of equal magnitude that must exist in any cubic equation. Labeling $z_2,z_3$ as $z_0(\theta) e^{\pm i \theta}$ for real $\theta$~\footnote{Here, we have not assumed that $\theta$ takes the role of a momentum index.}, we have either of the two possibilities:
\begin{enumerate}
\item $|z_1|>|z_0|$, such that the last row of $M$ in Eq.~\ref{MNNN} would be dominated by $z_1^{N+1}$. Then, in order to satisfy $M\bold c = \bold 0$, we must have $c_1\ll c_2,c_3$ i.e. $c_1\approx 0$. 
\item $|z_1|<|z_0|$, such that the last row of $M$ is dominated by $z_2^{N+1}, z_3^{N+1}\sim z_0^{N+1}$. While $c_1\gg c_2,c_3$ may satisfy the last two rows of $M\bold c = \bold 0$, it cannot possibly satisfy the first row $\frac{c_1}{z_1}+\frac{c_2}{z_2}+\frac{c_3}{z_3}=0$. Hence the only possibility is also that $c_1\approx 0$, but that $c_2$ and $c_3$ conspire to satisfy $M\bold c = \bold 0$ in a manner similar to the HN model earlier.
\end{enumerate}
In any case, since inverting the system mathematically amounts to $z\leftrightarrow 1/z$, we can safely assume that $|z_0|<|z_1|$. Hence, in any finite setting, the relative size of $c_1$ relative to $c_2,c_3$ must scale like $\left|\frac{z_0}{z_1}\right|^N$. This should rapidly diminish 
 i.e. lead to a non-fragmented doubly degenerate GBZ at (non-constant) $|z|=|z_0(\theta)|$, unless $|z_0|\approx |z_1|$ for some $\theta$. 

 To make further progress, note that $z^3+b-E_\text{OBC}z=(z-z_1)(z-z_2)(z-z_3)=0$ (Vieta's formula) gives $ -b = z_1z_2z_3 = z_1 z_0^2$ and $z_1+z_2+z_3 =z_1+2z_0\cos\theta= 0$, which together give
 \begin{equation}
 \frac{z_1}{z_0}=-\frac{b}{z_0^3}=-2\cos\theta.
 \label{NNNeq}
 \end{equation}
Since $\cos\theta$ is real, we arrive at the conclusion that $z_0$ can only take on branches with $\text{arg}\,z_0$ separated by $2\pi/3$. As such, the OBC spectrum $E_\text{OBC}=(z_1^3 + b)/z_1$ must similarly assume 3 branches, since $z_1$ is just a real multiple of $z_0$~\footnote{Note that unlike in previous works~\cite{lee2020unraveling,tai2023zoology} which derived this 3-fold branching in another way, this derivation does not assume that $|z_2|=|z_3|$ right from the start.}.

To further derive the exact (slight) dependence of $z_0$ and $\theta$ on the system size $N$, we invoke
\begin{equation}
\text{Det}M = \frac{z_1^Nz_2^N(z_2-z_1)}{z_3}+ \frac{z_2^Nz_3^N(z_3-z_2)}{z_1}+ \frac{z_3^Nz_1^N(z_1-z_3)}{z_2}=0.
\label{DetMNNN}
\end{equation}
Note that Eq.~\ref{DetMNNN} remains the same even if an additional NN hopping is introduced to give nonvanishing $H_1$. To proceed, we note that the middle term does not contain any power of $z_1^N$, and will thus be dominated by the other two terms at large $N$. Hence, we obtain $z_1^{N+1}z_2^{N+1}(z_2-z_1)= z_1^{N+1}z_3^{N+1}(z_3-z_1)$, which yields
\begin{align}
\frac{z_1}{z_0}&=\frac{\sin(N+2)\theta}{\sin(N+1)\theta}\notag\\
&=\cos\theta + \cot[(N+1)\theta] \sin\theta
\end{align}
which, with the help of Eq.~\ref{NNNeq}, gives the condition for the allowed $\theta$:
\begin{equation}
3\cot \theta = -\cot [(N+1)\theta].
\label{coteq}
\end{equation}
The transcendental equation can be approximately solved through the small angle cotangent approximation $\cot(n\pi + \delta x )\approx 1/\delta x$ for $|\delta x|\ll 1$. Writing $\theta = \frac{n\pi}{N+1}+\delta \theta$, we have 
$\frac{3}{\frac{n\pi}{N+1}+\delta \theta} \approx -\frac1{(N+1)\delta\theta}$, 
which finally gives
\begin{equation}
\theta =  \frac{n\pi}{N+1}+\delta \theta\approx \frac{n\pi}{N+\frac{4}{3}},
\label{thetaeq}
\end{equation}
with index $n$ running within the range where $|\theta|<\frac{\pi}{3}$. Interestingly, the denominator is not $N$ as in usual PBC Bloch waves, or $N+1$ as in the OBC Hatano-Nelson model, but $N+\frac{4}{3}$, which includes the shift meticulous obtained from the transcendental Eq.~\ref{coteq}. In this sense, it appears that $\theta$ can be regarded as a ``momentum" index corresponding to the wavenumbers of non-Bloch states in a slightly ``expanded" space. These states can still be consistent with the boundary conditions due to the presence of the small $c_1$ coefficient. 

\begin{figure*}
    \includegraphics[width=1\linewidth]{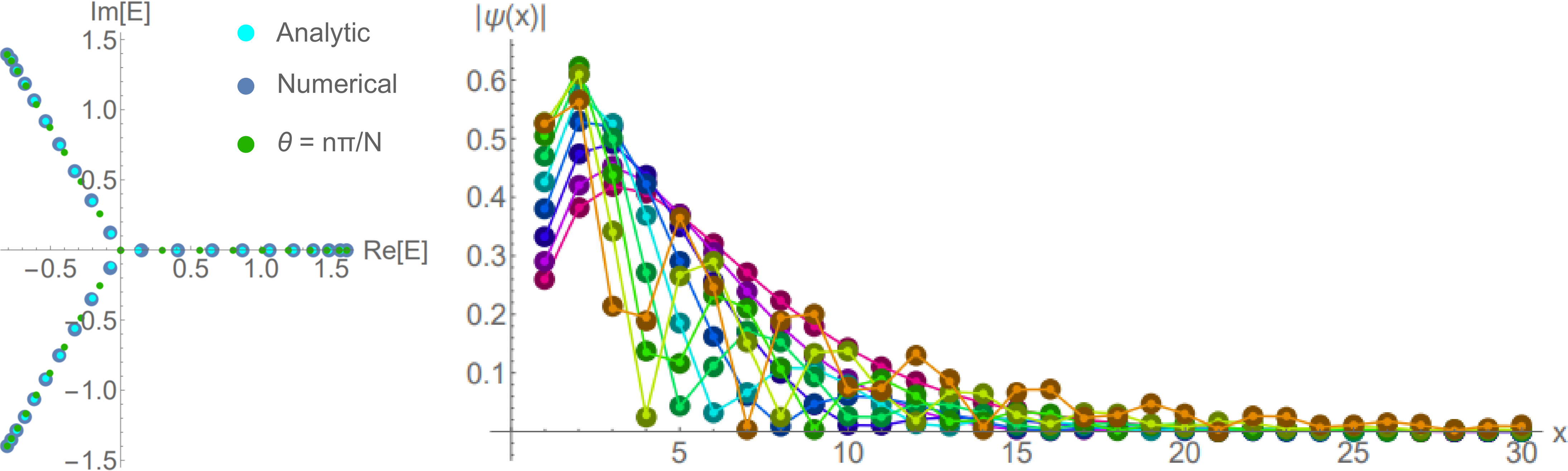}
    \caption{Analytic vs. numerical results for the NNN model Eq.~\ref{hNNN}, for $b=0.8$ and system size $N=30$. (a) Excellent agreement between the analytic OBC eigenspectrum from Eqs.~\ref{EOBCNNN}  and~\ref{thetaeq} (light cyan), and the numerically computed counterpart (dark blue). For comparison, the green dots using the naive choice of $\theta = n\pi/N$, $n\in \mathbb{Z}$ show considerable discrepancies. 
    (b) There is also excellent agreement between the analytical OBC eigenfunctions from Eq.~\ref{psiOBCNNN} (colored curves) and their numerical profiles ( corresponding darker disks). The colors magenta$\rightarrow$ blue$\rightarrow$ green$\rightarrow$ orange correspond to $n=1$ to $N/3-1$. 
}
    \label{fig:NNN}
\end{figure*}

For $N\gg 1$, Eq.~\ref{thetaeq} has been verified to be highly accurate in obtaining the exact GBZ, eigenspectrum and eigenstates. Substituting Eqs.~\ref{NNNeq} and~\ref{thetaeq} into the eigenenergy equation, it is straightforward to show that 
\begin{equation}
E_\text{OBC} = -\eta(1-4\cos^2\theta)\left(\frac{b}{2\cos\theta}\right)^{2/3},
\label{EOBCNNN}
\end{equation}
where $\eta =1$ or  $e^{\pm2\pi i /3}$ is a cube root of unitary, and $\theta$ is given by Eq.~\ref{thetaeq}. Its accuracy is demonstrated in Fig.~\ref{fig:NNN}(a): The  numerical eigenvalues (dark blue) exhibit excellent agreement with analytically computed eigenvalues from Eq.~\ref{EOBCNNN}, with $\theta$ given by Eq.~\ref{thetaeq} (light cyan). However, if $\theta$ is computed without replacing $N$ by $N+4/3$, then the computed eigenvalues (green) no longer coincide. 

Our results also allow for the accurate computation of the eigenstates. Assuming that $c_1\ll c_2,c_3$, the first row of Eq.~\ref{MNNN} gives $c_2/c_3 = - z_2/z_3$, such that 
\begin{eqnarray}
\psi_\text{OBC}(x) &\approx& c_2 z_2^x+c_3z_3^x \notag\\
&\propto& z_2^{x+1}-z_3^{x+1} \notag\\
&=&2i|z_0|^{x+1}\sin[\theta(x+1)]\notag\\
&\propto & \left(\frac{b}{2\cos\theta}\right)^{x/3}\sin[(N+1)x].
\label{psiOBCNNN}
\end{eqnarray}
In Fig.~\ref{fig:NNN}(b), Eq.~\ref{psiOBCNNN} (colored curves) is also shown to predict the numerical eigenstates (dark disks) very accurately for almost all the eigenstates.

\item \textbf{Coupled HN model:} Next, we consider one of the simplest models exhibiting GBZ fragmentation, the coupled Hatano-Nelson model
\begin{equation}
H_\text{coupled-HN}(z) = \left(\begin{matrix}
z+\frac{h}{z} & \Delta \\
\Delta & \frac1{z}+hz
\end{matrix}\right),
\label{cNHSEsupp}
\end{equation}
also with 2-atom unit cells ($b=2$), hopping ranges $p=q=1$ and hopping matrix coefficients $H_1=\left(\begin{matrix}
1 & 0\\
0 & h
\end{matrix}\right)$, $H_{-1}=\left(\begin{matrix}
h & 0\\
0 & 1
\end{matrix}\right)$ and $H_{0}=\left(\begin{matrix}
0 & \Delta\\
\Delta & 0
\end{matrix}\right)$.
While this model is well-known for exhibiting the critical NHSE (strong system size-dependence of its spectrum) under weak $\Delta$, we note that GBZ fragmentation is more general, occurring even in the absence of the critical NHSE i.e. in the regime of strong coupling $\Delta \sim h$. 

At very small $N$, the system essentially behaves like two opposite uncoupled HN chains, with a total of 4 constant GBZ solutions, $|z_+|=|z_-|$ and $|z_+|^{-1}=|z_-|^{-1}$. However, beyond a critical system size $N$, the GBZ solutions bifurcate (fragmentate) into non-degenerate solutions $z_+,z_-,1/z_+,1/z_-$, as shown in Fig. ~\ref{fig:cNHSE}(a). In general, these solutions then converge very slowly to their thermodynamic limit values, remaining fully fragmented unless $|z_-|$ hits unity (as in the illustrative example in Fig.~\ref{fig:cNHSE}(a)). Below, we derive these behavior in detail.

To compute $M$, Eq.~\ref{Mmunu} gives $M_{1\mu}=z^{-1-x_0}_\mu H_{-1}\varphi_\mu$ and $M_{2\mu}=z^{N-x_0}_\mu H_{1}\varphi_\mu$ where $N$ is the system size, which is similar in form to the previous case since we also have $p=q=1$. However, due to the very different $\varphi_\mu$ eigenvectors, $M$ would turn out very different. Writing $\varphi_\mu=\left(\begin{matrix}
 \varphi_\mu^\uparrow\\
 \varphi_\mu^\downarrow
\end{matrix}\right)$, $M\bold c=\bold 0$ is equivalently expressed as
\begin{eqnarray}
\sum_\mu c_\mu z_\mu^{-1-x_0}\left(\begin{matrix}
 h\varphi_\mu^\uparrow\\
 \varphi_\mu^\downarrow
\end{matrix}\right)=\bold 0,\\
\sum_\mu c_\mu z_\mu^{N-x_0}\left(\begin{matrix}
 \varphi_\mu^\uparrow\\
 h\varphi_\mu^\downarrow
\end{matrix}\right)=\bold 0.
\end{eqnarray}
Although this is a $4\times 4$ matrix equation, we can employ the symmetry of the system to reduce it to a $2\times 2$ problem. This can be manifested shown by setting $x_0=(N-1)/2$, and noting that the two sublattices are related by the inversion $z\rightarrow 1/z$. Hence we left with a rank-2 problem spanned by $\varphi_-=\left(\begin{matrix}
 \varphi_2^\uparrow\\
 \varphi_2^\downarrow
\end{matrix}\right)=\left(\begin{matrix}
 \varphi_3^\downarrow\\
 \varphi_3^\uparrow
\end{matrix}\right)$ and $\varphi_+=\left(\begin{matrix}
 \varphi_1^\uparrow\\
 \varphi_1^\downarrow
\end{matrix}\right)=\left(\begin{matrix}
 \varphi_4^\downarrow\\
 \varphi_4^\uparrow
\end{matrix}\right)$, with only two independent composition coefficients $c_+=c_1=c_4$ and $c_-=c_2=c_3$. The OBCs can hence be fully encapsulated by a reduced $M_\text{red}$ matrix equation:
\begin{equation}
M_\text{red}\left(\begin{matrix}
 c_+\\
 c_-
\end{matrix}\right)= \left(\begin{matrix}
 z_+^{-(x_0+1)}\varphi_+^\uparrow + z_+^{x_0+1}\varphi_+^\downarrow &  z_-^{-(x_0+1)}\varphi_-^\uparrow + z_-^{x_0+1}\varphi_-^\downarrow \\
 z_+^{-(x_0+1)}\varphi_+^\downarrow + z_+^{x_0+1}\varphi_+^\uparrow &  z_-^{-(x_0+1)}\varphi_-^\downarrow + z_-^{x_0+1}\varphi_-^\uparrow 
 \end{matrix}\right)\left(\begin{matrix} c_+\\ c_-\end{matrix}\right)=\left(\begin{matrix} 0\\ 0\end{matrix}\right),
 \label{M2by2}
 \end{equation}
such that the composition ratio $R=c_+/c_-$ is given by 
\begin{eqnarray}
R=\frac{c_+}{c_-}&=&\frac{z_-^{-(x_0+1)}\varphi_-^\downarrow + z_-^{x_0+1}\varphi_-^\uparrow }{ z_+^{-(x_0+1)}\varphi_+^\downarrow + z_+^{x_0+1}\varphi_+^\uparrow}\notag\\
&=&\frac{\varphi_-^\downarrow}{\varphi_+^\downarrow}\frac{z_-^{-(x_0+1)} + r_-z_-^{x_0+1} }{ z_+^{-(x_0+1)} + r_+z_+^{x_0+1}}\notag\\
&=&\sqrt{\frac{1+|r_+|^2}{1+|r_-|^2}}\frac{z_-^{-\frac{N+1}{2}} + r_-z_-^{\frac{N+1}{2}} }{ z_+^{-\frac{N+1}{2}} + r_+z_+^{\frac{N+1}{2}}}
\label{Rcpcm}
\end{eqnarray}
where we have defined $r_\pm = \frac{\varphi^\uparrow_\pm}{\varphi^\downarrow_\pm}$. 
The composition IPR [Eq.~\ref{cIPR}] is given by cIPR$=\frac1{2}-\left(\frac{R}{R^2+1}\right)^2$. Depending on whether $|z_+|$ and/or $|z_-|$ is greater or less than $1$, $R$ scales with the system size differently, as shown in the following table. 
\begin{table}[h]
\begin{tabular}{|c|c|c|}
\cline{1-3} 
\multicolumn{1}{|c|}{\textbf{}} & \textbf{$|z_-|<1$}    & \textbf{$|z_-|>1$}  \\ 
\cline{1-3}
\textbf{$|z_+|<1$}  & $R\sim \frac{\varphi_-^\downarrow}{\varphi_+^\downarrow}\left(\frac{z_+}{z_-}\right)^{\frac{N+1}{2}}$ & $R\sim r_-\frac{\varphi_-^\downarrow}{\varphi_+^\downarrow}\left(z_+z_-\right)^{\frac{N+1}{2}}$
  \\
  \cline{1-3}
\textbf{$|z_+|>1$}   & $R\sim \frac1{r_+}\frac{\varphi_-^\downarrow}{\varphi_+^\downarrow}\left(z_+z_-\right)^{-\frac{N+1}{2}}$ & $R\sim \frac{r_-}{r_+}\frac{\varphi_-^\downarrow}{\varphi_+^\downarrow}\left(\frac{z_-}{z_+}\right)^{\frac{N+1}{2}}$
  \\ 
 \cline{1-3}
\end{tabular}
\end{table}

Notwithstanding the constant coefficients, these four cases all obey $R\sim \text{Exp}\left[\left(|\log|z_-||-|\log|z_+||\right)\frac{N}{2}\right]$. This asymptotic exponential dependence is numerically verified in Fig.~\ref{fig:cNHSE}(b).

Eq.~\ref{M2by2} also implies that Det$M_\text{red}=0$, which must always hold for OBCs. With some algebra, it simplifies to the following constraint relating $r_\pm$ with $z_\pm^{N+1}$:
\begin{equation}
\frac{z_+^{N+1}-z_-^{N+1}}{1-(z_+z_-)^{N+1}}= \frac{r_--r_+}{1-r_+r_-}
\label{detMred1}
 \end{equation}
 or 
\begin{equation}
\tanh^{-1} r_--\tanh^{-1} r_+=\tanh^{-1}\left(z_+^{N+1} \right)- \tanh^{-1}\left(z_-^{N+1}\right).
\label{detMred2}
 \end{equation} 
Note that Eq.~\ref{detMred2} does \emph{not} necessarily imply that $r_\pm=z_\mp^{N+1}$, since it is a single constraint that would generically admit a continuum of solutions on the complex $E_\text{OBC}$ plane. 
From the form of Eq.~\ref{cNHSEsupp}, $r_\pm$ is given by
\begin{equation}
r_\pm = \frac{\Delta}{E_\text{OBC}-\frac1{z_\pm}-hz_\pm},
\label{rpm}
\end{equation}
which depends on $\Delta$, $E_\text{OBC}$ and $z_\pm$. A second constraint is just the energy eigenequation
\begin{equation}
\text{Det}(H_\text{coupled-HN}(z)-E_\text{OBC})=\left(z+\frac{h}{z}-E_\text{OBC}\right)\left(\frac1{z}+hz-E_\text{OBC}\right)-\Delta^2=0,
\label{DetHsupp}
\end{equation}
where $z_\pm$ are the two roots adiabatically connected to the two solutions of $\left(z+\frac{h}{z}-E_\text{OBC}\right)$ in Eq.~\ref{DetHsupp}. Together, Eq.~\ref{DetHsupp} and Eqs.~\ref{detMred1} or~\ref{detMred2} constitute two constraints that allow the $E_\text{OBC}$ spectrum to be solved in terms of $\Delta$ and $N$ (typically numerically). Specifically, for any provisional $E_\text{OBC}$ on the 2D complex plane, one can first find its $z_\pm$ solutions using Eq.~\ref{DetHsupp}. Then Eqs.~\ref{detMred1} or~\ref{detMred2} can be applied to check whether such a solution is indeed valid. This constraints the valid $E_\text{OBC}$ solutions to a co-dimension one subspace, which is just the set of OBC spectral curves. 

\begin{figure*}
    \subfloat[]{\includegraphics[width=.36\linewidth]{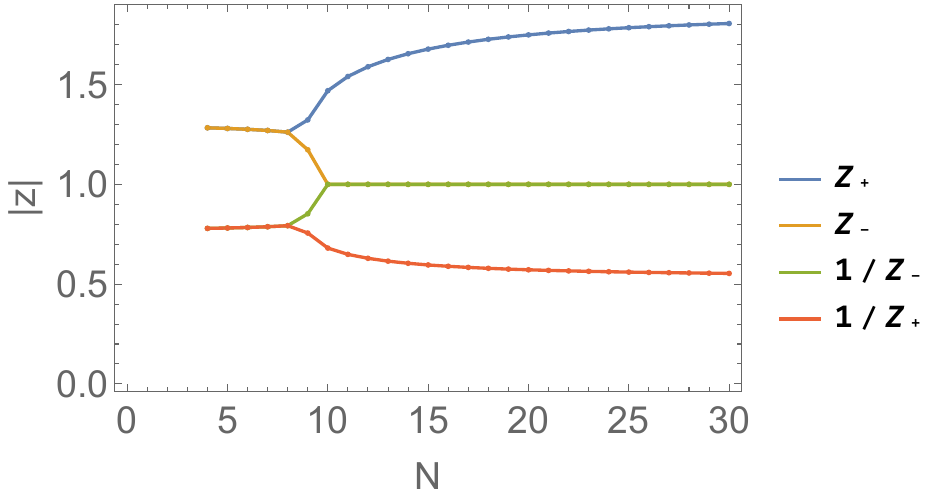}}
    \subfloat[]{\includegraphics[width=.32\linewidth]{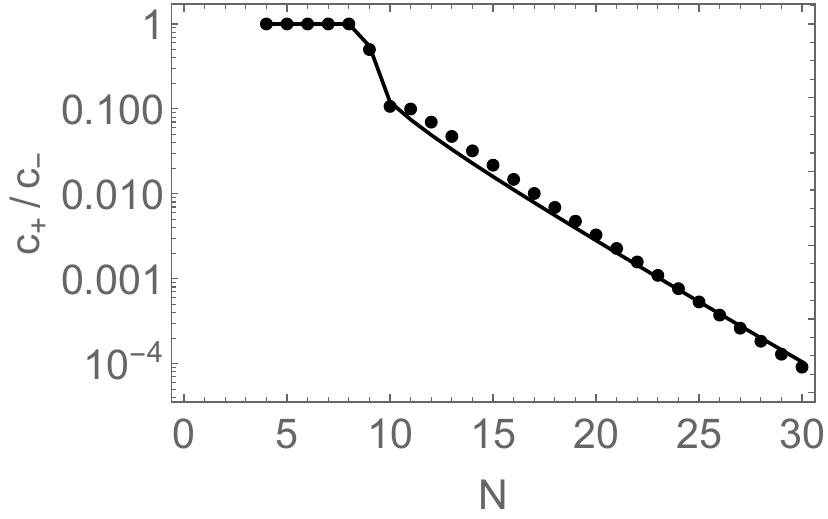}}
    \subfloat[]{\includegraphics[width=.31\linewidth]{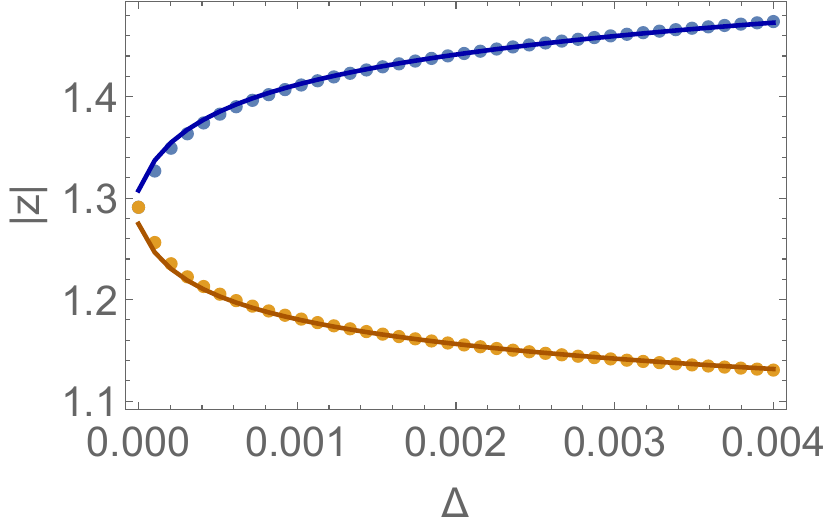}}
    \caption{GBZ fragmentation in the coupled-HN model [Eq.~\ref{cNHSEsupp}], all with $h=0.6$. (a,b) While the GBZ is constant and doubly degenerate (non-fragmented) at small $N$, it bifurcates and becomes fragmented (with $|z_+|\neq |z_-|$ ) beyond that. For the $\Delta=0.05$ coupling used, the $z_-$ solution subsequently becomes constant and doubly degenerate again after it meets with $1/z_-$. 
    (b) The composition ratio $R=c_+/c_-$ for the $z_+$ vs. $z_-$ solutions obeys an asymptotic exponential relationship with $N$, as theoretically given (black curve) in Eq.~\ref{Rcpcm} and its accompanying table, and numerically verified (black dots) here for the largest Re$(E)$ eigenstate with $\Delta=0.05$. 
    (c) Accurate prediction of the fragmentation of $z_+$(blue) and $z_-$(orange) for the illustrative $E_\text{OBC}=-1.512-0.0519i$ eigenstate at $N=30$. The numerical data (dots) closely align with their theoretical predictions (solid curves) $|z_+|=1.76 \Delta^{1/(N+1)}$, $|z_-|=0.947 \Delta^{-1/(N+1)}$ from Eq.~\ref{zpm}.  }
    \label{fig:cNHSE}
\end{figure*}

\subsection{GBZ solution for small interchain coupling $\Delta$}

So far, all the results above have been exact, but without explicit solutions. While exact solutions can only be obtained numerically under the most general scenarios, for small $\Delta$, distinct conclusions on GBZ fragmentation can be made. 

First, we establish that $r_\pm \sim \Delta$ for the two eigenstates $\varphi_\pm$ (the other two eigenstates observe $1/r_\pm$). This actually follows from the form of Eq.~\ref{cNHSEsupp}, as well as its eigenequation Eq.~\ref{DetHsupp}:
\begin{equation}
r_\pm=\frac{\Delta}{E_\text{OBC}-\frac1{z_\pm}-hz_\pm}=\frac{E_\text{OBC}-z_\pm-\frac{h}{z_\pm}}{\Delta}.
 \end{equation}
 Since $E_\text{OBC}\approx z_\pm+\frac{h}{z_\pm}$, with equality in the $\Delta\rightarrow 0$ limit, we see that the denominator on the LHS does not vanish. Hence $r_\pm \propto \Delta$ for small $\Delta$ (in practice, is approximate linear relationship may hold even above $\Delta \sim \mathcal{O}(10^{-1})$).  
 
Substituting $r_\pm \propto \Delta$ into Eq.~\ref{detMred1}, we note that $r_+r_-\ll 1$, such that the RHS is just proportional to $\Delta$. Since the LHS is dominated by the larger of $z_\pm^{N+1}$ (WLOG label it as $z_+^{N+1}$), we thus have $z_+\sim \Delta^{\frac1{N+1}}$. Since $z_+z_-\approx h$ in the $\Delta\rightarrow 0$ limit, we also expect $z_-\sim \Delta^{-\frac1{N+1}}$. To summarize, we have established that 
\begin{equation}
z_\pm \sim  \Delta^{\pm1/(N+1)}\sim 1\pm \frac1{N+1}\log\delta,
\label{zpm}
 \end{equation}
which resembles logarithmic behavior in the limit of large $N$. 
Very good agreement with numerics is demonstrated in Fig.~\ref{fig:cNHSE}(c) for an illustrative eigenstate with complex eigenvalue. Similar fits can be obtained for other eigenstates with complex energies; for real eigenergies, $|z_+|=|z_-|=\sqrt{h}$ remains constant, as if the chains are uncoupled.   

\subsubsection{Inevitability of GBZ fragmentation}

In the context of this model, GBZ fragmentation refers to the fact that $|z_+|\neq |z_-|$, such that there is no pair of roots among $z_\pm$, $1/z_\pm$ that have the same magnitude. Why can it happen here, and why does it not happen without the weak coupling $\Delta$?


To understand why, first note that GBZ fragmentation cannot occur at $\Delta=0$, when we only have uncoupled Hatano-Nelson chains with $E_\text{OBC}=\frac1{z_\pm}+hz_\pm$ i.e. $z_\pm = \frac1{2h}\left(E_\text{OBC}\pm \sqrt{E_\text{OBC}^2-4h}\right)$. 
 For real positive $h$, a complex $E_\text{OBC}$ would never give rise to a $\pi/2$ phase difference between $E_\text{OBC}$ and  $\sqrt{E_\text{OBC}^2-4h}$, which is necessary for $|z_+|=|z_-|$. Indeed, this GBZ condition $|z_+|=|z_-|$ only holds for real $E_\text{OBC}$ with $|E_\text{OBC}|<2\sqrt{h}$. 

 Next, let us understand how the $\Delta$ interchain coupling leads to GBZ fragmentation. In essence, that is because the OBC condition, which is entirely encapsulated in the condition Det$M=0$, now allows for $E_\text{OBC}$ solutions that are not allowed for uncoupled chains. For our specific model, they are energies outside of the real line segment $|E_\text{OBC}|<2\sqrt{h}$; in general, they are solutions that do not satisfy the GBZ condition for uncoupled chains. 
 
 Now, it is important to recognize that Det$M=0$ does \emph{not} in general imply that $|z_+|=|z_-|$ (GBZ condition) -- we have previously seen that this condition only arises in $M$ matrices that are constrained in particular ways. In our specific model Eq.~\ref{cNHSEsupp}, $M$ contains both positive and negative powers of $z_\pm$ [Eq.~\ref{M2by2}], resulting in the constraint equation Eq.~\ref{detMred1} or, equivalently, $\tanh^{-1}r_--\tanh^{-1}r_-+=\tanh^{-1}\left(z_+^{N+1}\right)-tanh^{-1}\left(z_-^{N+1}\right)$. Importantly, this constraint can in general be satisfied for $r_-\neq r_+$, which from Eq.~\ref{rpm} implies different values of $1/z_\pm-hz_\pm$ and hence $|z_\pm|$.

\subsubsection{Perturbative results with $\Delta$}

Below, we also end off with some analytic perturbative results of the GBZ with respect to $\Delta$. We employ a perturbative approach to Eq.~\ref{DetHsupp}, which relates the GBZ solutions $z=z_\mu$ (i.e. $z_\pm$ and $1/z_\pm$) and their corresponding 2-component eigenvectors $\varphi_\mu$ to $E_\text{OBC}$ and $\Delta$. While the eigensolutions to Eq.~\ref{DetHsupp} quickly become analytically intractable when substituted directly into Eqs.~\ref{M2by2} to~\ref{detMred2}, much analytic headway can be made in the small $\Delta$ limit, where $\mathcal{O}(\Delta^2)$ contributions can be neglected.

Throughout the discussion below, we shall take $E_\text{OBC}$ as a fixed value corresponding to one of the OBC eigenvalues when the interchain coupling is fixed at $\Delta\neq 0$. To derive the GBZ solutions, we first start from the tractable case when $\Delta$ is removed from Eq.~\ref{zUncoupled} (while keeping $E_\text{OBC}$ unchanged). Specifically, at this fixed $E_\text{OBC}$, we can easily write down the 4 ``uncoupled" GBZ solutions as $z_0$, $h/z_0$, $1/z_0$ and $z_0/h$, where $z_0$ is taken to be the larger of the two solutions of 
\begin{equation}
z_0+\frac{h}{z_0}=E_\text{OBC}.
\label{zUncoupled}
\end{equation}
Here $|z_0|\neq \sqrt{h}$, since $E_\text{OBC}$ is not an eigenenergy of the truly uncoupled Hatano-Nelson model. Note that while $h/z_0$ is automatically also a solution to both Eqs.~\ref{zUncoupled} and Eq.~\ref{DetHsupp}, the other two solutions $1/z_0$ and $z_0/h$ satisfy only Eq.~\ref{DetHsupp}.

When the small $\Delta$ is restored, we expect these GBZ solutions to be only slightly perturbed from these values. We take $z_+$ to be the perturbed version of $z_0$:
\begin{equation}
z_+=z_0(1+\delta).
\end{equation}
Substituting it into Eq.~\ref{DetHsupp} and doing a series expansion in $\delta$, we obtain
\begin{equation}
h\delta(2+\delta)\left(z_0^2+\frac1{z_0^2(1+\delta)^2}\right)+\delta(1+h)\left(z_0+\frac{h}{z_0}\right)\left(\frac1{z_0(1+\delta)}-z_0\right)=\Delta^2.
\end{equation} 
Keeping only the lowest order term in $\delta$, we obtain the closed-form solution
\begin{align}
z_+&\approx z_0\left(1-\frac{\Delta^2 z_0^2}{(1-h)(1-z_0^2)(h-z_0^2)}\right)\notag\\
&=z_0\left(1+\frac{\Delta^2 }{8h\sinh\left[\log\sqrt{h}\right]\sinh\left[\log z_0\right]\sinh\left[\log \frac{z_0}{\sqrt{h}}\right]}\right).
\label{zp}
\end{align} 
Evidently, even for small $\Delta$, $z$ can differ significantly from $z_0$ if $h\approx 1$, $z_0^2\approx 1$ or $z_0^2\approx h$. The first two conditions correspond the lack of the NHSE in each individual chain, a trivial scenario which we do not focus on. The last condition $z_0^2\approx h$ implies that the two solutions $z_0$, $h/z_0$ of Eq.~\ref{zUncoupled} are in fact almost equal, a scenario not generically satisfied by the GBZ condition (which requires $z_0$ and $h/z_0$ to have equal amplitudes but different phase).


By performing a similar perturbative analysis on $h/z_0$, we also obtain
\begin{align}
z_-&\approx \frac{h}{z_0}\left(1-\frac{h\Delta^2 z_0^2}{(1-h)(z_0^2-h^2)(h-z_0^2)}\right)\notag\\
&=\frac{h}{z_0}\left(1+\frac{\Delta^2 }{8h\sinh\left[\log\sqrt{h}\right]\sinh\left[\log \frac{h}{z_0}\right]\sinh\left[\log \frac{z_0}{\sqrt{h}}\right]}\right).
\label{zm}
\end{align} 
Evidently, when $\Delta\neq 0$, $z_+z_-\neq h$ and $z_\pm$ no longer satisfy Eq.~\ref{zUncoupled}:
\begin{equation}
z_+z_- \approx h\left(1+\frac{\Delta^2z_0^2(h+z_0^2)}{(h-z_0^2)(h^2-z_0^2)(1-z_0^2)}\right). 
\end{equation} 

\end{itemize}

\subsection{D. Impossibility of extended higher-fold GBZ degeneracies}

We now show why it is generically not possible for 3-fold degenerate GBZs to exist along a curve in the complex energy plane. Consider a general Hamiltonian $H(z)$, $z=e^{ik}$ where $k$ is the momentum that would be effectively complex-deformed under OBCs. Recall that, conventionally, a given $E$ belongs to the OBC skin spectrum if the dispersion polynomial $\text{Det}[H(z)-\mathbb{I}E]=0$ contains two $z,z'$ solutions with $|z|=|z'|$ i.e. has a 2-fold degenerate GBZ. Intuitively, since $|z|=|z'|$ constitutes a single constraint in the 2D complex $E$ plane, the solution set must mainly comprise 1D curves or line segments.

To understand why solution sets with triply (or higher) degenerate $|z|=|z'|=|z''|$ cannot in general form a 1D continuum on the 2D complex energy plane, we note that $\text{Det}[H(z)-E\mathbb{I}]=0$ takes the form of a Laurent polynomial in $z$, with coefficients being polynomials in $E$.  To have (usual) doubly degenerate $|z|=|z'|$, we just need their solution to be of the form $z_\pm \propto A(E)\pm i\sqrt{B(E)}$, where $A(E)$ and $i\sqrt{B(E)}$ are functions of $E$ with possessing e a relative phase of $\pi/2$. This is a single constraint which can be generically satisfied in a 1D subspace of the 2D complex $E$ plane.

However, to have higher degeneracies i.e. $|z|=|z'|$ and $|z'|=|z''|$, we require the coincidence of more than one such constraint. In general, that can only exist at the intersections of 1D solution curves, i.e. isolated points. To our knowledge, unlike in the quadratic case, there is no single constraint that can give rise to $3$ or more roots related by multiplication of the respective root of unity~\cite{hernandez2024introduction}.

\begin{figure}[h]
    \includegraphics[width=\linewidth]{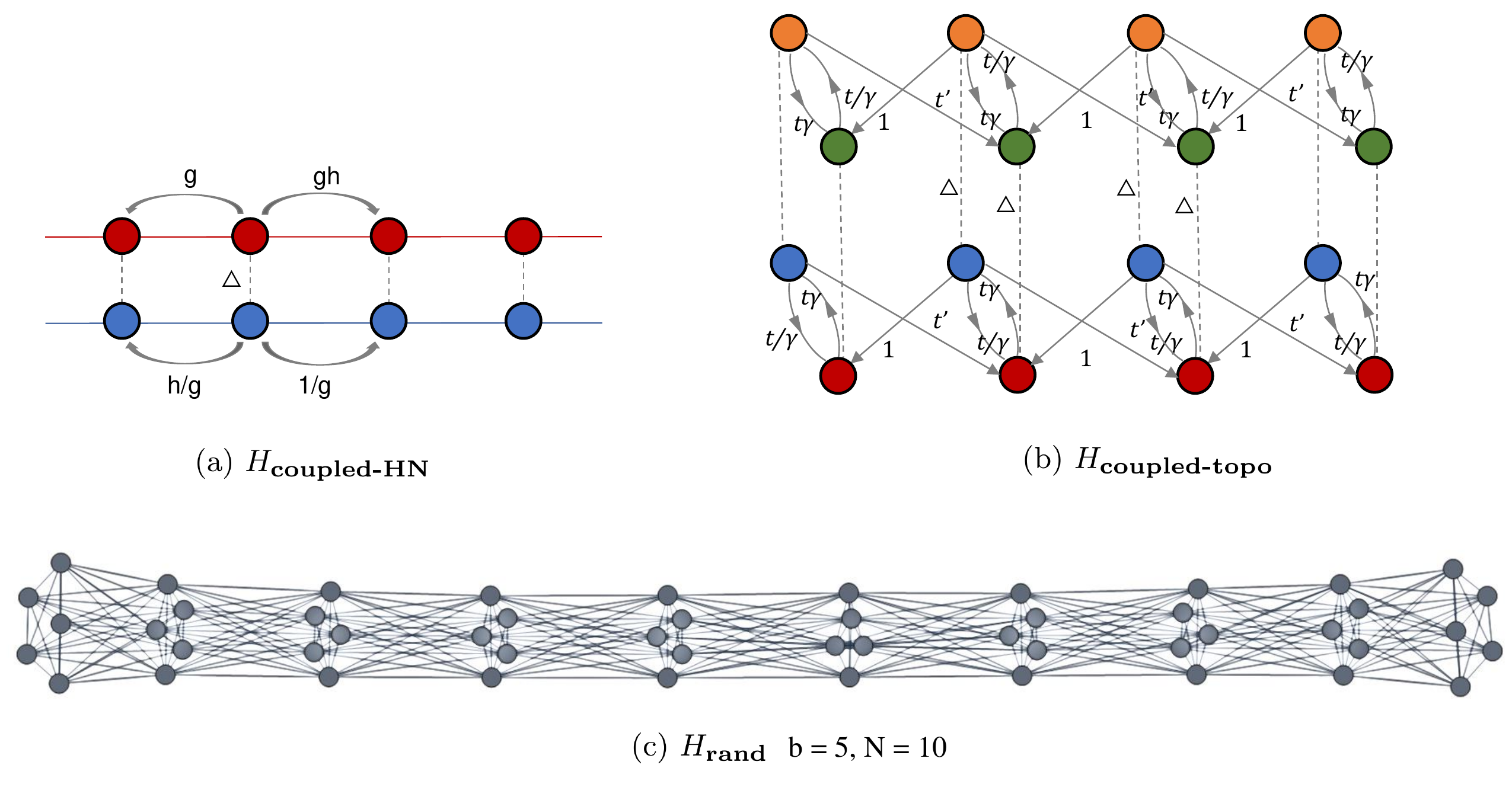}
    \caption{Sketches of the models corresponding to (a) Eq.~\ref{cNHSEgsupp}, (b) Eq.~\ref{Ctoposupp} and (c) Eq.~\ref{5bandsupp}, illustrated with a few unit cells for (a,b), and $N=10$ unit cells for (c). The bonds in (c) represent random asymmetric couplings. }
    \label{fig:models}
\end{figure}

\begin{figure}[h]
    \includegraphics[width=1\linewidth]{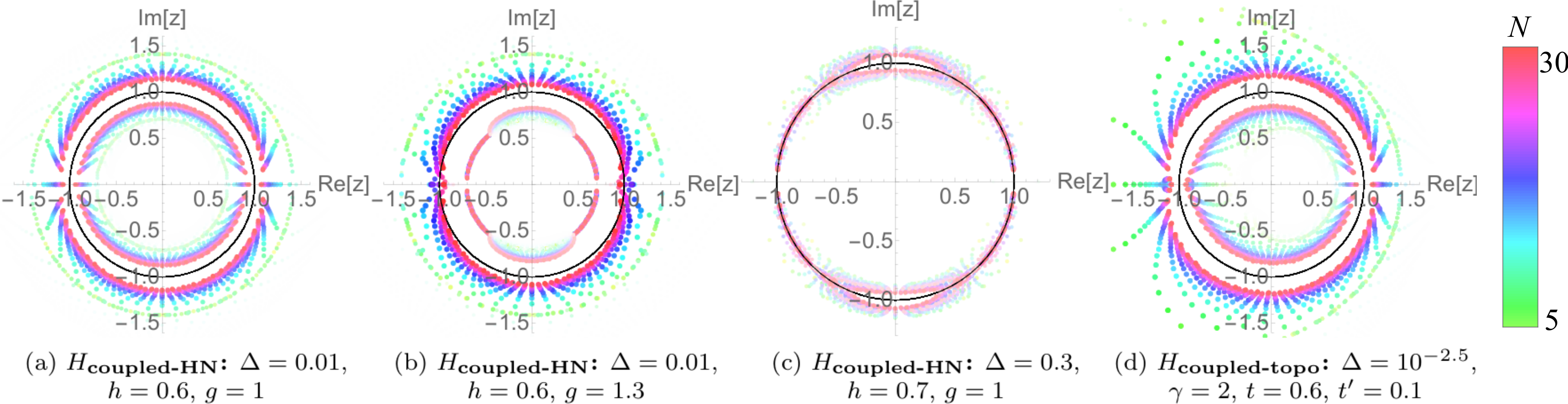}
   \caption{Evolution of the GBZ with system size $N$ for various models, from $N=5$ unit cells (green) to $N=30$ (pink).
   More intense colors correspond to greater GBZ weight $|c_\mu|^2$ computed at  $x_0=N/2$ [Eqs.~\ref{psixsupp} and~\ref{Mmunu}]. Shown are models (a-c) $H_\text{coupled-HN}$ [Eq.\ref{cNHSEgsupp}] and (d) $H_\text{coupled-topo}$ [Eq.\ref{Ctoposupp}], all containing subchains with antagonistic NHSE. All GBZs shown exhibit significant evolution with $N$, except in (c) where only the more fragmented (fainter) top and bottom regions shift with $N$. }
    \label{fig:sizeNvar}
\end{figure}

\section{II. Further results for GBZ fragmentation}
In this section, we showcase numerical results of paradigmatic models that exhibit fragmented GBZs, such as to elaborate on the anatomy of the GBZ composition, particularly on how the bulk solutions can conspire to satisfy the OBCs. The simplest model, which is also Eq.~6 in the main text, consists just of 2 oppositely-directed Hatano-Nelson (HN) chains coupled by $\Delta$ [Fig.~\ref{fig:models}(a)]:
\begin{equation}
H_\text{coupled-HN}(z) = \left(\begin{matrix}
g\left(z+h/z\right) & \Delta \\
\Delta & \left(1/z+hz\right)/g
\end{matrix}\right).
\label{cNHSEgsupp}
\end{equation}
Here $g$ controls the relative weight of the HN chains; the $g=1$ case [Eq.~\ref{cNHSEsupp}] has been analytically studied extensively in the previous section. Due to the antagonistic NHSE controlled by $g$ and the hopping asymmetry $h$, nontrivial competition exists between the GBZs inherited from the individual chains. The NHSE can never be completely canceled when $g\neq 1$, which has never been studied in the literature.

We also introduce a slightly more complicated key model $H_\text{coupled-topo}$, which we heavily use in this work to demonstrate how GBZ fragmentation interplays with topological modes and multiple bands. It consists of two coupled zig-zag chains which possess the same asymmetric hoppings across unit cells, as well as oppositely-directed $t$ hoppings within each unit cell (see Fig.~\ref{fig:models}(b)). 
Mathematically, it takes the form
\begin{equation}
H_\text{coupled-topo}(z)=\left(\begin{matrix}
0   & t\gamma +z+t'/z & \Delta & 0 \\
t/\gamma +1/z +t'z & 0 & 0 & \Delta\\
\Delta & 0 & 0 & t/\gamma +z +t'/z \\
0 & \Delta & t\gamma +1/z + t'z  & 0 
\end{matrix}\right),
\label{Ctoposupp}
\end{equation}
which resembles two extended non-Hermitian SSH chains coupled by $\Delta$.  Even though only the $t$-bonds are oppositely directed, they are sufficient to give rise to competing GBZ solutions. Each individual chain also possesses topological modes for certain combinations of $t$ and $t'$ -- while we do not need both $t$ and $t'$ to have the topological modes, they are both included such as to avoid complications from defectiveness of the $M$ matrix (which occurs if $t'=0$, as shown earlier in supplementary Sect. I.C).

\begin{figure}[h]

   \includegraphics[width=1\linewidth]{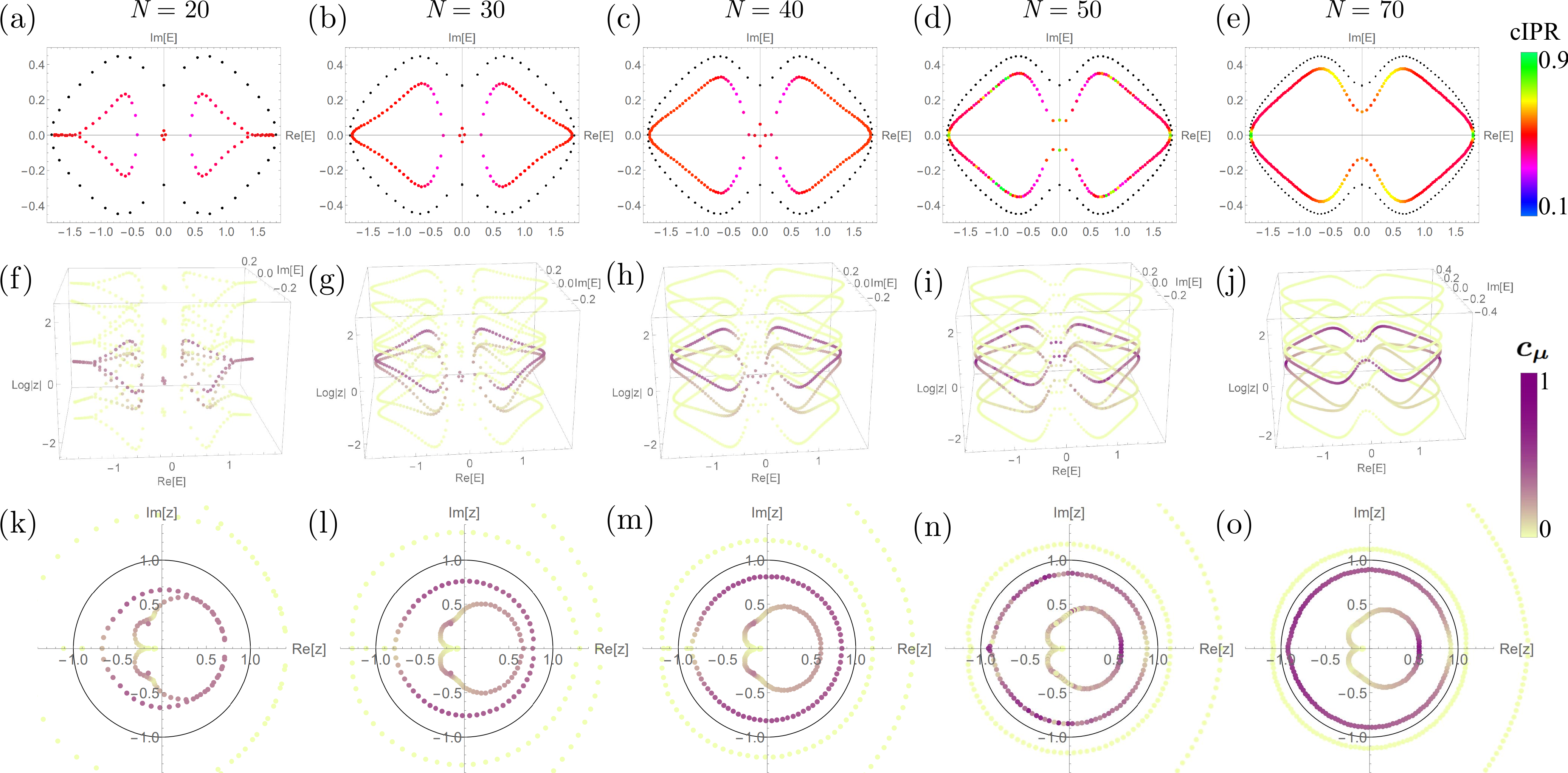}
   \caption{Variation of the spectra, GBZ composition and GBZ of $H_\text{coupled-topo}$ [Eq.~\ref{Ctoposupp}] across different system sizes $N$, with coupling $\Delta=10^{-4}$, hoppings $t=0.6$, $t'=0.1$ and asymmetry $\gamma=2$. Significant spreading i.e. fragmentation (yellow $\rightarrow$ purple) of the bulk contributions across different $|z|$ occur at all $N$, not just confined to the topological transition.}
    \label{fig:sizeNtopo}
\end{figure}

\subsection{A. Variation of the GBZ with system size $N$}

Since the $M$ matrix depends on $N$ [Eq.~\ref{Mmunu}], the allowed $E_\text{OBC}$ and $c_\mu$ composition coefficients also generically depend significantly on the system size $N$, unless the usual GBZ condition (where only two $z_\mu,z_\nu$ solutions dominate) is met. 

The system size dependence of the GBZs of $H_\text{coupled-HN}$ and $H_\text{coupled-topo}$ are shown in Fig.~\ref{fig:sizeNvar}, for illustrative parameters. In general, small inter-chain couplings $\Delta$ give rise to stronger variations with $N$ (a,b,d), even though the GBZs still do not converge into one non-fragmented ring at large $N$ (magenta). When the chains are weighted unequally by $g\neq 1$ (b), the GBZ rings also break up into qualitatively distinct segments with different $N$-dependent behavior, a result of the interplay between dissimilar hopping directions and strength. At the special choice of strong coupling $\Delta=0.3$ which prevents any amplification in the translation-invariant setting (such that the PBC spectrum becomes real), the GBZ is still not entirely Bloch (on the unit circle), and some fragmentation and variation with $N$ can be seen at the top and bottom regions.

The evolution of the GBZ with $N$ for $H_\text{coupled-topo}$ is further elaborated in Fig.~\ref{fig:sizeNtopo}, in the weak coupling regime of $\Delta=10^{-4}$. Its spectrum (Top Row) transitions from topologically gapped to a non-topological point gap (loop) as $N$ is increased, with some GBZ fragmentation (magenta) mostly observed near the gap crossing. Along these states, some significant distribution of bulk contributions $\phi_\mu$ across different $|z|$ can be observed (Middle Row). Correspondingly, they appear as region of fuzziness (faint yellow) in the GBZ loci (Bottom Row). These are also the same regions that "absorb" the topological states (isolated dots) as the topological transition is crossed.

Notably, the topological transition does not seem to occur at a well-defined fixed instance, but is instead spread out considerably around $N\approx 40$. In the next subsection, we will examine why and how GBZ fragmentation can fundamentally lead to the absence of a well-defined topological transition point.

\begin{figure}[h]
    \includegraphics[width=1\linewidth]{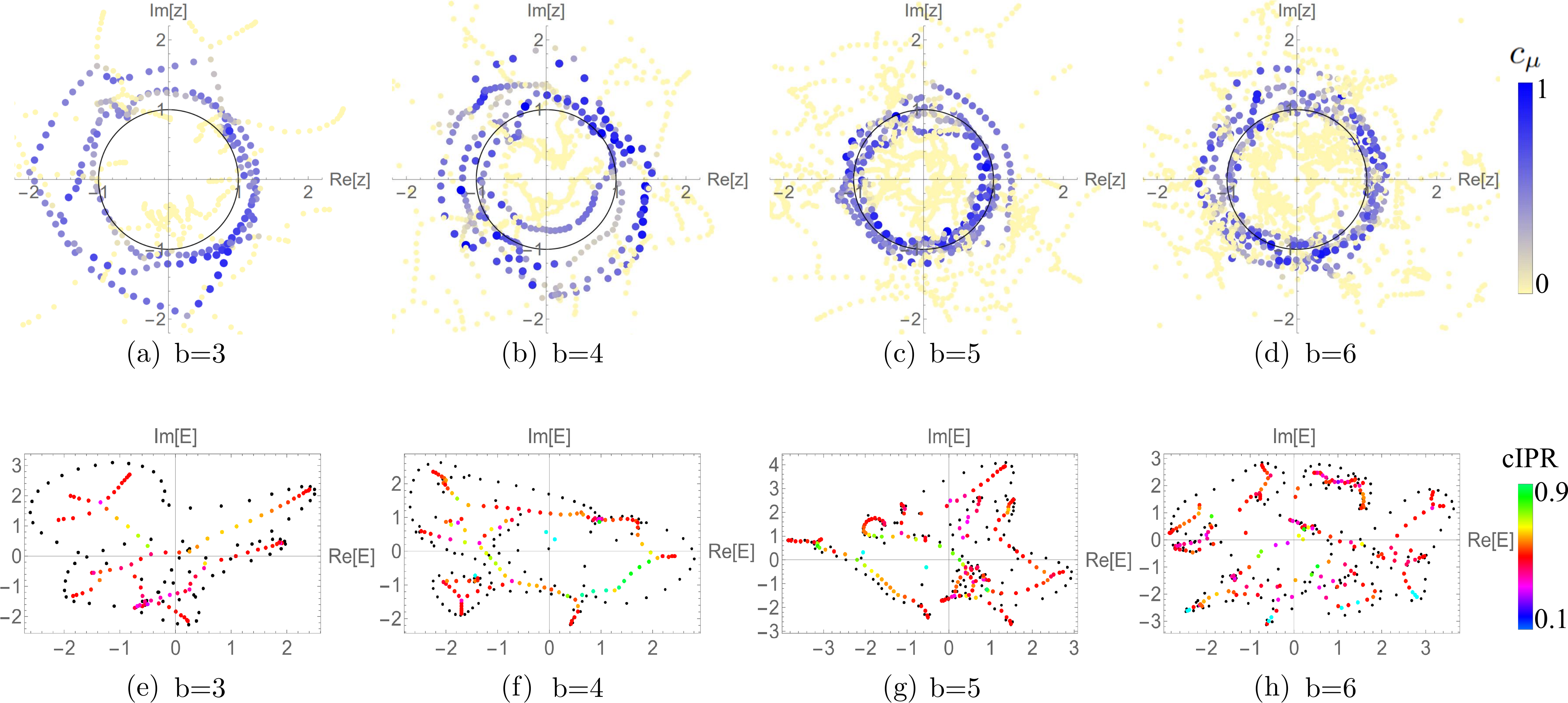}
    \caption{GBZ and spectra of illustrative $b$-component models $H_\text{rand}$ with randomly-generated hoppings across all atoms within and across NN unit cells [Eq.~\ref{5bandsupp}], all with $N=30$ unit cells. As $b$ increases, fragmentation (faint blue in the GBZ and non-red eigenvalues in the spectrum) becomes inevitable. }
    \label{fig:random3456}
\end{figure}

\begin{figure}[h]
    \includegraphics[width=.83\linewidth]{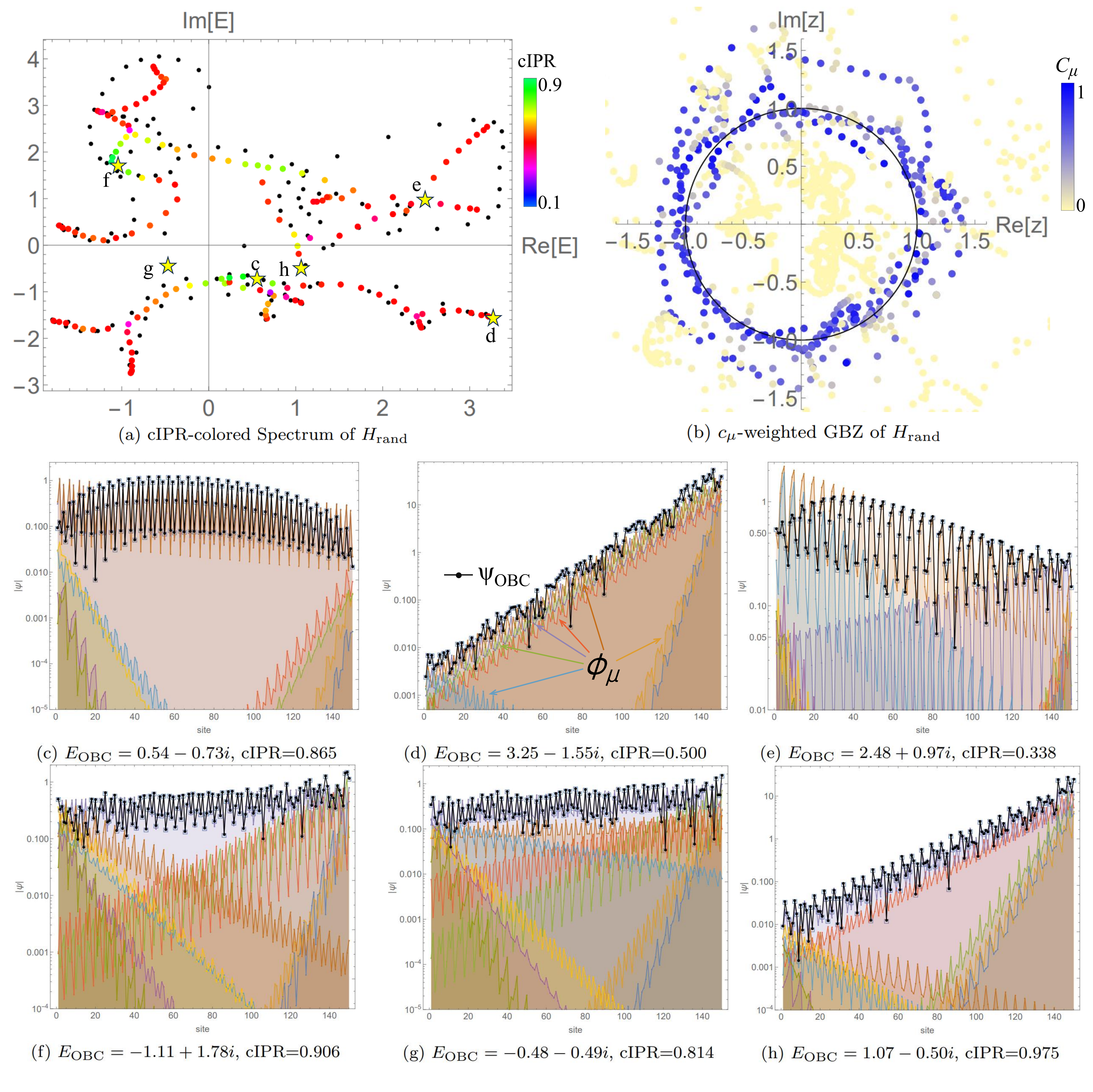}
    \caption{Spectrum, fragmented GBZ and the composition of six illustrative OBC eigenstates of $H_\text{rand}$ with $3b^2=75$ random hopping coefficients given by Eqs.~\ref{5bandsupp} to~\ref{A0}. The GBZ composition coefficients $c_\mu$ are computed with reference $x_0=N/2$.  Due to GBZ fragmentation, OBC eigenstates $\psi_\text{OBC}$ (black) are either dominated by one $\phi_\mu$ (colored) in the bulk and several of them at the boundary (c,f-h), or dominated by a few distinct $\phi_\mu$ (colored) in the bulk (e). For reference, an example state resembling a conventional GBZ state (d) is shown, with cIPR$=0.5$ and bulk solutions $\phi_\mu$ that decay exactly at the same rate as $\psi_\text{OBC}$. }
    \label{fig:random5}
\end{figure}

\subsection{B. Inevitable GBZ fragmentation in multi-component models with random hoppings}

GBZ fragmentation generically occur when a system contains subchains with different NHSE hopping asymmetries. As such, we expect it to inevitably occur in models with complicated unit cells, unless the hoppings are constrained by appropriate symmetries. To showcase this, we consider random models with $b$ atoms per unit cell, such that randomly asymmetric hoppings exist between every pair of atoms within each unit cell, as well as between neighboring unit cells [Fig.~\ref{fig:models}(c)]:
\begin{equation}
H_\text{rand}(z)=A_pz + \frac{A_m}{z} + A_0,
\label{5bandsupp}
\end{equation}
where $A_p$, $A_m$ and $A_0$ are $b\times b$ random matrices with complex elements uniformly distributed within the $[-1-i,1+i]$ rectangle. 

Shown in Fig.~\ref{fig:random3456} are the GBZs and spectra of four illustrative random realizations of $H_\text{rand}$, for $b=3$ to $b=6$ atoms per unit cells. While $b$ GBZ loops can still be vaguely made out for $b=3$, as $b$ increases, they become fuzzier. By $b=5$ and $b=6$, the GBZ essentially fragmentates into a fuzzy region with little discernible structure, with $c_\mu$ composition (color intensity) taking on a stochastic nature. The onset of GBZ fragmentation is also evident in the cIPR of the eigenenergies in the accompanying spectral plots (Bottom Row), whose departure from red (cIPR$=0.5$) indicates departure from the usual GBZ constraint $|z_\mu|=|z_\nu|$.

\begin{figure}[h]
    \includegraphics[width=\linewidth]{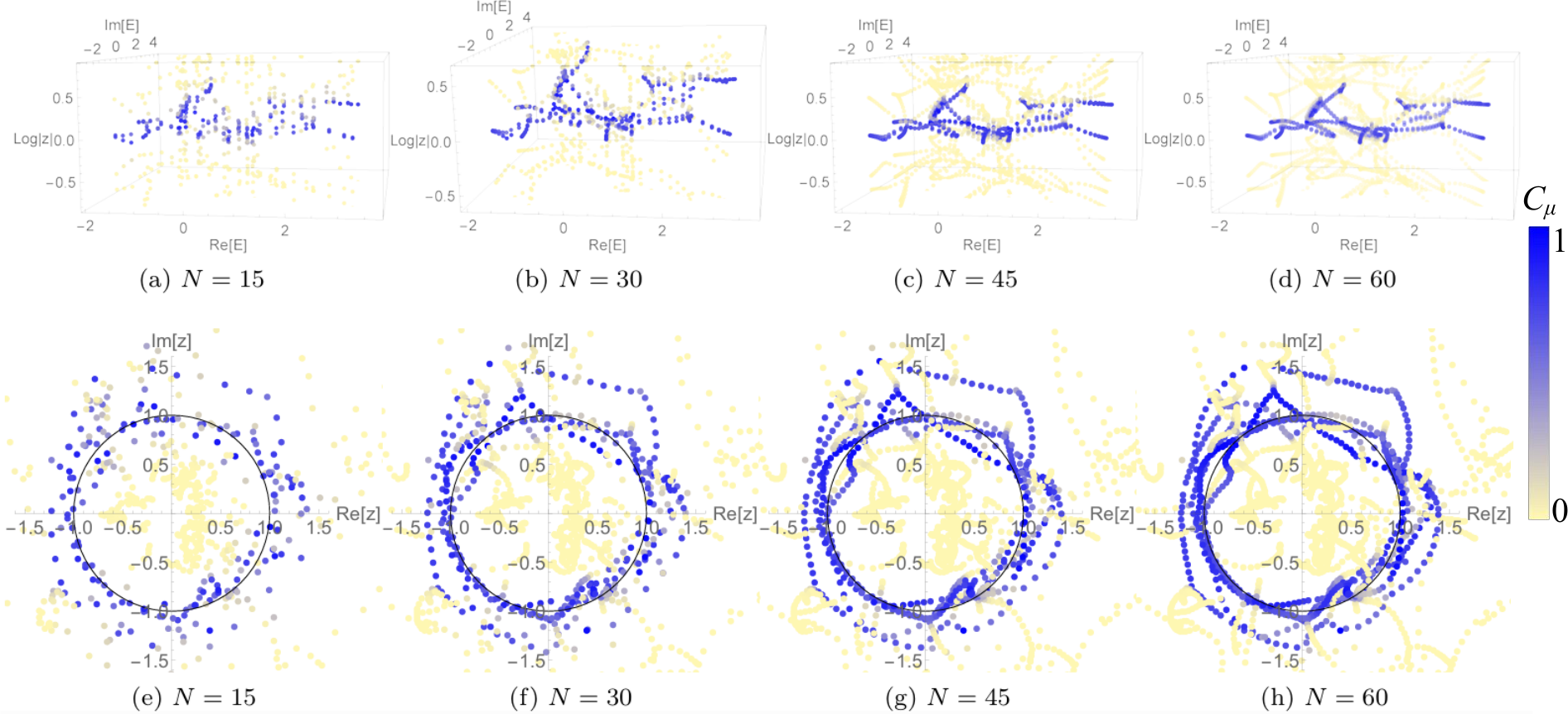}
    \caption{The GBZ composition of the spectrum (Top Row) and GBZ (Bottom Row) of the 5-component random model $H_\text{rand}$ as given by Eqs.~\ref{5bandsupp} to~\ref{A0}, colored by cIPR [Eq.~\ref{cIPR}].}
    \label{fig:random5size}
\end{figure}

We next closely examine the $b=5$-component random instance of $H_\text{rand}$ featured in main text Fig.~1, with coefficient matrices given by
\begin{equation}
A_p=\left(
\begin{array}{ccccc}
 0.49\, +0.679 i & -0.614+0.642 i & -0.268-0.409 i & -0.061-0.325 i & -0.196+0.102 i \\
 0.06\, -0.385 i & -0.311-0.487 i & -0.557-0.954 i & 0.78\, +0.434 i & -0.19-0.499 i \\
 0.256\, +0.905 i & 0.301\, -0.078 i & 0.419\, +0.234 i & -0.27+0.293 i & -0.241-0.86 i \\
 -0.839-0.542 i & 0.2\, +0.697 i & -0.682-0.125 i & -0.124-0.886 i & 0.182\, +0.457 i \\
 0.937\, -0.478 i & 0.893\, +0.364 i & -0.953+0.52 i & 0.637\, -0.328 i & 0.591\, -0.769 i \\
\end{array}
\right),
\end{equation}
\begin{equation}
A_m=\left(
\begin{array}{ccccc}
 -0.935-0.13 i & 0.377\, -0.252 i & 0.839\, -0.41 i & 0.33\, -0.863 i & -0.25+0.246 i \\
 0.026\, +0.62 i & 0.72\, +0.878 i & -0.738-0.949 i & -0.942-0.851 i & 0.075\, -0.244 i \\
 -0.314+0.697 i & -0.075+0.614 i & -0.981+0.228 i & -0.166+0.35 i & 0.07\, +0.082 i \\
 0.04\, -0.189 i & -0.157-0.368 i & 0.078\, +0.488 i & -0.06+0.358 i & 0.512\, +0.248 i \\
 0.542\, -0.75 i & 0.418\, +0.356 i & -0.835-0.417 i & 0.539\, -0.964 i & -0.368-0.557 i \\
\end{array}
\right),
\end{equation}
\begin{equation}
A_0=\left(
\begin{array}{ccccc}
 0.636\, +0.938 i & -0.885-0.381 i & -0.935+0.249 i & -0.505-0.124 i & 0.063\, +0.138 i \\
 0.308\, +0.038 i & 0.615\, -0.548 i & 0.599\, +0.7 i & -0.315-0.617 i & 0.781\, +0.579 i \\
 0.966\, -0.558 i & -0.212-0.681 i & -0.005+0.588 i & -0.403-0.707 i & -0.57+0.7 i \\
 0.792\, -0.133 i & -0.619+0.185 i & 0.482\, +0.714 i & 0.504\, +0.824 i & -0.655+0.163 i \\
 -0.268+0.566 i & -0.484+0.298 i & 0.701\, -0.899 i & 0.955\, +0.684 i & 0.326\, +0.203 i \\
\end{array}
\right).
\label{A0}
\end{equation}
Its cIPR-colored spectrum, GBZ and six illustrative composition-resolved OBC eigenstates are shown in Fig.~\ref{fig:random5}. The latter plots showcase how each eigenstate $\psi_\text{OBC}$ (black) is made up of its constituent bulk solutions $\phi_\mu(x)$ (colored). For very large cIPR $\approx 1$ (taken with respect to $x_0=N/2$ at the Center) i.e. (c), (f) and (h), only one $\phi_\mu$ dominates $\psi_\text{OBC}$ in the bulk, and the OBCs are satisfied at both ends with the help of several other bulk solutions $\phi_\mu$ that decay quickly into the bulk. This is different from the conventional GBZ scenario, as satisfied by eigenstate (d) with cIPR$=0.5 $ , where $\psi_\text{OBC}$ is contributed by two bulk solutions that also decay at the same rate. Another distinct scenario occurs where the GBZ fragmentates into 3 dissimilar constituent bulk solutions (cIPR$\approx 1/3$, as in (e)) -- in such cases, the 3 bulk solutions (blue, purple, orange) all decay very differently from the $\psi_\text{OBC}$. 

For completeness, the scaling behavior of the GBZ composition for this random $5$-component model is also plotted in Fig.~\ref{fig:random5size}. Even at large sizes $N$, some eigenstates maintain their fragmented GBZ nature, appearing blurry in $z$.

\section{III. Physical consequences of GBZ fragmentation}

\subsection{A. Biorthogonal current distribution and GBZ fragmentation}

Biorthogonal expectations are encountered in physical measurements of operators that are energetically weighted. That is because left and right energy eigenvectors are needed in expanding a state in the energy basis: $\langle ...\rangle \rightarrow \sum_j e^{-\beta E_j}|\phi_j^R\rangle\langle \phi_j^L|$, where $H|\phi_j^R\rangle=E_j|\phi_j^R\rangle$ and $H^\dagger|\phi_j^L\rangle=E^*_j|\phi_j^L\rangle$. Note that it is this biorthogonal basis that gives the correct form of the projection operator: $\left[\sum_j |\phi_j^R\rangle\langle \phi_j^L|\right]^2 = \sum_j |\phi_j^R\rangle\langle \phi_j^L|$, since $\langle \phi_j^L|\phi^R_{j'}\rangle=\delta_{jj'}$.

\noindent For a given eigenenergy $E_j$, its right OBC eigenstate is:
\begin{equation}
|\psi^R_\text{j}(x)\rangle = \sum_\mu c_{\mu,j} z_{\mu,j}^{x-x_0}|\varphi_{\mu,j}\rangle,
\end{equation}
and its left eigenstate (which experiences reversed NHSE) is given by
\begin{equation}
|\psi^L_\text{j}(x)\rangle = \sum_\mu c_{\mu,j} z_{\mu,j}^{x_0-x}|\varphi_{\mu,j}\rangle,
\end{equation}
such that $\langle \psi^L_\text{j}(x)|\psi^R_\text{j}(x)\rangle=\sum_{\mu\mu'}c_{\mu,j} c^*_{\mu',j}z_{\mu,j}^{x-x_0}z_{\mu',j}^{x_0-x}\langle \varphi_{\mu',j}|\varphi_{\mu,j}\rangle = \sum_\mu |c_{\mu,j}|^2=1$. 
Hence, the expectation of an observable $\hat O$ in a thermal ensemble of inverse temperature $\beta $ is
\begin{eqnarray}
\langle\hat O(x)\rangle &=& \frac1{Z}\sum_j e^{-\beta E_j}\langle\psi^L_\text{j}(x)|\hat O|\psi^R_\text{j}(x)\rangle\notag\\
&=&\frac1{Z}\sum_j\sum_{\mu\mu'}  e^{-\beta E_j}c_{\mu,j} c^*_{\mu',j}\left(\frac{z_{\mu,j}}{z_{\mu',j}}\right)^{x-x_0}\langle \varphi_{\mu',j}|\hat O|\varphi_{\mu,j}\rangle,
\label{hatOsupp}
\end{eqnarray}
where $Z=\sum_j e^{-\beta E_j}$. For conventional GBZs where the only non-negligible GBZ contributions are $z_{\mu,j},z_{mu',j}$ with $||z_{\mu',j}^{x_0-x}|$, the factor $\left(\frac{z_{\mu,j}}{z_{\mu',j}}\right)^{x-x_0}$ becomes just a phase, and $\langle\hat O(x)\rangle$ does \emph{not} contain any exponential spatial profile i.e. does not reveal any NHSE. However, for fragmented GBZs, there exists at least some pairs of $|z_{\mu,j}|\neq|z_{\mu',j}|$ with non-negligible composition. As such, 
\begin{equation}
\langle\hat O(x)\rangle\sim \text{Max}_{\mu',\mu,j}\left|\frac{z_{\mu,j}}{z_{\mu',j}}\right|^{x}
\end{equation}
is growth-dominated by the ratio of the largest and the smallest $|z_{\mu,j}|$ at very large $x$. For typical $x$, Eq.~\ref{hatOsupp} involves the competition between a large number of $\left(\frac{z_{\mu,j}}{z_{\mu',j}}\right)^{x-x_0}$, weighted by the composition coefficients $c_{\mu,j} c^*_{\mu',j}$, the overlap integral $\langle\varphi_{\mu',j}|\hat O|\varphi_{\mu,j}\rangle$, and of course the Boltzmann weight $e^{-\beta E_j}$. But the key conclusion that holds is that, \textbf{in the presence of GBZ fragmentation,  $\langle\hat O(x)\rangle$ becomes highly spatially non-uniform, typically exhibiting skin-like accumulation near the edges.}

\begin{figure}[h]
    \includegraphics[width=\linewidth]{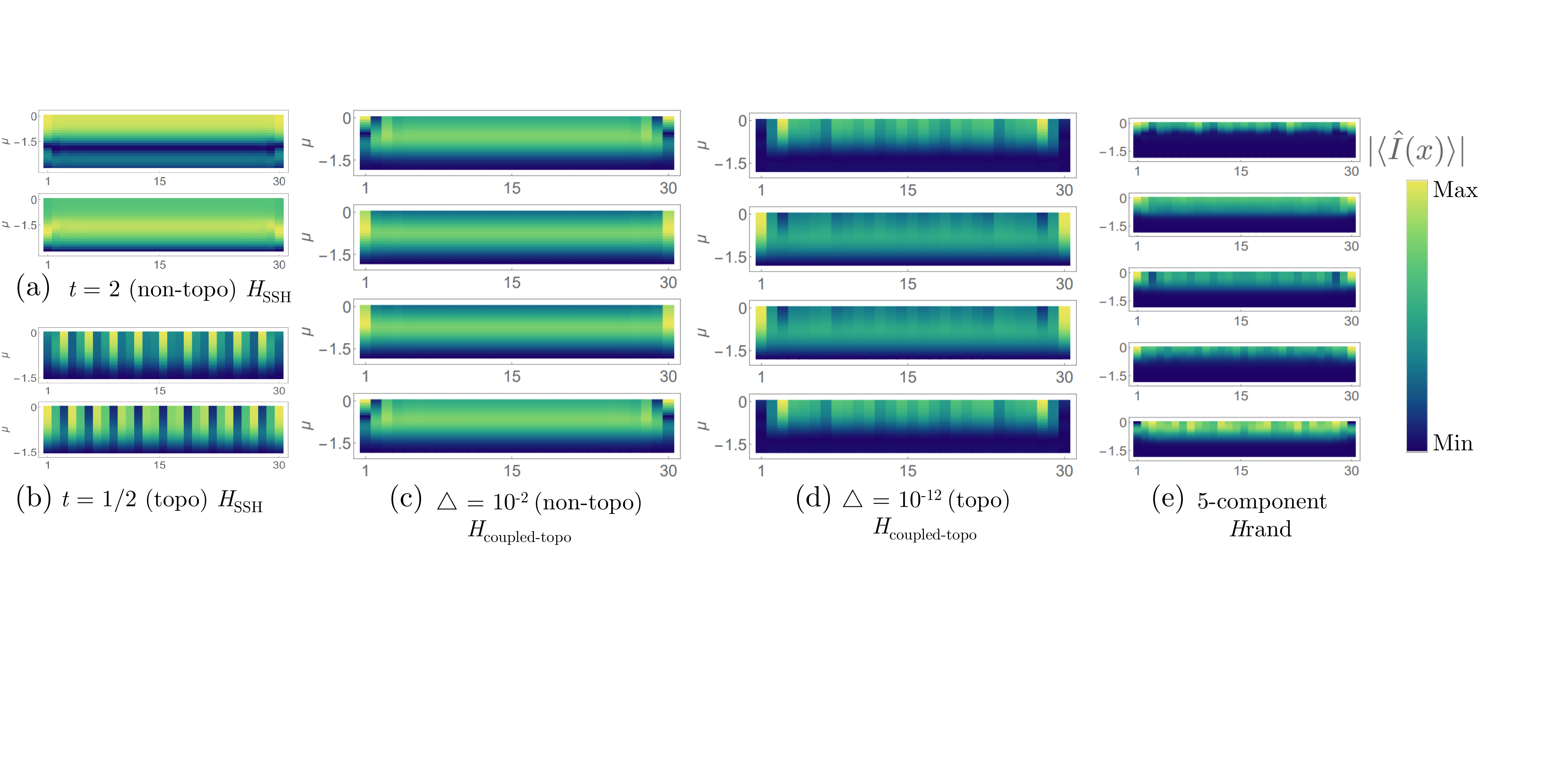}
    \caption{Tunneling current amplitude profile $|\langle\hat I(x)\rangle|$ [Eq.~\ref{currentsupp}], all at inverse temperature $\beta=5$ and system size $N=30$. Each row corresponds to one sublattice. Models are (a,b) Non-Hermitian SSH model [Eq.~\ref{SSH}] with $\gamma=2$, (c,d) Coupled extended SSH model [Eq.~\ref{Ctoposupp}] with $\gamma=2$, $t=0.6$ and $t'=0.1$, and (e) 5-component random hopping model $H_\text{rand}$ [Eq.~\ref{5bandsupp} to~\ref{A0}]. The SSH model in (a) and (b) do not exhibit GBZ fragmentation, and should rightly possess homogeneous $\langle\hat I(x)\rangle$, even though striations exist in (b) due to the breaking of 4-fold translation symmetry by the topological modes. The coupled extended SSH model in (c,d) exhibits distinctly amplified current $\langle\hat I(x)\rangle$ near the edges due to GBZ fragmentation. Notably, this edge amplification has nothing to do with topological edge states, appearing in both non-topological and topological cases. While not explicitly constructed to harbor NHSE competition, the random model in (e) contains $b^2=25$ asymmetric hopping channels and inevitably hosts some GBZ fragmentation. This is manifested in the edge amplification and spatial inhomogeneity of $\langle\hat I(x)\rangle$ in most of its sublattices.}
    \label{fig:current_supp}
\end{figure}

In Fig.~\ref{fig:current_supp}, we present the expected NN tunneling current  
\begin{equation}
\langle\hat I(x)\rangle =i\left\langle b^\dagger_{x+1}b_x - b^\dagger_{x}b_{x+1} \right\rangle
\label{currentsupp}
\end{equation}
for a few illustrative 1D models. As described in the captions, edge current accumulation exists whenever there is GBZ fragmentation, often non-uniformly across the sublattices, depending on where NHSE competition rages. These cases are contrasted with those systems possessing only topological modes, or non-competitive NHSE modes (i.e. the usual HN or SSH models), who invariably do not show any edge current accumulation.

\subsection{B. Topological transition with fragmented vs. non-fragmented GBZs}

It is well-established that the topological character of a 2-component model $H(z)=\left(\begin{matrix} 0 & b(z)\\ a(z) & 0 \end{matrix}\right)$ can be determined by the winding of $U(z)=\sqrt{\frac{a(z)}{b(z)}}$ as $z$ is cycled around the GBZ. This can be shown by directly evaluating the Berry connection,  
or by examining the off-diagonal terms of the biorthogonal band projector of $H(z)$.  
However, this conventional result assumes the existence of a well-defined GBZ loop $z$, where for each $\text{arg}(z)$, only contributions from an unique $|z|$ exist. Below, we propose the extension of this topological winding to systems with arbitrarily fragmented GBZs, and demonstrate its applicability on our $H_\text{coupled-topo}$ model.

To define any sort of winding in a system with fragmented GBZ, the GBZ solutions need to be consistently labeled by a periodic parameter $\theta$, such that they can be resolved into "bands" $z_\mu(\theta)$. Under the non-fragmented limit, $\theta$ should take the role of the lattice momentum. 
We propose defining $\theta$ via flux threading, which can be implemented by multiplying each nearest-neighbor hopping via a phase: $c^\dagger_ic_j \rightarrow e^{iN\theta}c^\dagger_ic_j$. Upon translating $\theta\rightarrow \theta + \frac{2\pi}{N}$, the spectral flow brings the set of eigenstates back to themselves, up to permutation. In practice, each eigenstate is mapped into the eigenstate at the "adjacent" momentum point (note that we are talking about OBC systems). Even though momentum bands become ill-defined with GBZ fragmentation, the $\theta$ spectral flow is always well-defined.

Hence, for a fragmented GBZ system $H(z)$ with GBZ "bands" $z_\mu(\theta)$ and winding $U(z)=\sqrt{\frac{a(z)}{b(z)}}$, we propose the winding number
\begin{eqnarray}
W&=&\frac1{2\pi i}\int_0^{2\pi}\sum_\mu \sigma_\mu |c_\mu(\theta)|^2 \frac{d}{d\theta}\left(\log\sqrt{\frac{a(z_\mu(\theta))}{b(z_\mu(\theta))}}\right)d\theta\notag\\
&=& \frac1{2\pi i } \sum_\mu \sigma_\mu\oint_{z=z_\mu(\theta)}|c_\mu(\theta)|^2  d[\log U(z_\mu(\theta))]
\label{Wsupp}
\end{eqnarray}
where $d[\log U(z_\mu(\theta))]$ picks up variations in the winding $U(z_\mu(\theta))$, and $\sigma_\mu=\pm $ is a sign correction ensures that bulk contributions to the GBZ are oriented constructively (for instance, $\sigma_\mu$ takes on opposite signs for the two $z_\mu$ contributions in the conventional GBZ of the HN or SSH model, since they are parametrized by $\theta=\pm k$.).
Importantly, $W$ is no longer quantized to be an integer, since the integral contains the composition weight $|c_\mu(\theta)|^2$ that can range from 0 to 1. For conventional (non-fragmented) GBZ cases, however, we have two GBZ solutions $z=z_\mu,z_\nu$ with equal contours, with unity combined weight $|c_\mu(\theta)|^2+|c_\nu(\theta)|^2=1$. 

\begin{figure}[h]
    \subfloat[$\Delta=0$]{\includegraphics[width=.16\linewidth]{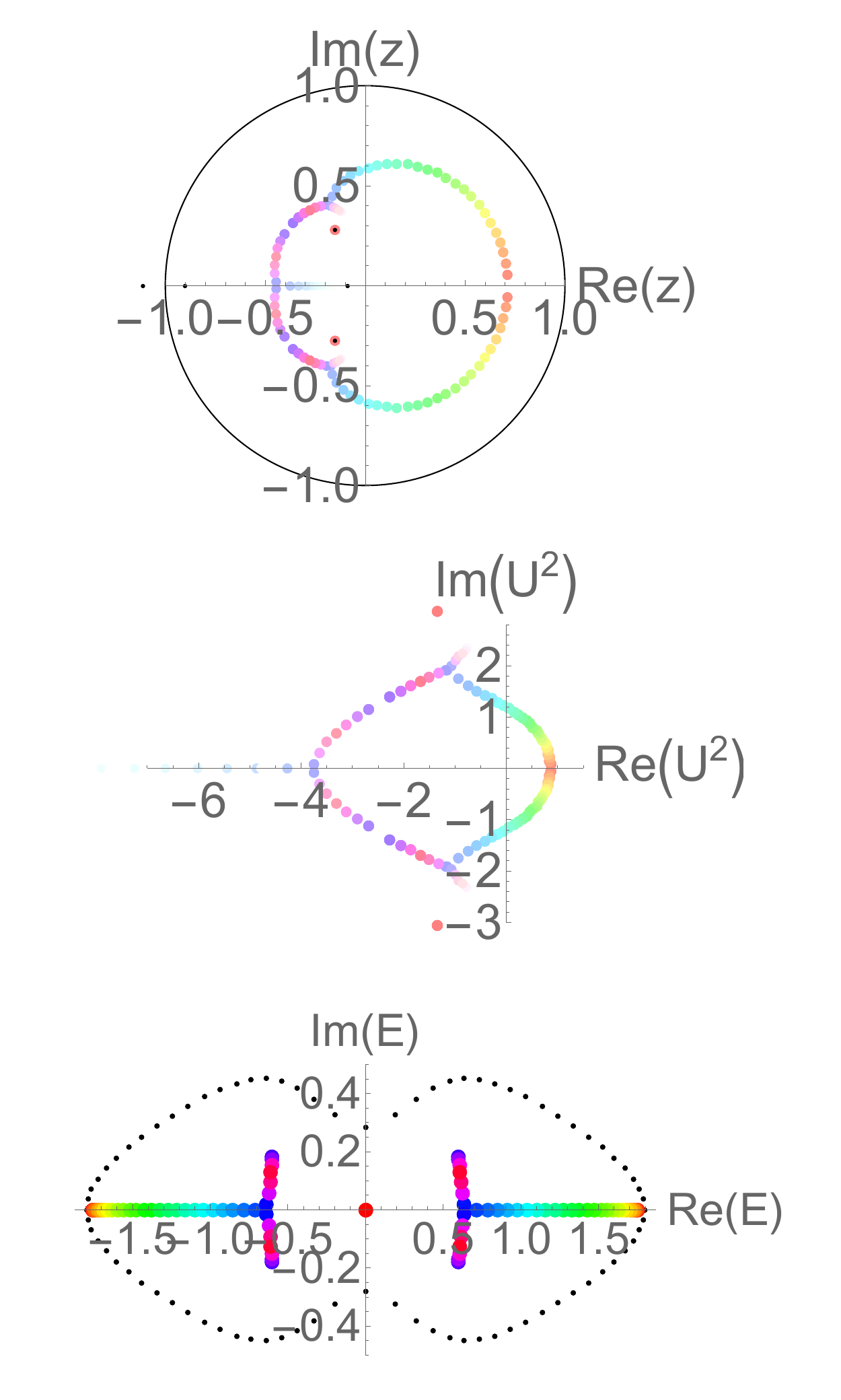}}
    \subfloat[$\Delta=10^{-12}$]{\includegraphics[width=.16\linewidth]{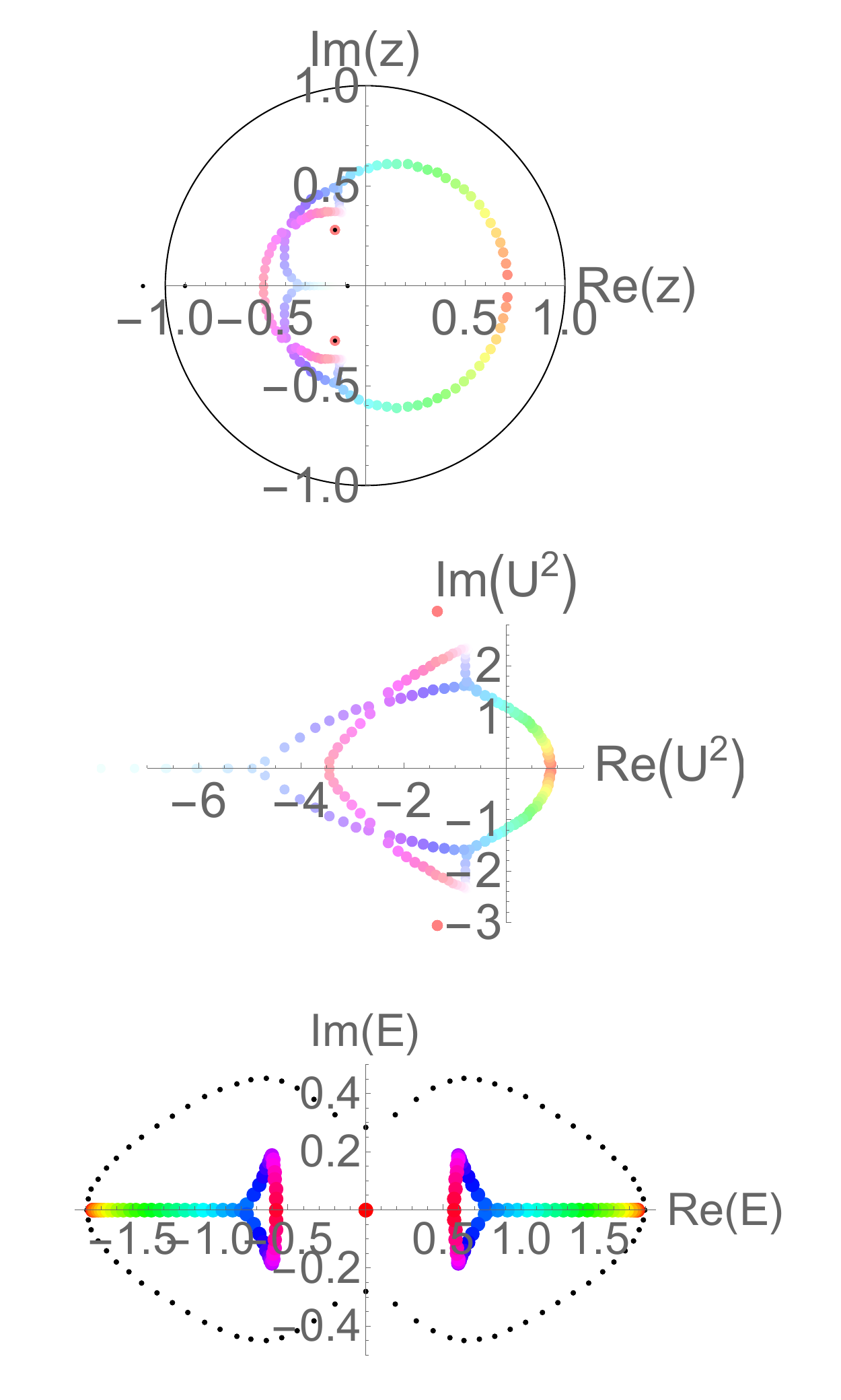}}
    \subfloat[$\Delta=10^{-10}$]{\includegraphics[width=.16\linewidth]{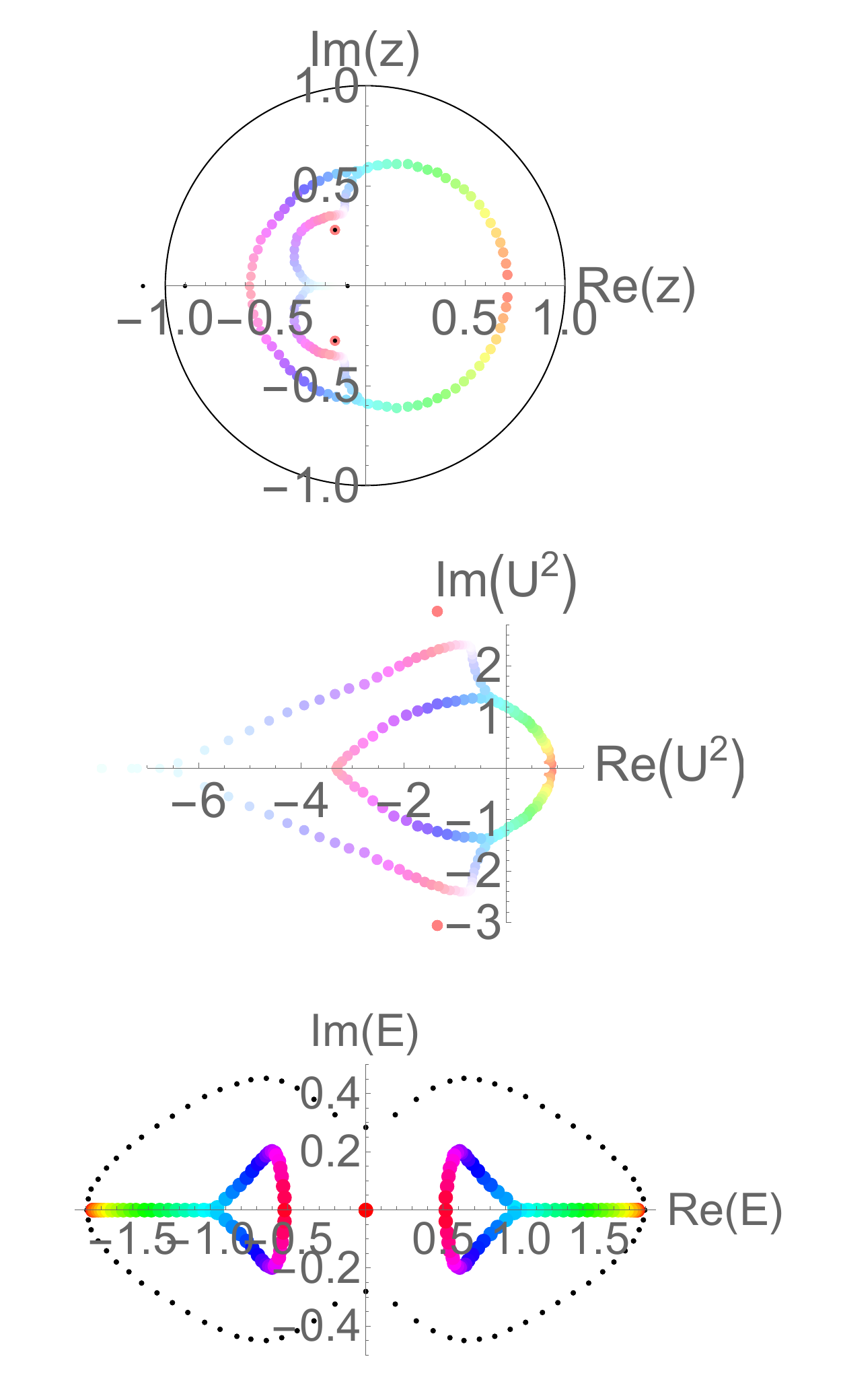}}
    \subfloat[$\Delta=10^{-7}$]{\includegraphics[width=.16\linewidth]{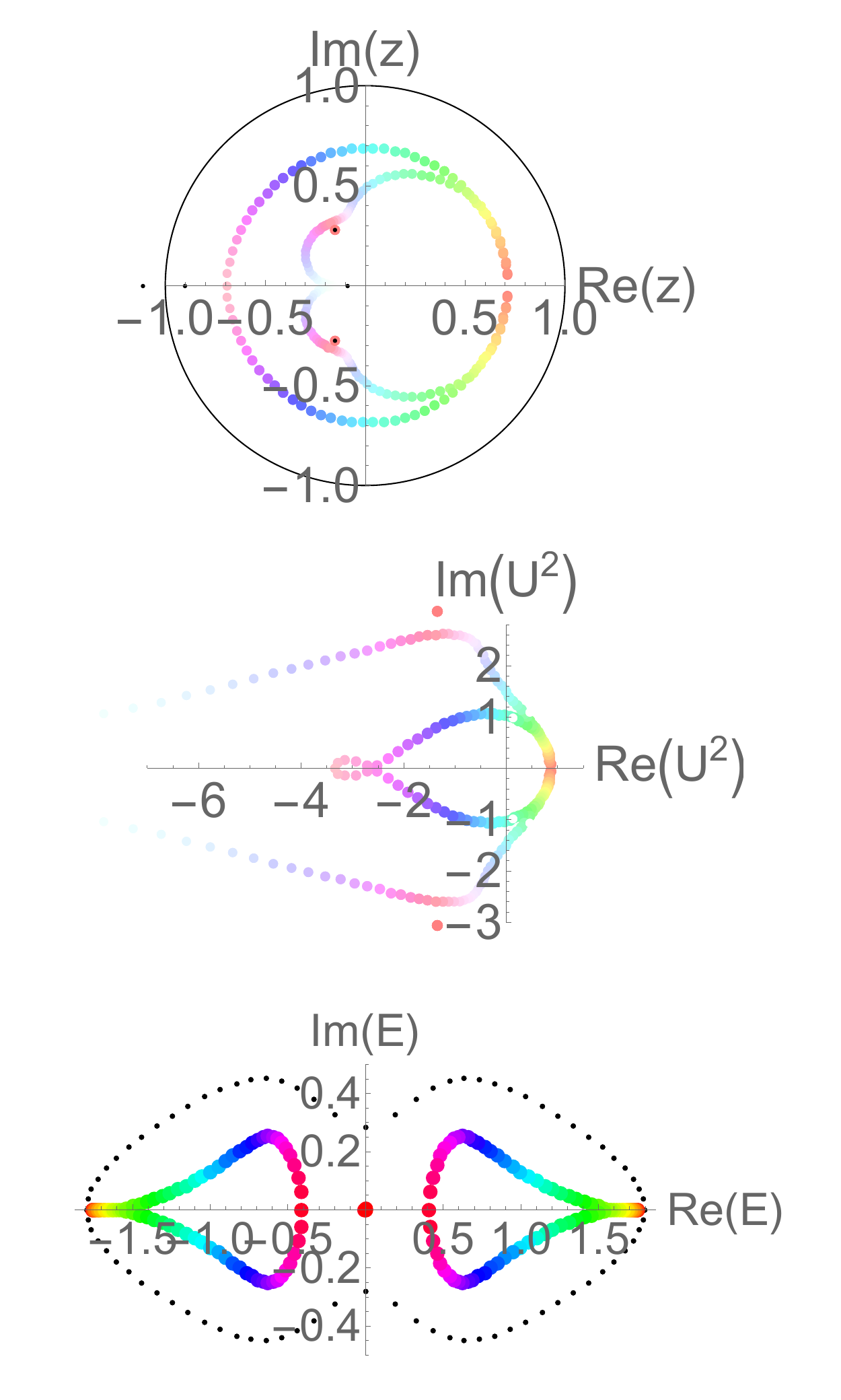}}
    \subfloat[$\Delta=10^{-5}$]{\includegraphics[width=.16\linewidth]{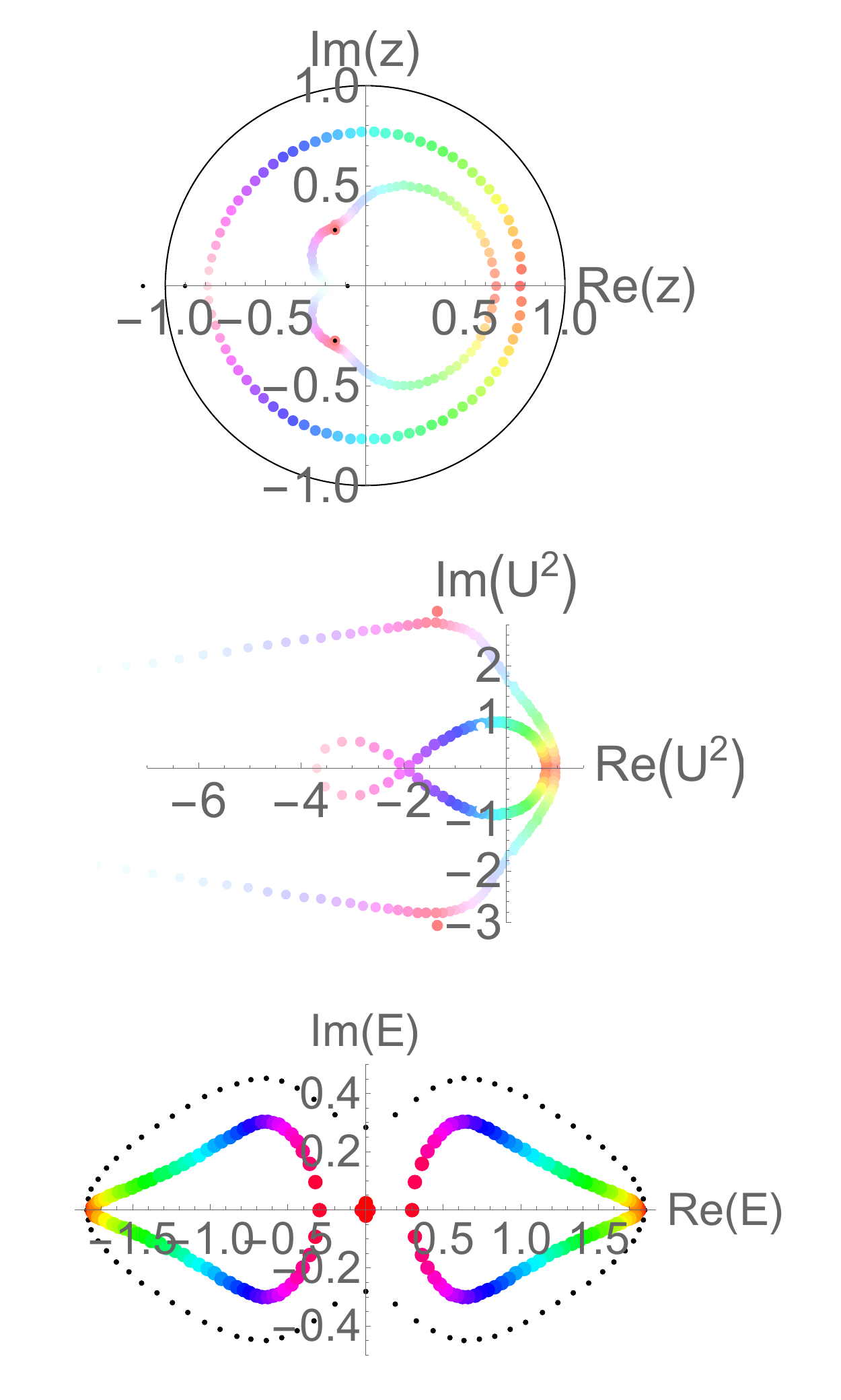}}
    \subfloat[$\Delta=10^{-2.5}$]{\includegraphics[width=.16\linewidth]{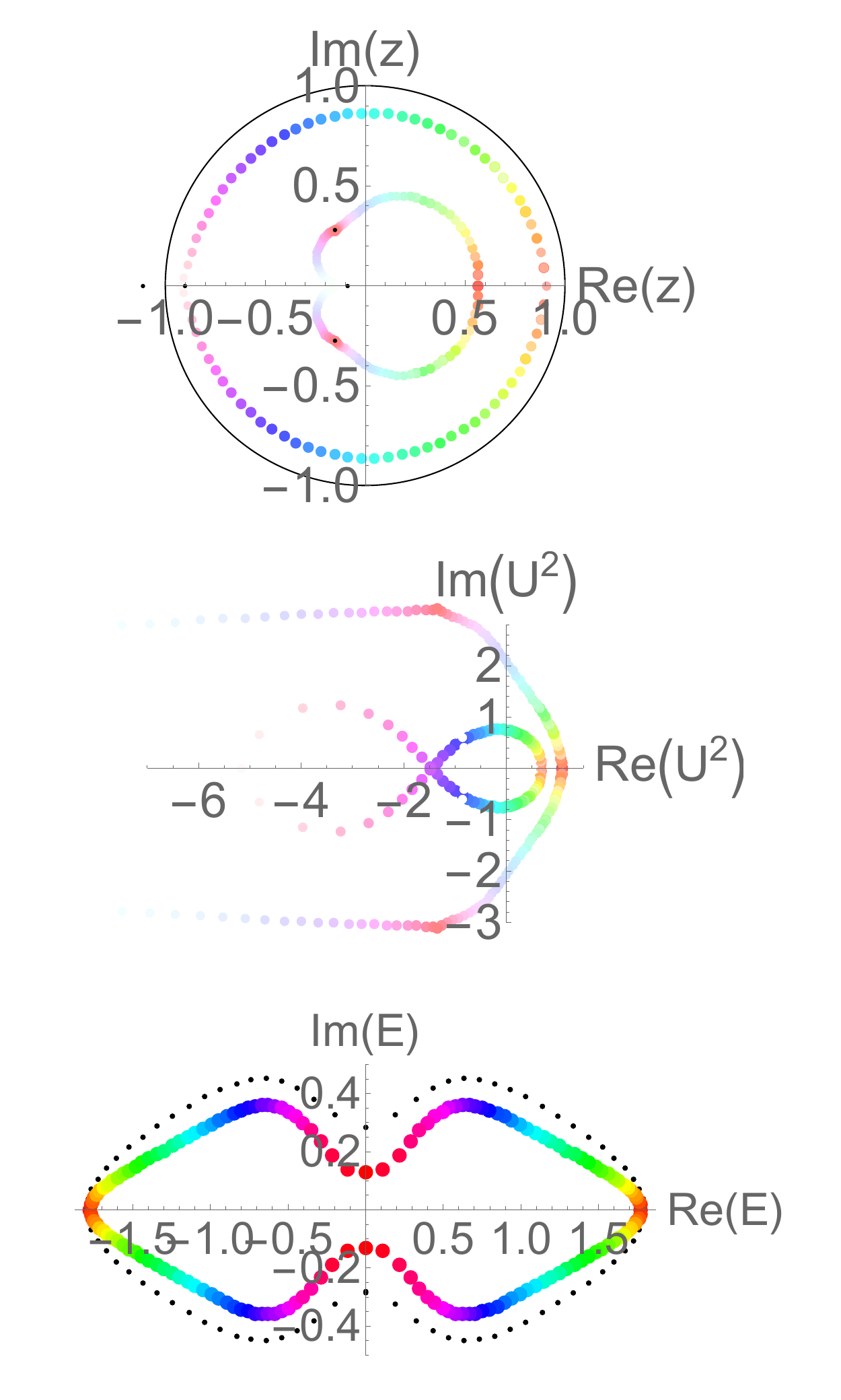}}
   \caption{Variation of the GBZ, squared winding $U^2(z)$ [Eq.~\ref{Uzsupp}] and spectrum of $H_\text{coupled-topo}$ [Eq.~\ref{Ctoposupp}], as shown in the Top, Middle and Bottom Rows respectively. Data points represent bulk contributions to the eigenstates and are colored by Re$(E)$, with intensity proportional to their $c_\mu$ weight. Black dots represent PBC eigenenergies. A continuous topological transition occurs as $\Delta$ is increased across a wide regime around $\Delta=10^{-5}$, where parts of the winding trajectory on the complex $U^2$-plane "melts" away. Hopping parameters are $t=0.6$, $t'=0.1$ with asymmetry $\gamma=2$, all computed with $N=40$ unit cells. 
    }
    \label{fig:Ctoposupp}
\end{figure}

\subsubsection{Continuous topological transition from GBZ fragmentation}

The non-quantization of the topological winding $W$ is fundamentally expected from the fuzzy occupation of the fragmented GBZ bands. This is concretely demonstrated for our $H_\text{coupled-topo}$ [Eq.~\ref{Ctoposupp}] in Fig.~\ref{Ctoposupp}, where increasing the inter-chain coupling $\Delta$ from $0$ to $\mathcal{O}(10^{-2})$ drives the OBC system from being topologically line-gapped to non-topologically point-gapped (Bottom Row). The corresponding eigenstates in the $z$ and $U^2(z)$ planes are color coded by Re$(E)$, with color intensity proportional to $c_\mu$ weight. 

Even though $H_\text{coupled-topo}$ is a 4-component model, the idea is that the non-perturbative effects of the small coupling $\Delta$ should be completely absorbed by its GBZ description. Hence, when expressed in its (fragmented) GBZ, the very small ($<\mathcal{O}(10^2)$) inter-chain couplings can be neglected, and the model is effectively $2\times2$ of the off-diagonal $H(z)$ form, and 
\begin{equation}
U(z)= \sqrt{\frac{\frac{t}{\gamma} +\frac1{z}+t'z}{t\gamma + z +\frac{t'}{z}}}.
\label{Uzsupp}
\end{equation}
From the GBZ plots (Top Row), the direction of $\theta$ generally follows the anticlockwise direction, in the order orange $\rightarrow$ yellow $\rightarrow$ green $\rightarrow$ blue. After that, the two loops exhibit distinct behaviors: while the outer loop exhibits a simple period-one winding around the $z$-plane, the inner loop actually consists of two periods of $\theta$, with the first period terminating faintly along the negative real line, such that there are 3 non-negligible $z_\mu$ solutions in total. This segment is faint because the $c_\mu(\theta)$ is small, which is necessary in facilitating the appearance of the two loops as $\Delta$ is increased from $0$ to $10^{-10}$. To see why, note that at the intermediate $\Delta=10^{-12}$ step, the middle curved triangular segment can exist independently only because it is actually a complete $\theta$ loop, albeit with very low $c_\mu(\theta)$ near the cusp along the negative real line, such that it looks like a small loop. Had GBZ fragmentation not occurred (no faint segments), the evolution from the decoupled, single-GBZ loop limit ($\Delta=0$) to the two-loop coupled regime could not have been smooth.

The role of GBZ fragmentation is even more crucial in explaining the topological transition, as evinced in the $U^2$-plane plots (Middle Row). While the topological transition only manifests in the $z$-plane plots as the absorption of the two zero modes (dots) as $\Delta$ increases, in the $U^2$-plane plots, it translates into the salient disappearance of complete $U^2$ winding. Deep in the topological regime ($\Delta=0$), $W=1$, as unmistakably seen in the well-defined (doubly degenerate) loop in the $U^2$-plane. As $\Delta$ increases, that winding initially does not change, since the splitting of that (doubly degenerate) loop into two only amounts to a relabeling of the $z_\mu$ segments. However, $\Delta$ increases, the left half of the outer loop fades away, and some parts of its right half also becomes fainter. That decreases its contribution to $W$ by more than 50\% (the winding from the inner loop remains unchanged). This continuous transition in $W$ corresponds to the gradual disappearance of the topological zero mode (Bottom Row) -- Note that because $W$ decreases continuously, it is not possible pinpoint the exact value of $\Delta$ where the topological transition occurs. This is corroborated by numerical observations that the central zero modes merge with the surrounding red bulk bands over a range of $\Delta$, and it is also not possible to specify the exact moment where the bulk gap closes.

\begin{figure}[h]
    \subfloat[$t=0.4$]{\includegraphics[width=.19\linewidth]{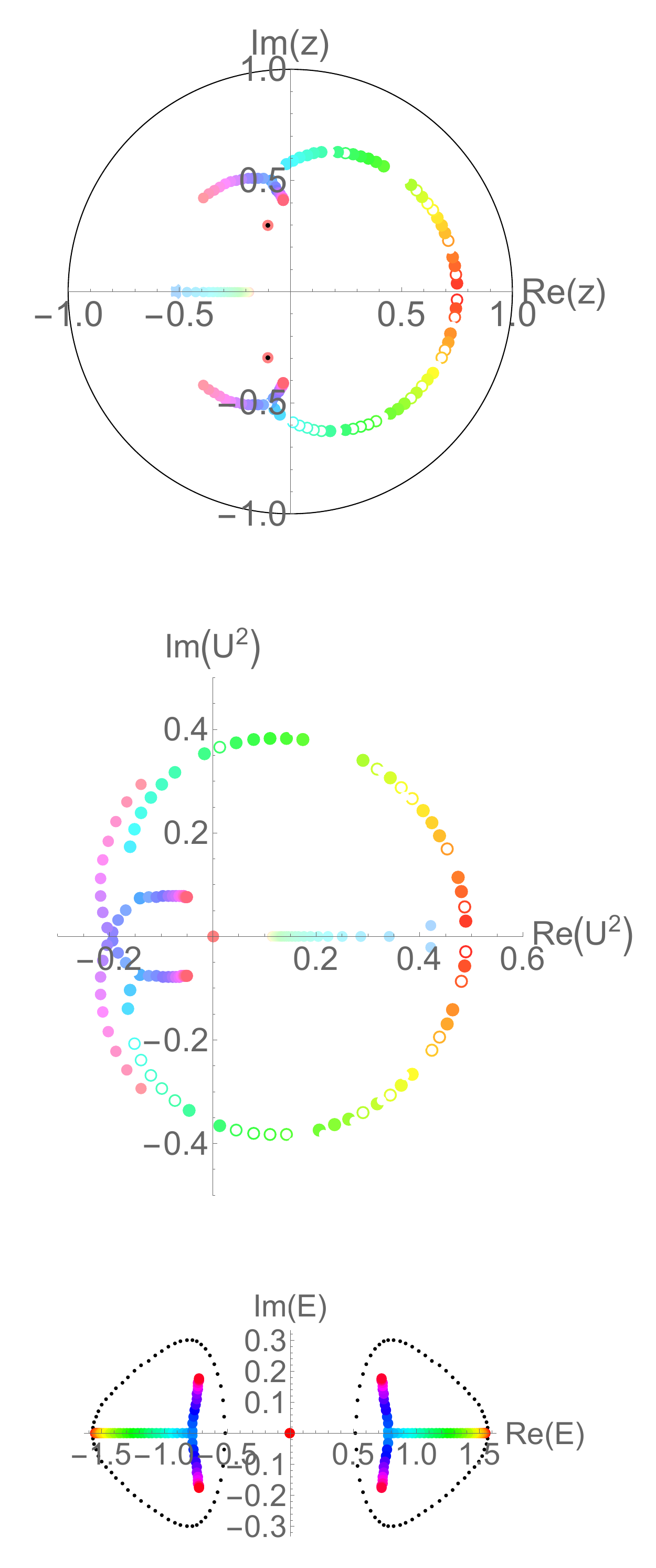}}
    \subfloat[$t=1.0$]{\includegraphics[width=.19\linewidth]{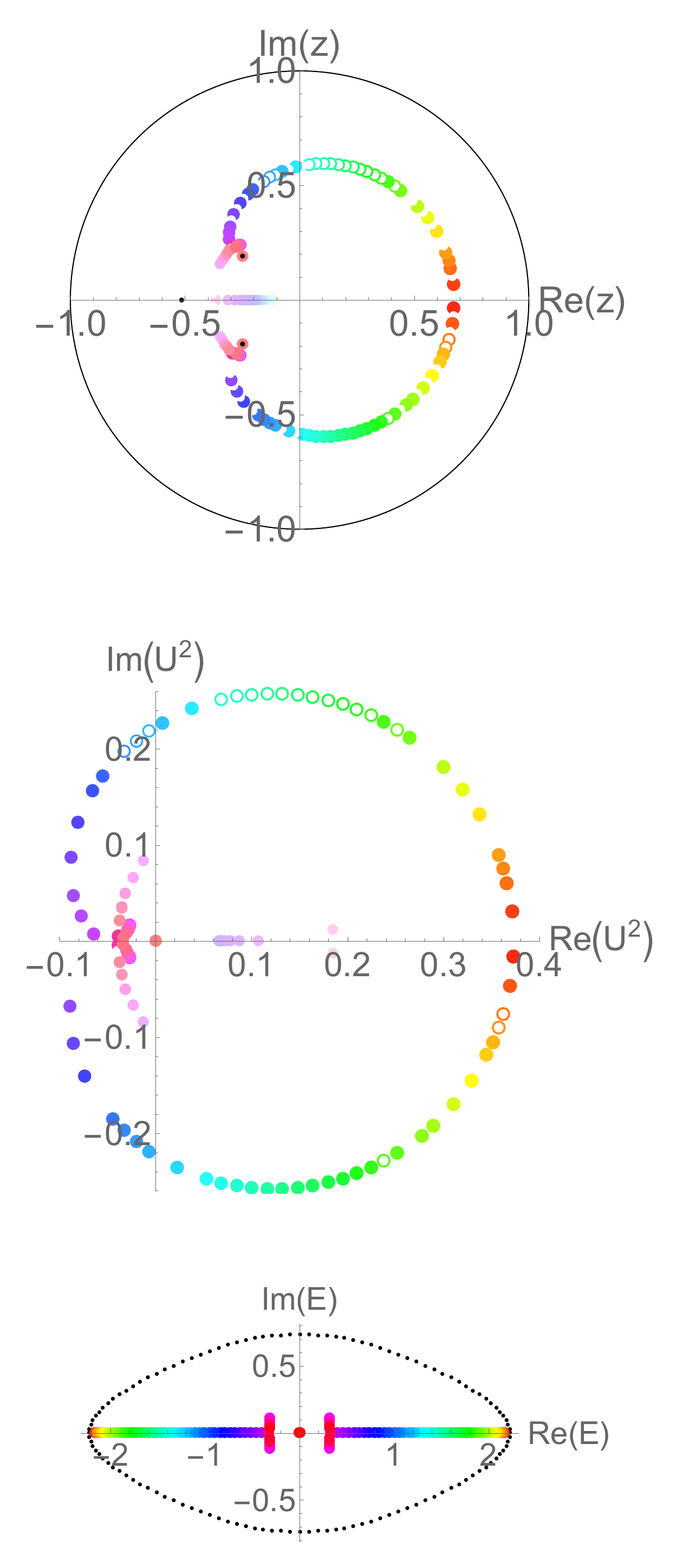}}
    \subfloat[$t=1.3$]{\includegraphics[width=.19\linewidth]{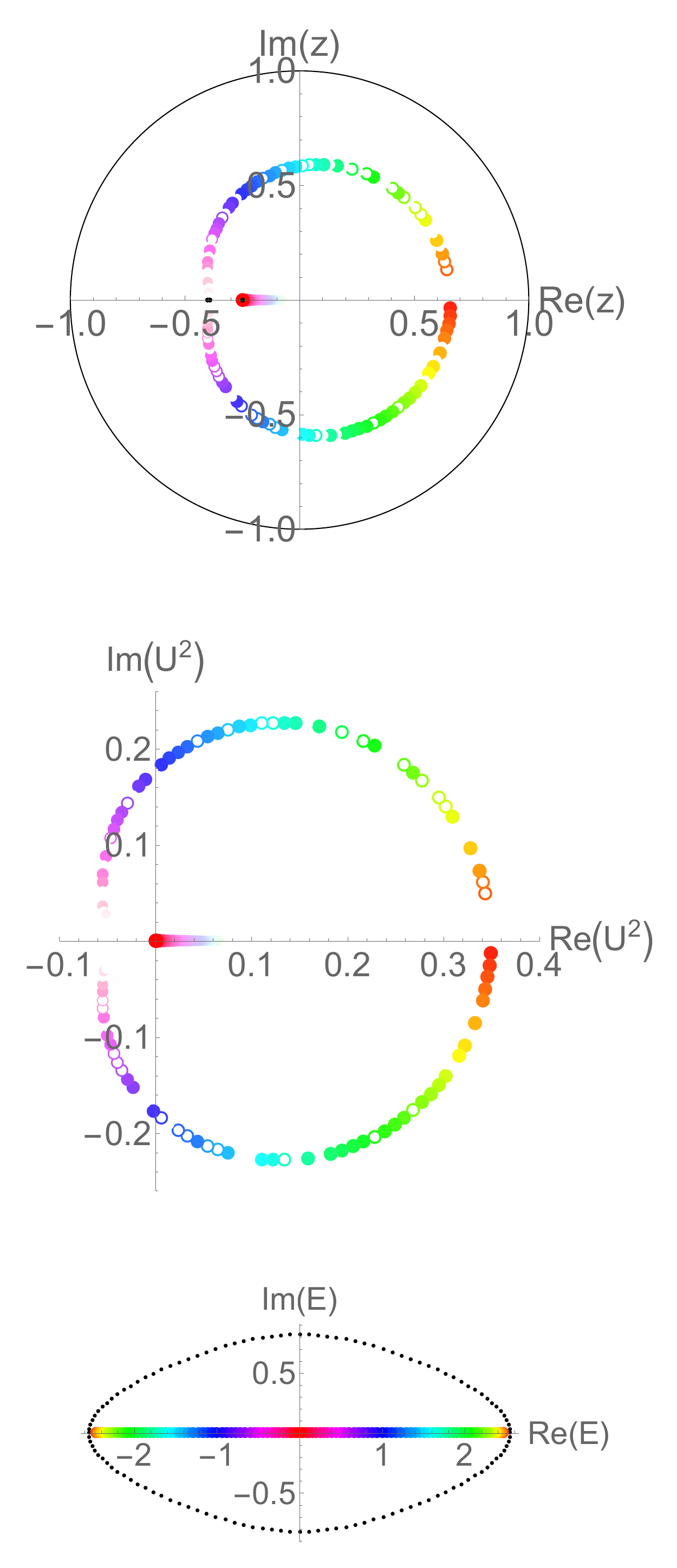}}
    \subfloat[$t=1.5$]{\includegraphics[width=.19\linewidth]{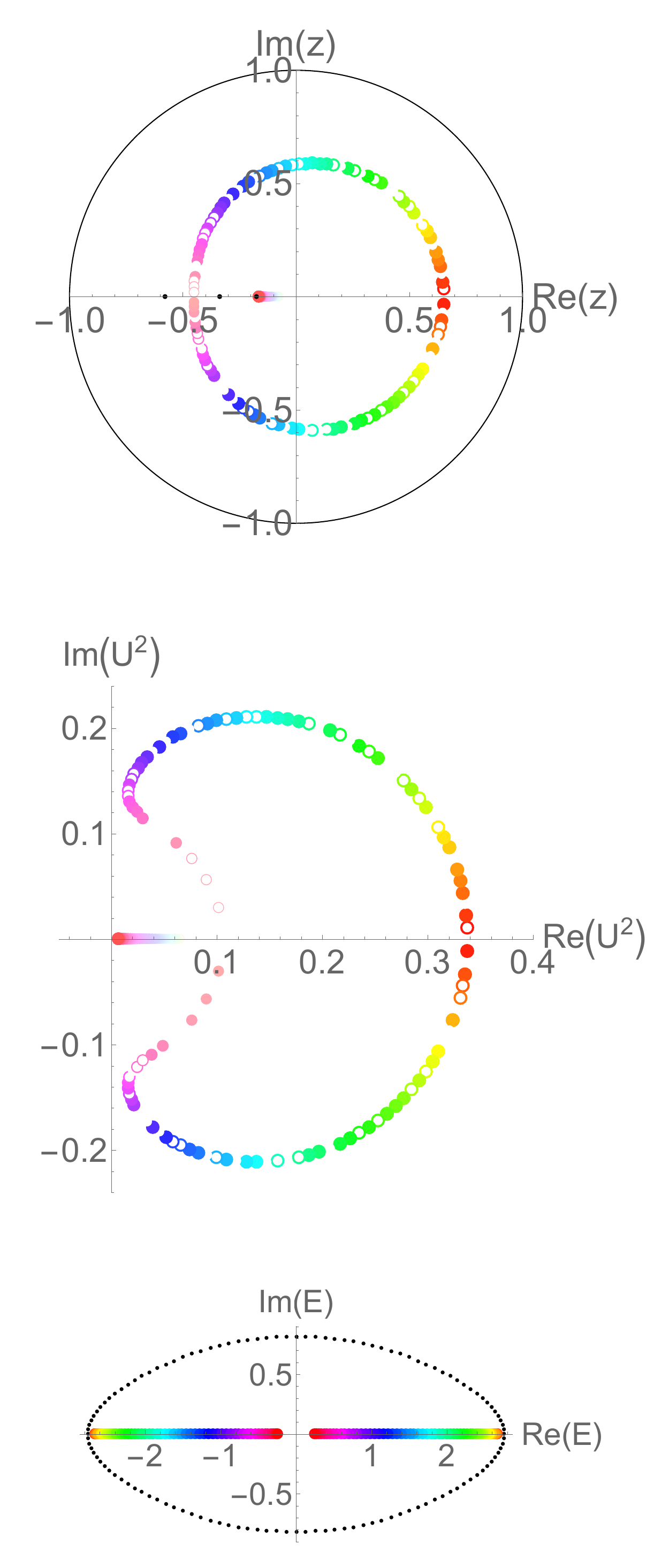}}
    \subfloat[$t=2.5$]{\includegraphics[width=.19\linewidth]{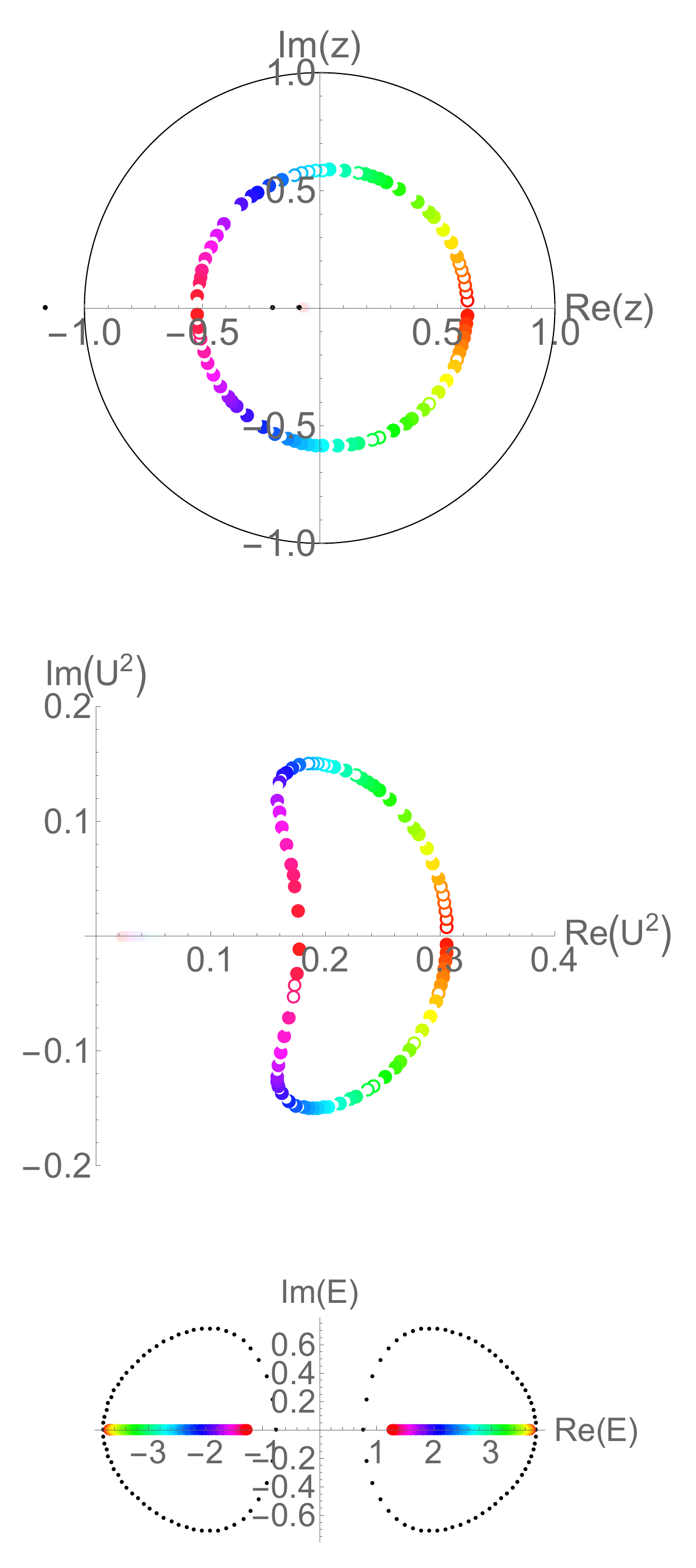}}
   \caption{
Variation of the GBZ, squared winding $U^2(z)$ [Eq.~\ref{Uzsupp}] and spectrum of $H^\text{ext-SSH}_{\gamma}$, as shown in the Top, Middle and Bottom Rows respectively. Data points represent bulk contributions to the eigenstates and are colored by Re$(E)$, with intensity proportional to their $c_\mu$ weight. Black dots represent PBC eigenenergies. Unlike in Fig.~\ref{fig:Ctoposupp2}, a discontinuous topological transition occurs as $t$ is increased beyond $1.3$, where the winding trajectory on the complex $U^2$-plane abruptly fades on the left of the origin and reappears on the right, with winding jumping from $W=1$ to $0$. Hopping parameters are $t'=0.1$ with asymmetry $\gamma=2$, all computed with $N=60$ unit cells. }
    \label{fig:Ctoposupp2}
\end{figure}

\subsubsection{Discontinuous topological transition from conventional GBZs}

It is instructive to contrast the above observations with a topological transition in the extended non-Hermitian SSH model $H^\text{ext-SSH}_{\gamma}(z) = \left[(1+t')\cos k +t\left(\gamma+\gamma^{-1}\right)/2\right]\sigma_x+ \left[(1-t')\sin k +it\left(\gamma-\gamma^{-1}\right)/2\right]\sigma_y$, which is just the uncoupled $\Delta=0$ limit of $H_\text{coupled-topo}$ whose GBZ is not fragmented (except very close to gap closure). As plotted in Fig.~\ref{fig:Ctoposupp2} analogously to Fig.~\ref{fig:Ctoposupp}, well-defined $z_\mu(\theta)$ loops in $U^2$ exist for most values of hopping parameter $t$. The topological transition clearly occurs just after $t=1.3$, when the winding changes discontinuously from $W=1$ to $W=0$ (note the double degeneracy). This abrupt transition occurs at gap closure (Bottom Row), where the leftmost part of the $U^2$ loop (Middle Row) opens up and re-emerges at Re$(U^2)\approx 0.1$. Some GBZ fragmentation (faint points) can be observed only immediately around the transition, unlike in the coupled case (Fig.~\ref{fig:Ctoposupp} Middle Row) where the GBZ loop fades away to give rise to a diminishing $W$. 

We would also like to contrast transitions in the GBZ morphology with the above-mentioned transitions in the band topology. The former correspond to transitions in the spectral graph topology~\cite{lee2020unraveling,qin2024kinked,yan2025hsg}, as from $t=0.4$ to $t=1.3$, where the segments of the GBZ can reorganize (Top Row) without a change in the $U^2$ winding $2W$. On the other hand, the latter corresponds to a change in $W$ which necessitate spectral gap closure, and may or may not be accompanied by a change in GBZ morphology.


\section{IV. Converting non-Hermitian spectra into distributions for comparison via their relative entropy}

Heavily fragmented GBZs resemble random distributions, and the extent of NHSE that they represent should be characterized statistically. 
To compare the similarity between two spectra, one can invoke the relative entropy. To do so, it is necessary to first convert them into ``probability" distributions. This can be done by replacing each eigenvalue $\omega_i$ in the 2D complex plane by a Gaussian "dot" of size i.e. radius $\delta$, and summing over all eigenvalues to obtain a positive distribution
\begin{eqnarray}
p(E) &=& \frac{\sum_j e^{-|E-\omega_j|^2/\delta^2}}{\int\sum_j e^{-|E-\omega_j^2/\delta^2}d^2E}\notag\\
&=& \frac1{\pi N_\text{tot}\delta^2}\sum_j e^{-\frac{|E-\omega_j|^2}{\delta^2}},
\label{pE}
\end{eqnarray}
where $N_\text{tot}$ is the total number of eigenvalues. Numerically, we implement these distributions on a $r\times r$ discretized grid (we call $r$ the resolution) in the spectral region of interest. For the photonic crystal spectra analyzed in this work (elaborated in the next section), this region is $\text{Re}[\omega a/2\pi c]\in [0.005,0.065]$ and $\text{Im}[\omega a/2\pi c]\in [2.3,2.7]$, where $E=\frac{\omega a}{2\pi c}$.

\begin{figure}
    \centering
    \includegraphics[width=1\linewidth]{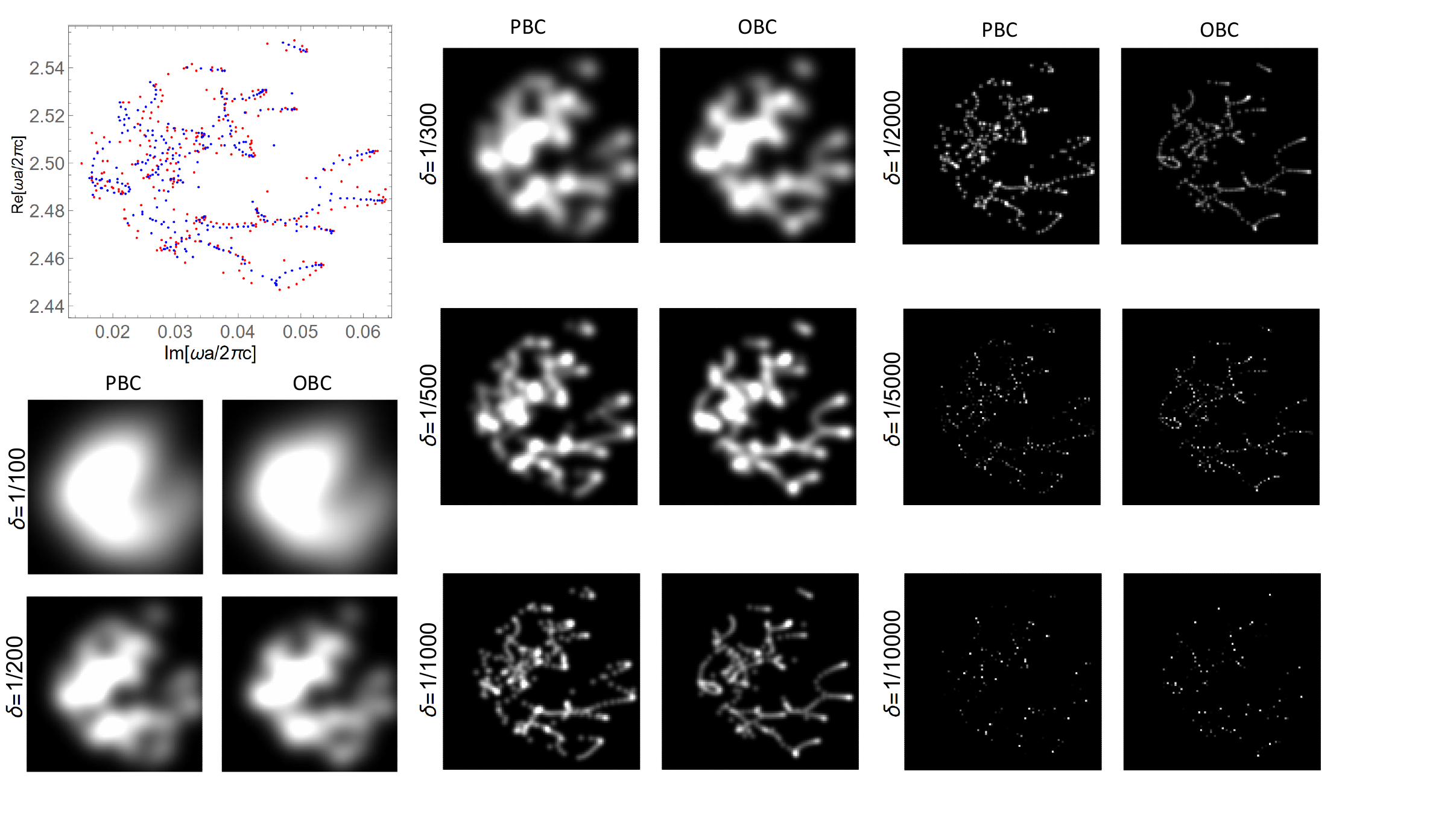}
    \caption{How the representation of complex spectra (Top Left) by its distribution $p(\omega a/2\pi c)$ [Eq.~\ref{pE}] is affected by the Gaussian dot size $\delta$. The illustrative PBC (red) and OBC (blue) photonic spectra represent the aligned dielectric configuration (see Fig S17) with $g = 1$, $N = 100$. Since the main features in this spectral plot are smaller than $0.01$, having $\delta\sim 10^{-2}$ only produces a featureless blur. Some features can be captured with $\delta=10^{-3}$, but the spectrum is only faithfully represented with a slightly finer dot size. Even smaller dot sizes of $\delta  \sim 10^{-4}$ fail to capture many of the eigenvalues unless the resolution $r$ is further increased.}
    \label{fig:droplets}
\end{figure}

Given any two distributions $p_1(E)$ and $p_2(E)$, their similarity can be quantified via the symmetrized relative entropy (also known as the symmetrized Kullback–Leibler divergence ~\cite{agrawal2019unifying,nielsen2019jensen,nielsen2021variational,fang2025efficient}) 
\begin{equation}
S=\int (p_1(E)-p_2(E))\log\frac{p_1(E)}{p_2(E)}dE.
\end{equation}
If $p_1$ and $p_2$ were identical, $S$ would vanish. Hence the magnitude of $S$ quantifies how different they are. In this work, we would be concerned with comparing the OBC and PBC spectra of various systems, such that $S$ gives a quantitative measure of the extent the NHSE i.e. bulk-boundary correspondence breaking. This provides a complementary diagnostic of the formation of a fragmented GBZ, which may not be amenable to any analytic characterization (or if the tight-binding model is not even known).  

\begin{figure}
    \centering
    \includegraphics[width=1\linewidth]{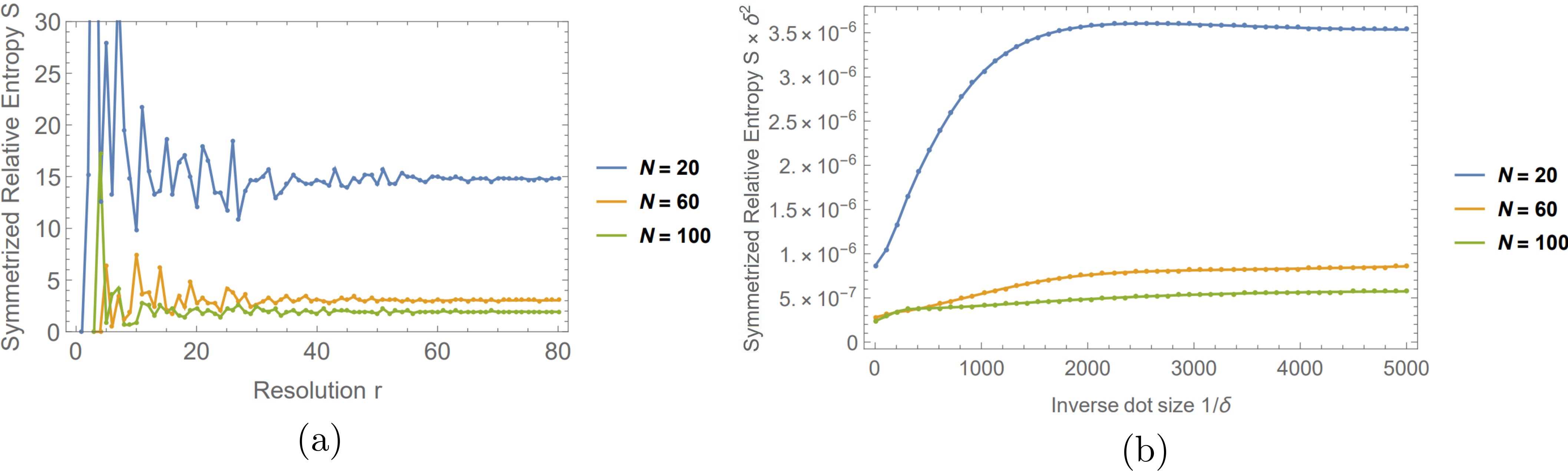}
    \caption{Calibration of $r$ and $\delta$ in the relative entropy computations. The calibration spectrum is based on the separation $g=1$ cases of the aligned photonic crystal configuration, with spectra with different number of unit cells $N$ exhibiting qualitatively similar behavior, despite differing slightly. (a) The symmetrized relative entropy $S$ fluctuates wildly at low resolutions $r$, but eventually stabilizes beyond $r=60$ for all cases. (b) The symmetrized relative entropy $S$ initially increases as Gaussian dots with smaller sizes $\delta$ represent the spectra more faithfully and better resolve their differences. However, after reaching a peak of $\delta \approx 1/2000$, $S$ starts to decrease for the $N=20$ case as some eigenvalues, particularly the isolated ones, start to disappear. This disappearance occurs earliest for the $N=20$ case because it has the fewest eigenvalues, which are hence least likely to clump together.}
    \label{fig:calibration}
\end{figure}

The chosen Gaussian dot size $\delta$ should be small enough to resolve each individual eigenvalue, or at least each spectral arc if the eigenvalues lie on a curve. However, they are lower-bound by the resolution limit set by $r$ -- below which some eigenvalues would fail to be captured. As shown in Fig.~\ref{fig:droplets} for an illustrative pair of OBC (blue) vs. PBC (red) photonic spectra (see next section), the resultant distribution become sharper as $\delta$ is decreased, optimally representing our representative spectral distribution (colored) for $\delta$ between $1/1000\sim 1/5000$. Beyond that, only the denser eigenvalue clusters survive.

Accurately representing the spectra as the distribution $p(z)$ therefore hinges on using an appropriate dot size $\delta$ and resolution $r$, which should also be kept constant across all spectra being compared. Appropriate values can be determined by computing the symmetrized relative entropy $S$ of a calibration spectrum across different $\delta$ and $r$, and observing the ranges in which $S\delta^2$ stabilizes (Note that $S\propto \delta^{-2}$ because the amount of relative information scales as the effective number of possible dot positions). Since different spectra samples may stabilize at slightly different $\delta$ and $r$, more than one calibration spectra may be needed to ensure their reliably representation. This is illustrated in Fig.~\ref{fig:calibration}, where the optimal window of $1/5000\leq \delta \leq 1/2000$ and $r>60$ was determined. In the spectra analyzed in this work, we fix these parameters to $\delta=1/2000$ and $r=100$. 

\begin{figure}
    \centering
     \includegraphics[width=1\linewidth]{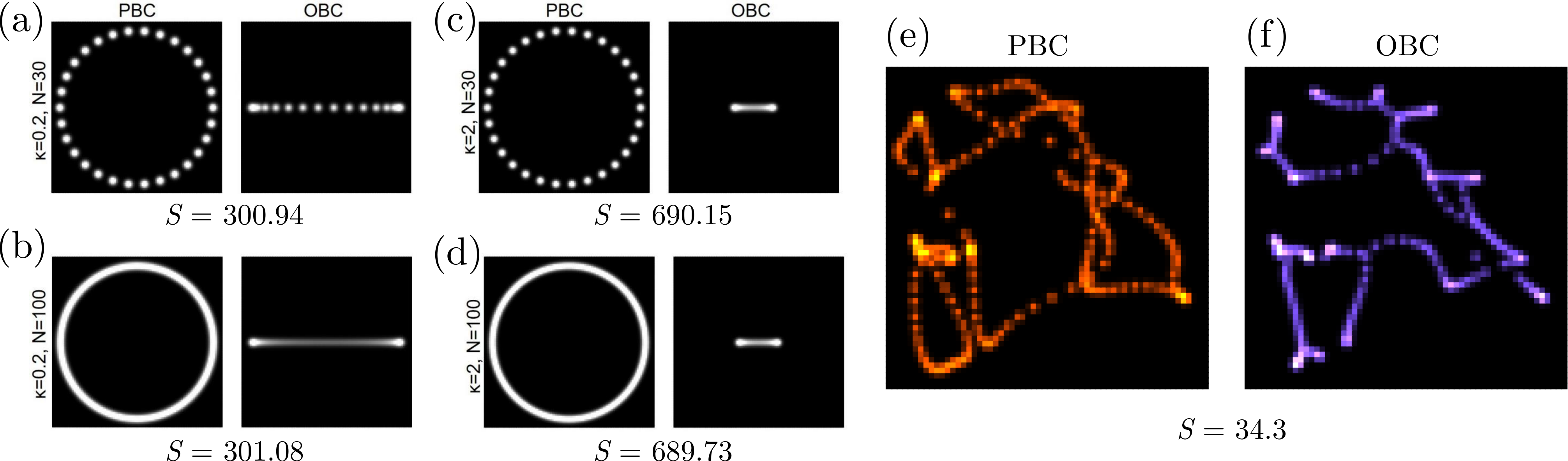}
    \caption{
    PBC and OBC spectral distributions of the Hatano-Nelson model $e^{\kappa}z+e^{-\kappa}/z$  (a-d) and a random instance of our $b=3$-component random hopping model $H_\text{rand}$ [Eq.~\ref{5bandsupp}], computed from numerical data via Eq.~\ref{pE} with Gaussian dot size $\delta=1/3$ and resolution $r=100$. These parameters should not be confused with the system size $N$, which controls the number of complex eigenvalues. In (a-d), $S$ is shown to depend only on $\kappa$ but not $N$, since the relative entropy compares the overall shapes of the spectra, rather than their discretization details. The cleanly distinct OBC and PBC spectra in (a-d) contrasts with those of (e-f), which are characteristic of a fragmented GBZ. }
    \label{fig:pz}
\end{figure}

Shown in Fig.~\ref{fig:pz} are the spectral density distributions $p(E)$ of a few illustrative systems. We immediately observe that the relative entropy $S$ is far larger in models with resolutely distinct PBC vs. OBC spectra, for instance in the Hatano-Nelson model $e^{\kappa}z+e^{-\kappa}/z$ (Left), with $S\sim 10^2$ to $10^3$. But contrast, in the random 3-component model $H_\text{rand}$ [Eq.~\ref{5bandsupp}] (Right, colored), $S$ is at least one order of magnitude lower. 

Importantly, under usual circumstances with conventional GBZ, $S$ changes negligibly with the system size, shifting by less than 1\% in the Hatano-Nelson model shown. This is because $S$ is insensitive to the discretization of the eigenvalue points, as long as the OBC spectrum does not change qualitatively with system size. To have a sense of what can change $S$, we observe that $S$ approximately doubles when the non-Hermitian asymmetry $\kappa$  is changed from $0.2$ to $2$, such that the inner OBC spectrum is much further away from the PBC spectral ring. As such, by measuring $S$, one can obtain a qualitative sense of how much the PBC and OBC spectra differ, independently of its implementation details i.e. resolution or lattice discretization.

\begin{figure}
    \centering
    \includegraphics[width=1\linewidth]{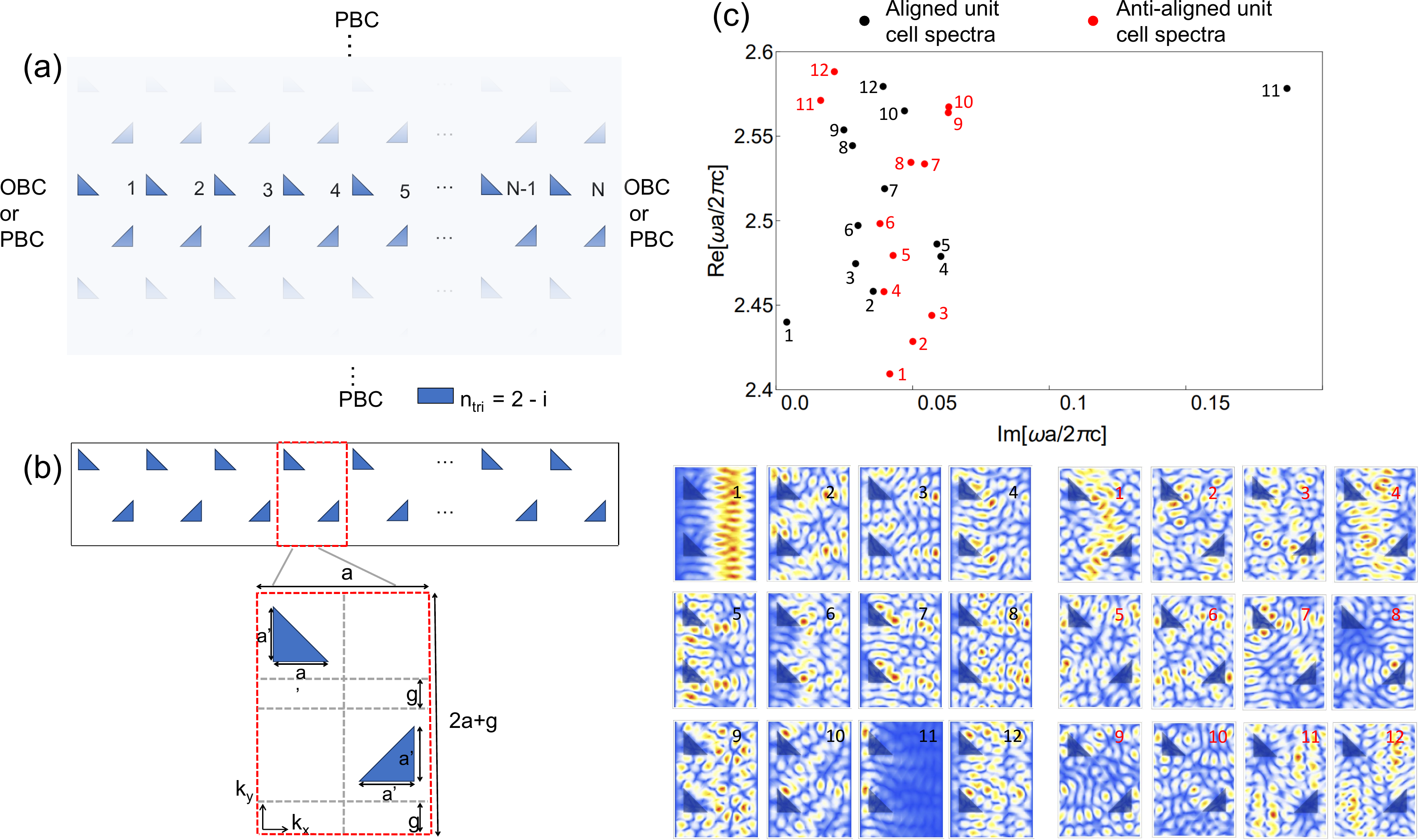}
    \caption{Physical description of our photonic crystals. 
    (a) Our photonic crystal is a 2D array of triangular dielectric rods separated by air, with $N$ rods in the x-direction and unbounded in the y-direction. Non-Hermitian loss is incorporated in the imaginary part of the refractive index, which we set as $n_{tri} = 2 - i$. OBCs, when desired, are implemented as perfect electrical conductors (PECs) in COMSOL. 
    (b) Upon enforcing PBCs in the y-direction and fixing the y-momentum (which is always set to $k_y = 0.8\pi/a$ for definiteness), our photonic crystal reduces to an effectively 1D system with two coupled chains (rows). Shown is the configuration where the dielectric structures in the two chains are anti-aligned; we also consider the aligned configuration, as in Fig.~4 of the main text. A unit cell (red) consists of two triangular structures of sides $a'=0.3a$, with an additional transverse spacing $g$ for controlling the inter-chain coupling.    
    (c) A non-exhaustive set of illustrative eigenmodes and corresponding complex eigenfrequencies for a single unit cell, for both aligned and anti-aligned configurations, with eigenmode amplitude (color) representing the transverse electric field. Taking a variety of complicated forms and not confined to the dielectrics, these eigenmodes give rise to an effectively pseudo-random lattice model with a large number of components, reminiscent of $H_\text{rand}$ [Eq.~\ref{5bandsupp}]. 
     }
\label{fig:PC_system}    
\end{figure}


\section{V. Photonic crystal setup that exhibits GBZ fragmentation}


In the text, it was established that GBZ fragmentation requires the competition between different subsystems that contain different NHSE directions or strengths. Furthermore, the fragmentation becomes more substantial and unavoidable if the unit cell becomes more complicated.

Photonic crystals thus come in as ideal candidates that can be designed to host fragmented GBZs, with their experimental maturity in engineering loss, and intrinsic existence of multiple eigenmodes per orbital that leads to complicated multi-orbital effective lattice models. 
While reciprocal photonic media with gain/loss cannot replicate the requisite asymmetric couplings in a NHSE chain, they can host "hidden" identical but oppositely-directed NHSE chains, reminiscent of reciprocal 2D NHSE setups that contain different NHSE pumpings in different transverse momentum sectors\cite{li2020topological, helbig2019band,helbig2020generalized,poddubny2024mesoscopic,zhong2023eigenenergy}. 

Employing this strategy, we design 2D photonic crystals that possess not just one but two effectively antagonistic NHSE channels that compete and lead to GBZ fragmentation. Furthermore, features in our photonic unit cells are deliberately spatially spaced out such as to support a relatively large number of competing eigenmodes that couple in a somewhat delocalized manner.

\subsection{A. Photonic crystal description}


To elucidate the role of antagonistic NHSE in GBZ fragmentation, we design photonic crystals in two different staggered configurations with aligned or anti-aligned structures that break x- and y-mirror symmetry, as sketched in Fig.~4(a) of the main text, and further elaborated in Fig.~\ref{fig:PC_system} (which explicitly shows the anti-aligned configuration). Our photonic crystal consists of a 2D array of equicrural right-angled triangular structures (blue) with complex lossy dielectric constant, which we set as $n_\text{tri}=2-i$~\cite{yokomizo2022non,zhong2021nontrivial}. As shown in Fig.~\ref{fig:PC_system}(b), each unit cell consists of two dielectric structures placed at its top left and bottom right halves; they are flipped relative to each other (anti-aligned) here, but we also consider another configuration (aligned) without the flip (see Fig.~4(a) of the main text). Each triangle is set to be of length $a'=0.3a$, where $a$ is the horizontal unit cell length i.e. lattice constant, and separated from other triangular structures by air. We also introduce an additional tunable transverse spacing $g$ between each row to control their transverse spacing.

To realize the NHSE in this otherwise reciprocal 2D medium, we employ the approach in~\cite{zhong2021nontrivial}, which is to focus on a fixed 1D momentum sector. Specifically, we stipulate the y-direction to be always periodic, and focus on the single $k_y=0.8\pi/L_y$ momentum sector, which is non-reciprocal by definition. The resultant effective 1D system, shown in Fig.~\ref{fig:PC_system}(b), consists of two rows of triangular structures which are now coupled by effectively non-reciprocal couplings. Each row is thus expected to behave like an multi-component NHSE chain, since each unit cell supports a number of distinct eigenmodes [Fig.~\ref{fig:PC_system}(c)]. As illustrated, the rows consist of anti-aligned triangular elements, and should thus mimic two coupled chains with oppositely-directed NHSE. But in the alternative aligned configuration (see Fig.~4(a) of the main text), the effective model would consist of two coupled chains with parallel-directed NHSE. Note that in the frequency range chosen in Fig.~\ref{fig:PC_system}(c), which is also examined in Figs.~\ref{fig:aligned},~\ref{fig:antialigned} and the main text, the eigenmodes exhibit significant localization outside of the triangular dielectric structures, and will thereby couple strongly with many other eigenmodes in nearby unit cells.

The effective degrees of freedom are captured by the tranverse electric (TE) eigenmodes of the photonic crystal, which is effectively a waveguide. These TE modes exhibit in-plane magnetic fields and out-of-plane electric fields, the latter which take the role of the eigenstates in the effective lattice model. We use COMSOL Multiphysics to numerically solve Maxwell's equations through finite-element methods, such as to compute the photonic band structure (spectrum) and eigenstates. Shown in Fig.~\ref{fig:PC_system}(c) are the illustrative spectra and eigenstates of a \emph{single} unit cell, for both aligned and anti-aligned configurations. The complexity of these (non-exhaustive) eigenstates suggests that the effective lattice model must contain a large number of orbital components and very complicated non-Hermitian in-orbital couplings, resembling $H_\text{rand}$ [Eq.~\ref{5bandsupp}] with large number of components $b$. To implement the $x$-OBCs, perfect electric conductor (PEC) boundary conditions are utilized along the x-direction boundaries. The PEC boundary condition in COMSOL assumes an idealized surface with perfect conductivity and without loss. Mathematically, the electric field satisfies $\mathbf{E}$ = 0 at the surface, which from Maxwell's equation implies that the tangential component of electric field vanishes at the surface i.e. $\mathbf{n} \times \mathbf{E}$ = 0, where $\mathbf{n}$ is the surface normal vector.  This condition imposes electric field antisymmetry and magnetic field symmetry across the boundary. It supports the existence of induced surface currents, which automatically balance any incoming electric currents at the boundary, including prescribed volume, surface, or edge currents.


\begin{figure}
    \centering
    \includegraphics[width=1\linewidth]{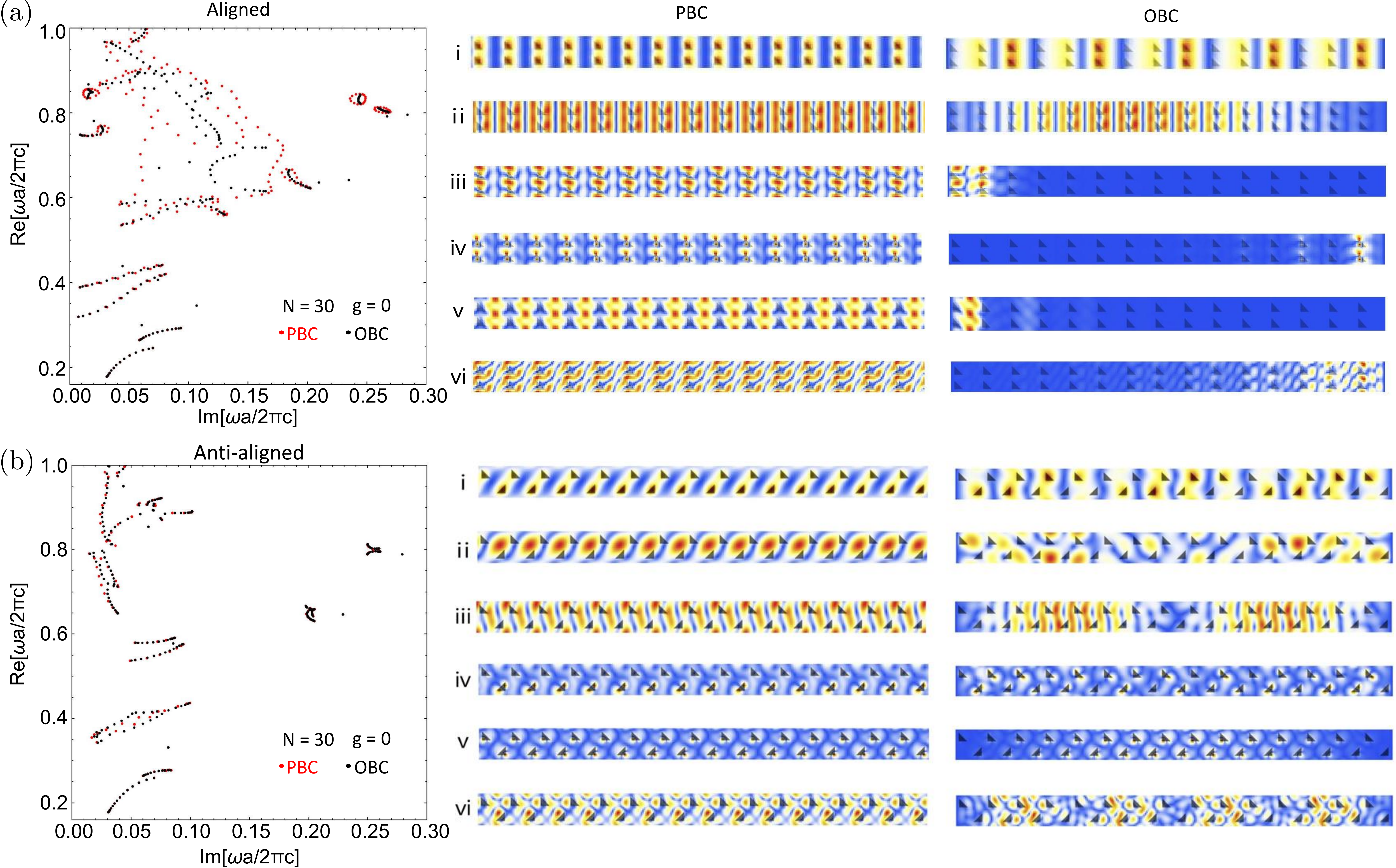}
    \caption{
    Complex band structure of our photonic crystal in the ``strongly coupled" limit under PBCs and OBCs, and the anatomy of the eigenstates of its $k_y=0.8\pi/a$. Setting $g=0$ brings the two rows of dielectric structures in the effective 1D system [Fig.~\ref{fig:PC_system}(b)] very close together, such that the NHSE survives only in the aligned case (a), almost canceling out in the anti-aligned case (b). This is reflected in the different/similar PBC (red) and OBC (black) spectra in (a)/(b). For each case, representative eigenstates numbered (i) to (vi) are plotted on the right, colored according to the transverse electric field density. OBC eigenstates that are far from any PBC eigenstate generally exhibit strong edge-skin localization. However, due to the complexity of the effective lattice models, PBC and OBC eigenstates can look very different even if they possess very similar eigenfrequencies. 
    }
\label{fig:PCspec}
 \end{figure}

\subsection{B. Photonic crystal setups under OBCs and PBCs -- spectra and eigenstates}

Here, we showcase the OBC and PBC spectra of our photonic crystal, for both aligned and anti-aligned configurations. While the complexity of the orbitals makes deriving an accurate lattice model difficult, the extent of NHSE and GBZ fragmentation can be captured by the relative entropy discussed in the previous section, as demonstrated in Fig.~4 of the main text.

\subsubsection{Occurrence and cancellation of the NHSE}

First, we establish that our photonic crystal indeed exhibits the NHSE, and that the anti-alignment of the dielectric structures really leads to an antagonistic mechanism that cancels the NHSE. To do so, we consider the $g=0$ case, where the rows of dielectric structures are placed very close apart [Fig.~\ref{fig:PC_system}(b)]. Indeed, by fixing $k_y=0.8\pi/a$, we see clearly different PBC and OBC eigenfrequency spectra in the aligned case where the triangular structures are oriented in the same direction [Fig.~\ref{fig:PCspec}(a)]. Specifically, at least at higher Re$[\omega]$, the PBC spectrum forms loops, while the OBC spectrum form arcs or trees in their interior, qualitatively similar to that from our random hopping model $H_\text{rand}$ [Eq.~\ref{5bandsupp}] with larger number of components $b$. However, this breakdown of spectral bulk-boundary correspondence does not occur in the regime of lower Re$[\omega]$ (which we shall henceforth exclude), purportedly because longer wavelength waveguide modes are less sensitive to the details of the triangular structures. That said, because of the complexity of the inter-mode couplings (unit cell connectivity of the effective lattice model), the spatial mode profile is still somewhat affected by the boundary conditions [Fig.~\ref{fig:PCspec}(a)(i) and (ii)], even though the center of mass is unaffected. However, the higher Re$[\omega]$ states (iii-vi) are clearly localized only in one direction.

In the anti-aligned case in [Fig.~\ref{fig:PCspec}(b), however, the PBC and OBC spectra are almost identical, indicating the failure of any overall non-local NHSE influence. However, the boundary conditions can still exert some influence on the eigenstates locally, except for Fig.~\ref{fig:PCspec}(b)(v) which lies in a flat band (dot). Due to the complexity of eigenmode couplings, such influences typically span across several unit cells, even if no directional NHSE pumping occurs.


\subsubsection{Variation of spectra with system length $N$ and transverse spacing $g$}

\begin{figure*}[h]
    \centering
     \includegraphics[width=1\linewidth]{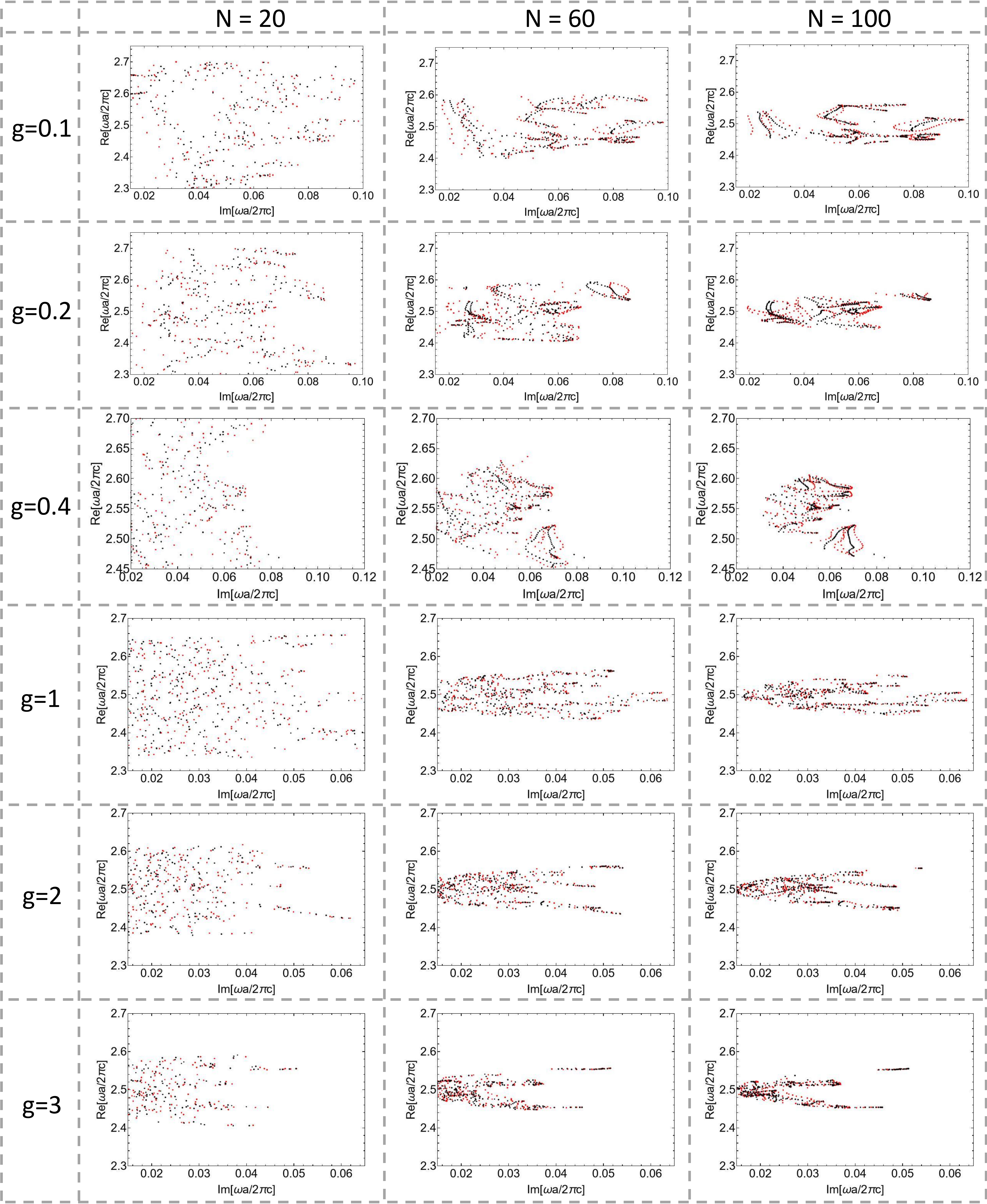}
    \caption{PBC (red) and OBC (black) complex eigenfrequency spectra of our photonic crystal with aligned triangular dielectric structures, for various system lengths $N$ and transverse spacing $g$ between the dielectric rows. Evidently, the PBC and OBC spectra are most different at large $N$ and small $g$, where the two chains couple constructively and have their NHSE enhanced most effectively. 
        }
    \label{fig:aligned}
\end{figure*}

\begin{figure*}[h]
    \centering
    \includegraphics[width=1\linewidth]{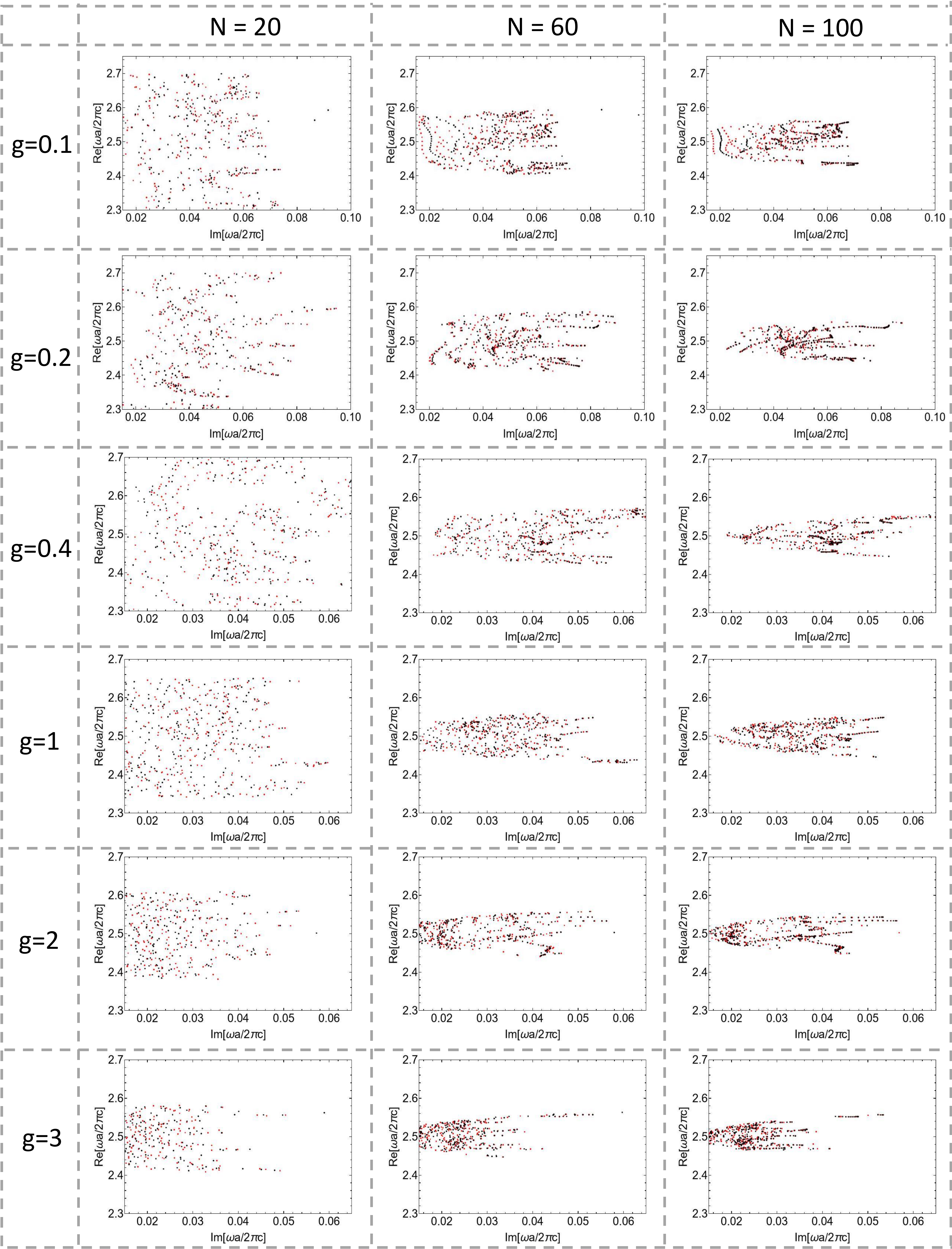}
   \caption{PBC (red) and OBC (black) complex eigenfrequency spectra of our photonic crystal with anti-aligned triangular dielectric structures, for various system lengths $N$ and transverse spacing $g$ between the dielectric rows. While the OBC and PBC spectra are evidently different at small $N=20$, by $N=100$, they have become more similar due to stronger NHSE cancellation through antagonistic mechanisms.
      }
    \label{fig:antialigned}
\end{figure*}

Finally, in Figs.~\ref{fig:aligned} and~\ref{fig:antialigned}, we vary the number of unit cells $N$ per row, such as to check whether our photonic crystal exhibits the spectral scaling characteristic of GBZ fragmentation. Also, in order to control the influence of inter-chain coupling, we vary the transverse spacing between the dielectric rows through $g$ [Fig.~\ref{fig:PC_system}(b)]. All other parameters will be kept fixed relative to lattice constant $a$, which does not affect the physics due to the scale invariance of Maxwell's equations.


Most saliently, for small spacing $g<1$, the PBC and OBC spectra differ much more in the aligned [Fig.~\ref{fig:aligned}] than in the anti-aligned [Fig.~\ref{fig:antialigned}] structure. This is particularly prominent at larger number of unit cells $N$, i.e. for $g=0.4$, $N=100$. Ostensibly, this is because the antagonistic NHSE mechanism is strongest when the chains have a long runway, and when the NHSE channels are oppositely directed. 

Note that due to the complex tangle of effective hoppings that must exist in the effective model, some antagonistic NHSE still exists even in the aligned cases, such that the PBC and OBC spectra generally becomes more similar at larger $N$. Another subtlety is that as the spacing $g$ gets larger, the PBC and OBC spectra eventually becomes more similar. This is because while increasing $g$ weakens the NHSE cancellation initially,  large $g$ also leads to more compact spectral dispersion (flatter bands) in the complex plane, which are less affected by the boundaries.


\end{document}